\documentclass[12pt]{article}
\usepackage{amsmath, xcolor}
\usepackage{graphicx,psfrag,epsf,amssymb}
\usepackage{epstopdf}
\usepackage{subfigure}
\usepackage{enumerate}
\usepackage{natbib}
\usepackage{url} 
\usepackage[pdftex,colorlinks=true,hypertexnames=false]{hyperref}
\definecolor{darkblue}{rgb}{0,0,.6}
\hypersetup{citecolor=darkblue,linkcolor=darkblue,urlcolor=darkblue}
\usepackage{threeparttable,multirow}
\usepackage{amsmath}
\usepackage{multirow}
\usepackage{diagbox}

\DeclareMathOperator*{\argmin}{arg\,min}
\usepackage{epsfig}
\usepackage{graphics}
\usepackage{palatino,eulervm}
\usepackage{amsthm,thmtools}
\declaretheorem{theorem}
\declaretheorem{lemma}
\makeatletter
\def\th@newremark{\th@remark\thm@headfont{\bfseries}}
\makeatletter
\theoremstyle{newremark}
\newtheorem{remark}{Remark}
\newtheorem{prop}{Proposition}
\newtheorem{corollary}{Corollary}
\newtheorem{assumption}{Assumption}
\declaretheoremstyle[
  spaceabove=6pt, spacebelow=6pt,
  headfont=\bfseries,
  notefont=\mdseries, notebraces={(}{)},
bodyfont=\normalfont,
  postheadspace=0.5em,
]{mystyle}

\allowdisplaybreaks[4]

\newcommand{\blind}{0}
\begin{document}

\def\spacingset#1{\renewcommand{\baselinestretch}%
{#1}\small\normalsize} \spacingset{1}


\if0\blind
{
  \title{\Large\bf Functional-Coefficient Quantile Regression for Panel Data with Latent Group Structure
  }
  \author{\normalsize Xiaorong Yang\thanks{School of Statistics and Mathematics and Collaborative Innovation Centre of Statistical Data, Engineering Technology and Application, Zhejiang Gongshang University, China.},\quad Jia Chen\thanks{Department of Economics and Related Studies, University of York, United Kingdom},\quad Degui Li\thanks{Department of Mathematics, University of York, United Kingdom},\quad Runze Li\thanks{Department of Statistics, Pennsylvania State University, United States. The corresponding author, email address: rzli@psu.edu.}\vspace{0.5cm}\\
{\small\em Zhejiang Gongshang University, University of York, Pennsylvania State University}}
\date{}
  \maketitle
} \fi

\if1\blind
{
  \bigskip
  \bigskip
  \bigskip
  \begin{center}
    {\LARGE\bf Title}
\end{center}
  \medskip
} \fi

\bigskip
\begin{abstract}

\spacingset{1.2}

This paper considers estimating functional-coefficient models in panel quantile regression with individual effects, allowing the cross-sectional and temporal dependence for large panel observations. A latent group structure is imposed on the heterogenous quantile regression models so that the number of nonparametric functional coefficients to be estimated can be reduced considerably. With the preliminary local linear quantile estimates of the subject-specific functional coefficients, a classic agglomerative clustering algorithm is used to estimate the unknown group structure and an easy-to-implement ratio criterion is proposed to determine the group number. The estimated group number and structure are shown to be consistent. Furthermore, a post-grouping local linear smoothing method is introduced to estimate the group-specific functional coefficients, and the relevant asymptotic normal distribution theory is derived with a normalisation rate comparable to that in the literature. The developed methodologies and theory are verified through a simulation study and showcased with an application to house price data from UK local authority districts, which reveals different homogeneity structures at different quantile levels.
\end{abstract}

\spacingset{1.2}
\noindent%
\vfill

\newpage
\spacingset{1.6} 
\section{Introduction}\label{sec1}
\renewcommand{\theequation}{1.\arabic{equation}}
\setcounter{equation}{0}

The quantile regression models and their estimation have received increasing attention since the seminal work by \cite{KB78}. They have been widely applied in various disciplines including economics, finance, health science and social science. In contrast to classic mean regression, the quantile regression provides a more comprehensive picture in capturing the relationship between the response and explanatory variables, and serves as a robust alternative. Various parametric methods with theoretical treatment and empirical applications have been extensively studied for quantile regression, see \cite{Ko05} and \cite{KCHP17} for comprehensive reviews. Due to wide availability of panel/longitudinal data in many areas, it is natural to extend parametric linear quantile regression from independent data to more general panel data. Subject-specific individual effects are often incorporated in linear quantile panel models to reflect location shift effects on the quantile regression and describe heterogeneity over subjects. The number of these ``incidental parameters" diverges as the number of subjects $N$ increases, affecting estimation accuracy of the common quantile regression coefficients (in particular when the number of observations per subject $T$ is fixed). Various quantile estimation and inferential techniques have been proposed in the literature \citep[e.g.,][]{Ko04, Ca11, KGM12, GLL13, GK16, GGV20} for large panel data, i.e., both $N$ and $T$ are large. However, the aforementioned literature relies on a pre-specified parametric linear model assumption, which may be too restrictive in quantile regression and is often rejected in practical data analysis. In this paper, we adopt a nonparametric panel modelling approach, allowing data to ``speak for themselves" and thus providing more reliable numerical performance in quantile regression than the parametric one.

Nonparametric quantile regression estimation has been systematically studied in the literature for independent cross-sectional data or weakly dependent time series data \citep[e.g.,][]{YJ98, Ca02, YL04, CX08, LLR13, BCCF19, LLL21}. In recent years, there have been some attempts to study nonparametric quantile regression for panel data with individual effects. For example, \cite{YL18} introduce a three-step nonparametric conditional quantile estimation method combining series approximation, first-order difference (to remove incidental parameters) and deconvolution; and \cite{Ch21} proposes two local-linear-based methods to estimate quantile partial effects and further derives the asymptotic distribution theory under the large panel setting. Among various nonparametric quantile regression models, the functional-coefficient model is one of the most commonly-used frameworks. It is a natural extension of the linear quantile regression model, avoiding the so-called ``curse of dimensionality" problem in nonparametric estimation when the number of covariates is large. For the classic independent or time series data setting, the functional-coefficient quantile model and its generalised version have been extensively studied in the literature \citep[e.g.,][]{Ho04, Ki07, CX08, WZZ09, KLZ11, TSWZ13}. In particular, the functional-coefficient quantile regression allows the dynamic quantile relationship to vary smoothly over a state variable, and is thus connected to the functional linear quantile regression \citep[e.g.,][]{K12}, but the latter assumes the covariate takes a functional value and bases the estimation methodology on some dimension reduction techniques (such as the functional principal component analysis). For the panel data with subject-specific fixed effects, \cite{SH16} combine the series approximation and instrumental variable quantile regression \citep[e.g.,][]{CH06} to estimate functional coefficients under a large-$T$ framework, whereas \cite{CCF18} use a kernel weighted quasi-likelihood to estimate semiparametric functional-coefficient quantile regression under a fixed-$T$ framework.

A typical assumption imposed on nonparametric quantile regression for panel data in the existing papers is that the main nonparametric regression structure (after removing subject-specific location shift effects) is invariant over subjects, indicating that the dynamic relationship between the dependent and explanatory variables is the same for all subjects. However, such an assumption is often too restrictive in panel data studies when subjects involved have very different characteristics. An example is the house price data from UK local authority districts that we consider in Section \ref{sec5.2}. Due to differences in their location, population, and the socio-economic backgrounds of their population, the effects of factors, such as population growth and personal income growth, on house price growth are very unlikely to be homogeneous. As a result, in this paper, we relax the homogenous panel model assumption, allowing the functional-coefficients in nonparametric quantile regression to vary over subjects. 

However, for the heterogenous functional-coefficient panel quantile regression without imposing any structural restriction on the subject-specific coefficients, we can only rely on the sample information from an individual subject to estimate the subject-specific dynamic relationship, which leads to slow convergence of the estimated functional coefficients and unstable numerical performance of the estimates in finite samples. To address this problem, we assume that there exists a latent group structure on the subject-specific functional coefficients at each quantile level. If the underlying group structure is known or can be consistently estimated, more efficient coefficient estimates can be obtained by pooling information belonging to the same group. From an empirical perspective, some panel data studies, such as \cite{PS07} and \cite{HM10}, have found group structures for the panel models they use.  For the UK house price data in Section \ref{sec5.2}, we identify two or five homogeneous groups (depending on the quantile level) for the 335 local authority districts. Such homogeneous groups may exist due to similarities in the characteristics of many districts (e.g., type of district, i.e., urban or rural, and socio-economic background of the majority of the population). Hence, it is both beneficial and reasonable to assume a group structure in some panel studies.

There has been increasing interest on estimating latent group structure in mean regression models for panel data in recent years. For example, \cite{KLZ16} use a binary segmentation technique to identify the latent group structure in linear regression models for panel data, whereas \cite{SSP16} introduce a penalised method via the so-called classifier-LASSO. \cite{VL17, VL20} propose a kernel-based classification of univariate nonparametric regression functions in panel data, which is further extended by \cite{Ch19} to estimate the group structure in time-varying coefficient panel data models. Other relevant developments can be found in \cite{BM15}, \cite{AB17}, \cite{SWJ19}, \cite{LSZZ20}, \cite{WS20} and \cite{LQZ21}. In contrast, there is sparse literature on quantile regression models for panel data with latent group structures. \cite{CLP16} study IV panel quantile regression with group-specific coefficients defined as a linear regression with observable group-level covariates, but the ``groups" in their paper are essentially the subjects in the context of this paper. \cite{ZZW19} propose an $L_1$-penalised estimation method to identify the group structure on the intercept in linear median regression; \cite{GV19} estimate linear quantile regression for panel data with a latent group structure on the subject-specific fixed effects; and \cite{ZWZ19} introduce an iterative algorithm using an idea similarly to the classic k-means clustering to estimate groups of units in panel data with heterogeneous slope coefficients. These estimation methods and algorithms rely on the parametric linear model assumption in quantile regression and cannot be directly applied to estimate the latent structure in nonparametric panel quantile regression.

In this paper, we aim to consistently estimate the group structure, the group number and the group-specific functional coefficients, all of which are allowed to vary over quantile levels. As there is no prior information on the latent groups, we start with a preliminary local linear quantile estimation of the subject-specific functional coefficients and the incidental parameter, only using the sample information from one subject. Based on the preliminary estimates of the functional coefficients, we compute the distance matrix between the subjects and subsequently use a classic agglomerative clustering algorithm to estimate the unknown group structure for the heterogenous functional coefficients. The resulting estimate is shown to be consistent (once the group number is pre-specified). Then, we introduce a simple ratio criterion to consistently estimate the group number. As the preliminary quantile estimates have rather slow convergence rates, we further propose a post-grouping local linear smoothing method to estimate the group-specific functional coefficients using the consistently estimated group structure, and derive the asymptotic normal distribution theory for the developed estimate with a convergence rate comparable to that in the literature. In the asymptotic analysis, we focus on the large panel setting with both $N$ and $T$ diverging to infinity. The panel observations are allowed to be temporally dependent and cross-sectionally correlated, relaxing the commonly-used cross-sectional independence restriction for panel quantile estimation \citep[e.g.,][]{KGM12, CCF18, Ch21}.

We apply the proposed method to the house price data from UK local authority districts over the period Q1/1997--Q4/2016 and discover different group structures at different quantiles.  At the lower quartile and median, we find more homogeneity in the effects of population and income growth on house price growth across districts, while at the upper quartile, more groups (i.e., five) are identified. By allowing the group structure to vary with the quantile level, we uncover a clearer picture about the relationship between population and income growth and house price growth across the distribution of house price growth.

The rest of the paper is organised as follows. Section \ref{sec2} introduces the model and latent group structure. Section \ref{sec3} describes the clustering algorithm and the ratio criterion for estimating the latent structure, and the post-grouping local linear quantile estimation. The technical assumptions and main asymptotic properties are provided in Section \ref{sec4}. Section \ref{sec5} reports both the simulation and empirical studies. Section \ref{sec6} concludes the paper. Proofs of the main theorems and technical lemmas, extensions of the developed methods and theory and additional simulation studies are available in a supplement.


\section{Model structure}\label{sec2}
\renewcommand{\theequation}{2.\arabic{equation}}
\setcounter{equation}{0}

Suppose that we collect the panel random observations $\left(Y_{it}, {\mathbf X}_{it}\right)$, $i=1,\cdots,N$, $t=1,\cdots,T$, and time series random observations $Z_t$, $t=1,\cdots,T$, where $Y_{it}$ and $Z_t$ are univariate and ${\mathbf X}_{it}$ is $d$-dimensional. Let $\alpha_i$ be a subject-specific effect which may be correlated with $X_{it}$ and $Z_t$. At a given quantile level $0<\tau<1$, the conditional quantile function for the $i$-th subject has the following functional-coefficient regression form:
\begin{equation}\label{eq2.1}
{\sf Q}_{\tau, i}\left(Y_{it} | {\mathbf X}_{it}, Z_t, \alpha_i\right)= {\mathbf X}_{it}^{^\intercal} {\boldsymbol\beta}_{\tau,i}(Z_t)+\alpha_{\tau,i},
\end{equation}
where ${\boldsymbol\beta}_{\tau,i}(\cdot)$ is a $d$-dimensional vector of subject-specific functional coefficients. Both ${\boldsymbol\beta}_{\tau,i}(\cdot)$ and $\alpha_{\tau,i}$ are allowed to depend on the quantile level $\tau$. It is worth stressing that model (\ref{eq2.1}) is different from the random-coefficient quantile regression model \citep{KX06}, see the discussion in Appendix C.1 of the supplement. The index variable $Z_t$ does not play any role in our model identification. The functional coefficients ${\boldsymbol\beta}_{\tau,i}(Z_t)$ capture smooth changes of the dynamic quantile relationship (over $Z_t$) between $Y_{it}$ and ${\mathbf X}_{it}$ at a fixed quantile level. Without loss of generality, we assume $Z_t$ has a compact support $[0,1]$. In practical applications, we may replace the random index variable $Z_t$ in (\ref{eq2.1}) by the fixed scaled time, $t/T$, or a variable $Z_{it}$ that varies over both $i$ and $t$, which would lead to the following functional-coefficient quantile regressions:
\begin{equation}\label{eq2.2}
{\sf Q}_{\tau, i}\left(Y_{it} | {\mathbf X}_{it}, \alpha_i\right)= {\mathbf X}_{it}^{^\intercal} {\boldsymbol\beta}_{\tau,i}(t/T)+\alpha_{\tau,i},\ \ \ \mbox{or}\ \ \ {\sf Q}_{\tau, i}\left(Y_{it} | {\mathbf X}_{it}, Z_{it},\alpha_i\right)= {\mathbf X}_{it}^{^\intercal} {\boldsymbol\beta}_{\tau, i}(Z_{it})+\alpha_{\tau, i}.
\end{equation}
With slight modification, the methodology and theory to be developed in Sections \ref{sec3} and \ref{sec4} are still applicable to the above two model variants. Models in (\ref{eq2.1}) and (\ref{eq2.2}) can be seen as an extension of the functional-coefficient/time-varying panel data models studied by \cite{LCG11}, \cite{Ch19}, \cite{SWJ19} and \cite{PW22} from mean regression to quantile regression.

In this paper, we further assume that there exists a partition of the index set $\{1,2,,\cdots,N\}$, denoted by $\boldsymbol{{\cal G}}_\tau=\{{\cal G}_1^\tau,{\cal G}_2^\tau,\cdots,{\cal G}_{R_{\tau,0}}^\tau\}$ such that
\begin{equation}\label{eq2.3}
{\cal G}_j^\tau\cap {\cal G}_k^\tau=\emptyset\ \ {\rm for} \ 1\leq j\neq k\leq R_{\tau,0},\ \ {\rm and}\ \ {\boldsymbol\beta}_{\tau, i}(\cdot)={\boldsymbol\gamma}_{\tau,j}(\cdot)\ \ {\rm for}\ i\in{\cal G}_j^\tau,
\end{equation}
where ${\boldsymbol\gamma}_{\tau,j}(\cdot)$ denotes a $d$-dimensional vector of group-specific functional coefficients that may also depend on $\tau$. Neither the group membership nor the group number is known a priori. Combining (\ref{eq2.1}) and (\ref{eq2.3}), we readily have that
\begin{equation}\label{eq2.4}
{\sf Q}_{\tau, i}\left(Y_{it} | {\mathbf X}_{it}, Z_t, \alpha_i\right)= {\mathbf X}_{it}^{^\intercal} {\boldsymbol\gamma}_{\tau,j}(Z_t)+\alpha_{\tau,i},\ \ i\in{\cal G}_j^\tau,\ \ j=1,\cdots,R_{\tau,0}.
\end{equation}
Note that the total number of unknown functional coefficients in (\ref{eq2.4}) is $dR_{\tau,0}$, which is much smaller than $dN$, the number of heterogenous functional coefficients in (\ref{eq2.1}).

Two remarks are in order here. First, the group membership $\boldsymbol{\mathcal{G}}_\tau$ and the group number $R_{\tau,0}$ are allowed to vary over $\tau$, which implies that the latent group structure can change over quantile levels. This makes the proposed quantile regression model framework more flexible and applicable than the mean regression one for practical research. Second, although we assume a group structure on the functional coefficients, ${\boldsymbol\beta}_{\tau,i}(\cdot)$, no group structure is imposed on the individual effects $\alpha_{\tau,i}$. This means that the subjects belonging to the same group are still allowed some degree of heterogeneity, as represented by their individual specific effects, albeit having the same functional slope coefficients.

The main interest of this paper lies in estimation of $\boldsymbol{{\cal G}}_\tau$, $R_\tau$ and ${\boldsymbol\gamma}_{\tau,j}(\cdot)$, $j=1,\cdots,R_{\tau,0}$. For notational simplicity, we next write ${\boldsymbol\beta}_{\tau,i}(\cdot)={\boldsymbol\beta}_{i}(\cdot)=\left[\beta_{i,1}(\cdot),\cdots,\beta_{i,d}(\cdot)\right]^{^\intercal}$, ${\boldsymbol\gamma}_{\tau,j}(\cdot)={\boldsymbol\gamma}_{j}(\cdot)=\left[\gamma_{j,1}(\cdot),\cdots,\gamma_{j,d}(\cdot)\right]^{^\intercal}$, $\alpha_{\tau,i}=\alpha_i$, $R_{\tau,0}=R_0$ and $\boldsymbol{{\cal G}}_\tau=\boldsymbol{{\cal G}}=\{{\cal G}_1,{\cal G}_2,\cdots,{\cal G}_{R_{0}}\}$, suppressing their dependence on $\tau$.


\section{Estimation methodology}\label{sec3}
\renewcommand{\theequation}{3.\arabic{equation}}
\setcounter{equation}{0}



\subsection{Preliminary local linear estimation and clustering algorithm}\label{sec3.1}

Note that model (\ref{eq2.1}) is a semiparametric functional-coefficient quantile model by treating $\alpha_i$ as an incidental parameter for fixed $i$. Assume that the unknown functional coefficients have continuous second-order derivatives. For $z\in [0,1]$, with the sample information from the $i$-th subject, we define
\begin{equation}\label{eq3.1}
	\sum_{t=1}^T\rho_\tau\left(Y_{it}-{\mathbf X}_{it}^{^\intercal}{\mathbf b}_1-a_1-(Z_t-z){\mathbf X}_{it}^{^\intercal}{\mathbf b}_2-(Z_t-z)a_2\right)K_h(Z_t-z),
\end{equation}
where $\rho_\tau(\cdot)$ is the quantile check function defined by $\rho_\tau(z)=z\left[\tau-I(z\leq0)\right]$ with $I({\cal A})$ being the indicator function of the event ${\cal A}$, $K_h(u)=K(u/h)$, $K(\cdot)$ is a kernel function and $h$ is a bandwidth. The local linear estimates $\widehat{\boldsymbol\beta}_i(z), \widehat{\boldsymbol\beta}_i^\prime(z), \widehat{\alpha}_i(z),\widehat{\alpha}_i^\prime(z)$ are obtained as the solution to minimise the objective function in (\ref{eq3.1}). 

Let ${\boldsymbol\Delta}$ be an $N\times N$ distance matrix among the true functional coefficients ${\boldsymbol\beta}_{j}(\cdot)$, $j=1,\cdots,N$. The diagonal elements of ${\boldsymbol\Delta}$ are zeros, whereas the off-diagonal elements $\Delta(j,k)$, $1\leq j\neq k\leq d$, are defined by
\begin{equation*}
	\Delta(j,k)=\int_0^1 \left\Vert {\boldsymbol\beta}_j(z)-{\boldsymbol\beta}_k(z)\right\Vert f(z) dz,
\end{equation*}
where $f(\cdot)$ is the density function of $Z_t$ and $\Vert\cdot\Vert$ denotes the Euclidean norm. With the preliminary local linear quantile estimates, we have the following estimate of $\Delta(j,k)$:
\begin{equation*}
	\widehat\Delta(j,k)=\frac{1}{T}\sum_{t=1}^T\left\Vert \widehat{\boldsymbol\beta}_j(Z_t)-\widehat{\boldsymbol\beta}_k(Z_t)\right\Vert.
\end{equation*}
With $\widehat\Delta(j,k)$, we obtain $\widehat{\boldsymbol\Delta}$, an $N\times N$ estimated distance matrix of ${\boldsymbol\Delta}$. The $(j,k)$-entry of $\widehat{\boldsymbol\Delta}$ is $\widehat\Delta(j,k)$ and the diagonal elements of $\widehat{\boldsymbol\Delta}$ are zeros. Using the estimated distance matrix, we may apply the agglomerative clustering method which has been widely used in the literature of cluster analysis \citep[e.g.,][]{ELLS11,RC12}. Recently, such a method, combined with the kernel-based smoothing technique, has been applied to estimate the homogeneity/group structure in nonparametric mean regression models \citep[e.g.,][]{Ch19, VL20, CLWZ21}. However, so far as we know, there is virtually no work on applying the kernel-based agglomerative clustering method to quantile regression models with a latent group structure. We next introduce the clustering algorithm when the group number is assumed to be $R$. 

\begin{enumerate}

\item Start with $N$ clusters, each of which corresponds to one of the $N$ subjects. Search for the smallest off-diagonal element in $\widehat{\boldsymbol\Delta}$ which is the smallest distance estimate.

\item Merge the two clusters with the smallest distance. Consequently, the cluster number reduces from $N$ to $N-1$. Update the estimated distance matrix for the $N-1$ clusters. Here the distance between two clusters ${\cal A}_1$ and ${\cal A}_2$ is calculated via the complete linkage, i.e., compute the farthest distance between an element in ${\cal A}_1$ and that in ${\cal A}_2$.

\item Repeat the previous steps with the updated distance matrix, and stop the algorithm when the number of clusters reaches $R$. 

\end{enumerate}

Let $\widehat{\cal G}_{1|R},\cdots, \widehat{\cal G}_{R |R}$ be the estimated clusters for a given group number $R$. If the true number $R_0$ is known a priori, we denote the estimated groups by $\widehat{\cal G}_r=\widehat{\cal G}_{r|R_0}$, $r=1, \cdots,R_0$, whose consistency property is given in Theorem \ref{thm:4.1}.

\subsection{Estimation of the group number}\label{sec3.2}

We next introduce a ratio criterion to consistently estimate the group number $R_0$. For a given number $R$, with the estimated groups $\widehat{\cal G}_{r|R}$ defined in Section \ref{sec3.1}, we may pool the estimated functional coefficients $\widehat{\boldsymbol\beta}_j(\cdot)$, $j\in\widehat{\cal G}_{r|R}$, and obtain the following estimate:
\[
\widehat{\boldsymbol\beta}_{r|R}(z)=\frac{1}{\left\vert\widehat{\cal G}_{r|R}\right\vert}\sum_{j\in\widehat{\cal G}_{r|R}}\widehat{\boldsymbol\beta}_j(z),\ \ r=1,\cdots,R,
\]
where $\vert {\cal A}\vert$ denotes the cardinality of a set ${\cal A}$. Then, we calculate the average deviation for $\widehat{\boldsymbol\beta}_j(\cdot)$ if the group number is assumed to be $R$:
\begin{equation}\label{eq3.2}
{\sf D}(R)= \frac{1}{TR}\sum_{r=1}^R\frac{1}{\left\vert\widehat{\cal G}_{r|R}\right\vert}\sum_{j\in\widehat{\cal G}_{r|R}}\sum_{t=1}^T\left\Vert\widehat{\boldsymbol\beta}_j(Z_t)-\widehat{\boldsymbol\beta}_{r|R}(Z_t)\right\Vert.
\end{equation}
It follows from Theorem \ref{thm:4.1} that the functional-coefficient quantile panel regression model is either correctly- or over-fitted (with probability tending to one) when $R\geq R_0$, and ${\sf D}(R)$ is thus convergent to zero. On the other hand, the model is under-fitted when $R<R_0$, and at least two groups are falsely merged. Consequently ${\sf D}(R)$ is strictly larger than a positive constant (using Assumption \ref{ass:5}(ii) in Section \ref{sec4.1}). Hence, it is sensible to determine $R_0$ via the following simple ratio criterion:
\begin{equation}\label{eq3.3}
\widehat{R}=\argmin_{1\leq R\leq \overline{R}} \frac{{\sf D}(R)}{{\sf D}(R-1)},
\end{equation}
where $\overline{R}$ is a pre-specified positive integer larger than $R_0$, and we set ${\sf D}(1)/{\sf D}(0)=1$, ${\sf D}(R)=0$ if ${\sf D}(R)$ is smaller than $\omega_{NT}$, a threshold satisfying some mild restrictions, and define $0/0\equiv1$. A similar ratio criterion is also used by \cite{LY12} and \cite{AH13} to find the number of latent factors in approximate factor models, and by \cite{LRS20} to determine the dimension of dominant sub-space in functional time series. Theorem \ref{thm:4.2} in Section \ref{sec4.2} below shows that $\widehat{R}$ is a consistent estimate of $R_0$. With $\widehat{R}$, we may extract the estimated groups $\widetilde{\boldsymbol{\mathcal G}}=\{\widetilde{\cal G}_1, \widetilde{\cal G}_2,\cdots, \widetilde{\cal G}_{\widehat{R}}\}$ by terminating the agglomerative clustering algorithm when $R=\widehat{R}$.


\subsection{Post-grouping local linear estimation}\label{sec3.3}

Note that the preliminary functional coefficient estimates defined in Section \ref{sec3.1} only make use of the sample information from one subject, resulting in a relatively slow uniform convergence rate, see Lemma A.2 in Appendix A of the supplement. The numerical performance of these estimates may be unstable in finite samples in particular when $T$ is not sufficiently large. With the consistent estimates of the group number and membership constructed in Sections \ref{sec3.1} and \ref{sec3.2}, we next propose a post-grouping local linear quantile estimation method for the group-specific functional coefficients ${\boldsymbol\gamma}_j(\cdot)$, $j=1,\cdots,R_0$, improving the convergence rate of the preliminary functional coefficient estimates. From Corollary \ref{cor:4.1} to be given in Section \ref{sec4.2}, for any $j=1,\cdots,R_0$, there exists $1\leq j_\ast\leq \widehat{R}$ such that ${\sf P}\left({\cal G}_j=\widetilde{\cal G}_{j_\ast}\right)\rightarrow1$. Without loss of generality, we let $j_\ast=j$ in the rest of the section. Define the post-grouping local linear weighted objective function:
\begin{equation}\label{eq3.4}
	\sum_{i\in\widetilde{\cal G}_j}\sum_{t=1}^T\rho_\tau\left(Y_{it}-{\mathbf X}_{it}^{^\intercal}{\mathbf b}_1-a_{i1}-(Z_t-z){\mathbf X}_{it}^{^\intercal}{\mathbf b}_2-(Z_t-z)a_{i2}\right)K_{h_1}(Z_t-z),
\end{equation}
where $K_{h_1}(z)=K(z/h_1)$, $K(\cdot)$ is the kernel function and $h_1$ is a bandwidth which may be different from $h$ used in the preliminary local linear estimation. The post-grouping local linear estimates $\widetilde{\boldsymbol\gamma}_j(z), \widetilde{\boldsymbol\gamma}_j^{\prime}(z),\widetilde{\alpha}_i(z), \widetilde{\alpha}_i^\prime(z)$, $i\in\widetilde{\cal G}_j$, are obtained as a solution to minimise the objective function in (\ref{eq3.4}). As the fixed effects $\alpha_i$ are treated as ``nuisance parameters", our primary interest lies in $\widetilde{\boldsymbol\gamma}_j(z)$ whose asymptotic distribution theory will be derived in Section \ref{sec4.2} below.


\section{Main asymptotic theory}\label{sec4}
\renewcommand{\theequation}{4.\arabic{equation}}
\setcounter{equation}{0}



\subsection{Technical assumptions}\label{sec4.1}

For $i=1,\cdots,N$, we let
\begin{equation}\label{eq4.1}
	{\boldsymbol\Omega}_{i}(z)=f(z)\cdot{\sf E}\left[f_{ie}(0|{\mathbf X}_{it},Z_t)\left(
	\begin{array}{c}
		1\\
		{\mathbf X}_{it}
	\end{array}\right)
	\left(1,{\mathbf X}_{it}^{^\intercal}\right)\big|Z_t=z\right],
\end{equation}
when $f(\cdot)$ is the density of $Z_t$ and $f_{ie}(\cdot |{\mathbf x},z)$ is the conditional density of $e_{it}=Y_{it}-{\mathbf X}_{it}^{^\intercal} {\boldsymbol\beta}_{i}(Z_t)-\alpha_i$ given ${\mathbf X}_{it}={\mathbf x}$ and $Z_t=z$. Assumptions \ref{ass:1}--\ref{ass:5} below are sufficient to prove the consistency properties for the estimated group membership and number.

\begin{assumption}\label{ass:1}
	
{\em For each $i$, the process $\{(Y_{it}, {\mathbf X}_{it}, Z_t) \}$ is stationary and $\alpha$-mixing dependent with the mixing coefficient $\alpha_i(\cdot)$ satisfying $\max_{1\leq i\leq N}\alpha_i(s)\asymp \rho^s$, $0<\rho<1$. }
	
\end{assumption}	
	
\begin{assumption}\label{ass:2}

{\em (i) The conditional density function $f_{ie}(\cdot |{\mathbf x},z)$ is continuous and has a bounded first-order derivative. In addition, $f_{ie}(0|{\mathbf x},z)$ is continuous with respect to $z$ and satisfies that}
		\[\max_{1\leq i\leq N}\sup_{{\mathbf x},z}f_{ie}(0|{\mathbf x},z)\leq\overline{c}_e<\infty.\]
	
{\em (ii) The density function $f(\cdot)$ has continuous first-order derivative, and is bounded away from zero and infinity. }

\end{assumption}
	
\begin{assumption}\label{ass:3}

{\em (i) The matrix ${\boldsymbol\Omega}_i(z)$ defined in (\ref{eq4.1}) is continuous (with respect to $z$) and positive definite with all the eigenvalues bounded away from zero and infinity uniformly over $z\in[0,1]$ and $1\leq i\leq N$. Furthermore,}
		\begin{equation}\label{eq4.2}
			\max_{1\leq i\leq N}{\sf E}\left[\Vert {\mathbf X}_{it}\Vert^{\kappa+\epsilon} |Z_t\right]<\infty\ \ a.s.,\ \ 4<\kappa<\infty,\ \ \epsilon>0.
		\end{equation}
	
{\em (ii) The subject-specific coefficient functions ${\boldsymbol\beta}_i(\cdot)$ are twice continuously differentiable. In addition, there exists a positive constant $c_\beta$ such that
		\[\max_{1\leq i\leq N}\sup_{0\leq z\leq 1}\left\Vert{\boldsymbol\beta}_{i}^\prime(z)\right\Vert+\max_{1\leq i\leq N}\sup_{0\leq z\leq 1}\left\Vert{\boldsymbol\beta}_{i}^{\prime\prime}(z)\right\Vert\leq c_\beta,\]
		where ${\boldsymbol\beta}_{i}^{\prime}(z)$ and ${\boldsymbol\beta}_{i}^{\prime\prime}(z)$ are the first-order and second-order derivatives of ${\boldsymbol\beta}_{i}(z)$, respectively. Similar conditions also hold for the group-specific coefficient functions ${\boldsymbol\gamma}_j(\cdot)$.}
	
\end{assumption}

\begin{assumption}\label{ass:4} 

{\em (i) $K(\cdot)$ is a bounded, Lipschitz continuous and symmetric probability density function with a compact support $[-1,1]$.}
	
{\em (ii) The bandwidth $h$ satisfies that
		\begin{equation}\label{eq4.3}
			h^5=o\left(\frac{\log (T\vee N)}{T}\right),\ \ \frac{Th}{(NT)^{4/\kappa}\log^5 (N\vee T)}\rightarrow\infty,
		\end{equation}
		where $\kappa$ is defined in Assumption \ref{ass:3}(i). In addition, assume that}
	\begin{equation}\label{eq4.4}
		\xi_{NT}^2:=\frac{\log (N\vee T)}{Th}=o\left(\zeta_{NT}^2\right),\ \ \zeta_{NT}=\min_{1\leq j\neq k\leq R_\tau}\int_0^1 \left\Vert{\boldsymbol\gamma}_{j}(z)-{\boldsymbol\gamma}_{k}(z)\right\Vert f(z)dz.
	\end{equation}

\end{assumption}

\begin{assumption}\label{ass:5}	
	
{\em (i) Let the group number $R_0$ be fixed and there exist $0<\underline{c}_g\leq \overline{c}_g<1$ such that}
	\[\underline{c}_g N\leq \min_{1\leq r\leq R_0}\left\vert {\cal G}_r\right\vert \leq \max_{1\leq r\leq R_0}\left\vert {\cal G}_r\right\vert\leq \overline{c}_gN.\]
	
{\em (ii) For any $k$ different (true) groups ${\cal G}_{r_1},\cdots, {\cal G}_{r_k}$, when they are falsely merged, define
		\[\overline{\boldsymbol\gamma}_\ast(z)=\frac{1}{\left\vert {\cal G}_{r_1}\bigcup\cdots\bigcup{\cal G}_{r_k}\right\vert}\left[\left\vert {\cal G}_{r_1}\right\vert{\boldsymbol\gamma}_{r_1}(z)+\cdots+\left\vert{\cal G}_{r_k}\right\vert {\boldsymbol\gamma}_{r_k}(z)\right]. \]
		There exist $j\in\{r_1,\cdots,r_k\}$ and a positive constant $c_\ast$ such that}
		\[\int_0^1\left\Vert \overline{\boldsymbol\gamma}_\ast(z)-{\boldsymbol\gamma}_j(z)\right\Vert f(z)dz>c_\ast.\]
	
{\em (iii) Let $\omega_{NT}$ satisfy that $\omega_{NT}=o(1)$ and $\xi_{NT}=o(\omega_{NT})$, where $\xi_{NT}$ is defined in (\ref{eq4.4}).}

\end{assumption}

\renewcommand{\theremark}{4.\arabic{remark}}\setcounter{remark}{0}

\begin{remark}\label{re:4.1} 

Most of the above regularity conditions are mild and justifiable. Assumption \ref{ass:1} shows that the panel data are temporally dependent over $t$. The $\alpha$-mixing dependence is one of the weakest mixing dependence conditions, which is satisfied for some commonly-used time series models (such as a vector ARMA process). The smoothness conditions on the (conditional) density functions and functional coefficients in Assumptions \ref{ass:2} and \ref{ass:3} are needed due to application of the local linear smoothing technique to estimate the unknown functions in quantile regression \citep[e.g.,][]{CX08}. The relatively strong moment condition in Assumption \ref{ass:3}(i) is crucial to derive the uniform Bahadur representation and uniform consistency for the local linear quantile estimates, see Lemmas A.1 and A.2 in Appendix A of the supplement. It is worthwhile to point out that when $\kappa$ is larger (indicating a stronger moment condition on ${\mathbf X}_i$), we may relax the bandwidth restriction and allow $N$ to diverge at a faster polynomial rate of $T$. Letting $h\propto T^{-1/5}$, $N$ diverge at a polynomial rate of $T$ and $\kappa$ be sufficiently large, we may show that the two conditions in (\ref{eq4.3}) are satisfied. Similar to Assumption 4(iii) in \cite{Ch19} and Assumption 4(ii) in \cite{CLWZ21}, the condition (\ref{eq4.4}) indicates that the minimum Euclidean distance between distinct coefficient functions is allowed to converge to zero.  When $\zeta_{NT}>\underline c_\ast>0$, (\ref{eq4.4}) would be automatically satisfied. Assumption \ref{ass:5} is mainly used to derive the consistency for the group number estimate stated in Theorem \ref{thm:4.2}. Assumption \ref{ass:5}(i) shows that the latent groups have similar sizes (with the same divergence rate), whereas Assumption \ref{ass:5}(ii) is crucial to prove that $D(R)$ defined in (\ref{eq3.2}) would be strictly larger than a positive constant when the model is under-fitted (i.e., $R<R_0$). In fact, Assumption \ref{ass:5}(ii) can be verified by using Assumption \ref{ass:5}(i) and assuming $\zeta_{NT}>\underline c_\ast>0$. 

\end{remark}

\medskip

With the latent structure (\ref{eq2.3}), we write $e_{it}=Y_{it}-{\mathbf X}_{it}^{^\intercal}{\boldsymbol\gamma}_j(Z_t)-\alpha_i$ for $i\in{\cal G}_j$ and let
\[b_{it}(z)={\mathbf X}_{it}^{^\intercal}\left[{\boldsymbol\beta}_i(Z_t)-{\boldsymbol\beta}_i(z)-{\boldsymbol\beta}_i^\prime(z)(Z_t-z)\right]={\mathbf X}_{it}^{^\intercal}\left[{\boldsymbol\gamma}_j(Z_t)-{\boldsymbol\gamma}_j(z)-{\boldsymbol\gamma}_j^\prime(z)(Z_t-z)\right].\] 
Re-write ${\boldsymbol\Omega}_{i}(z)$ defined in (\ref{eq4.1}) in the block-matrix form:
\begin{equation}\label{eq4.5}
	{\boldsymbol\Omega}_{i}(z)=\left(
	\begin{array}{cc}
		\omega_i^\alpha(z)&{\boldsymbol\Omega}_{i}^{\alpha\gamma}(z)\\
		{\boldsymbol\Omega}_{i}^{\gamma\alpha}(z)& {\boldsymbol\Omega}_{i}^\gamma(z)
	\end{array}
	\right),
\end{equation}
where $\omega_i^\alpha(z)$ is univariate and ${\boldsymbol\Omega}_{i}^\gamma(z)$ is a $d\times d$ matrix. For $j=1,\cdots,R_0$, we define
\begin{eqnarray}
	&&{\boldsymbol\Omega}(z;{\cal G}_j)=\frac{1}{\vert{\cal G}_j\vert}\sum_{i\in{\cal G}_j}\left[{\boldsymbol\Omega}_i^\gamma(z)-{\boldsymbol\Omega}_i^{\gamma\alpha}(z){\boldsymbol\Omega}_i^{\alpha\gamma}(z)/{\omega}_i^{\alpha}(z)\right],\label{eq4.6}\\
	&&\Gamma_{t0}({\cal G}_j)=\sum_{i\in{\cal G}_j}\eta_{it}(z),\ \ {\boldsymbol\Gamma}_{t1}({\cal G}_j)=\sum_{i\in{\cal G}_j}\eta_{it}(z)\left[{\mathbf X}_{it}-{\boldsymbol\Omega}_i^{\gamma\alpha}(z)/\omega_i^\alpha(z)\right],\label{eq4.7}
\end{eqnarray}
where $\eta_{it}(z)=\tau-I\left(e_{it}\leq -b_{it}(z)\right)$. To derive the asymptotic distribution theory for the post-grouping local linear quantile estimation, we need some additional conditions.

\begin{assumption}\label{ass:6}
	
{\em (i) The joint process $\{({\mathbb Y}_{t}, {\mathbb X}_{t}, Z_t) \}$ is stationary and $\alpha$-mixing dependent with the mixing coefficient satisfying $\alpha(s)\asymp \rho^s$, where $\rho$ is defined as in Assumption \ref{ass:1}, ${\mathbb Y}_t=\{Y_{it}:\ i=1,2,\cdots\}$ and ${\mathbb X}_t=\{{\mathbf X}_{it}:\ i=1,2,\cdots\}$.}
	
{\em (ii) The bandwidth condition in (\ref{eq4.3}) holds when $h$ is replaced by $h_1$, the bandwidth used in the post-grouping local linear estimation. For any $j=1,\cdots,R_0$, }
\[N_j=|{\cal G}_j|=o\left((Th_1)^{1/2}/(\log T)^{3/2}\right).\]
	
{\em (iii) For $j=1,\cdots,R_0$, ${\boldsymbol\Omega}(z; {\cal G}_j)$ defined in (\ref{eq4.6}) is positive definite with all the eigenvalues bounded away from zero and infinity.}
	
\end{assumption}

\begin{assumption}\label{ass:7}	
	
{\em (i) The joint density function of $(Z_t,Z_s)$ exists and is bounded for any $t\neq s$.}
	
{\em (ii) There exists $\iota>4$ such that }
	\[{\sf E}\left[\left|\Gamma_{t0}({\cal G}_j)-{\sf E}\left[\Gamma_{t0}({\cal G}_j)\right]\right|^{\iota}\right]=O\left(N_j^{\iota/2}\right),\ \ {\sf E}\left[\left\Vert{\boldsymbol\Gamma}_{t1}({\cal G}_ju)-{\sf E}\left[{\boldsymbol\Gamma}_{t1}({\cal G}_j)\right]\right\Vert^{\iota}\right]=O\left(N_j^{\iota/2}\right).\]
	
{\em (iii) There exists a $d\times d$ matrix ${\boldsymbol\Lambda}(z;{\cal G}_j)$ such that, as $N_j\rightarrow\infty$,}
		\[\frac{1}{N_j}{\sf E}\left[\left({\boldsymbol\Gamma}_{t1}({\cal G}_j)-{\sf E}\left[{\boldsymbol\Gamma}_{t1}({\cal G}_j)\right]\right)\left({\boldsymbol\Gamma}_{t1}({\cal G}_j)-{\sf E}\left[{\boldsymbol\Gamma}_{t1}({\cal G}_j)\right]\right)^{^\intercal}|Z_t=z\right]\rightarrow {\boldsymbol\Lambda}(z;{\cal G}_j).\]
			
\end{assumption}

\begin{remark}\label{re:4.2}

Assumptions \ref{ass:6}(i) and \ref{ass:7}(ii)(iii) show that the panel observations are allowed to be temporally correlated over $t$ and cross-sectionally dependent over $i$. The high-level conditions are easy to justify when $Y_{it}$, ${\mathbf X}_{it}$ and $Z_t$ are independent over $i$ and $t$. The condition $N_j=o\left((Th_1)^{1/2}/(\log T)^{3/2}\right)$ in Assumption \ref{ass:6}(ii) indicates that the number of subjects (in each group) needs to be much smaller than the time series length in order to derive the limit distribution theory in Theorem \ref{thm:4.3} with root-($N_jTh_1$) convergence. Analogous restrictions can also be found in \cite{KGM12}, \cite{GK16} and \cite{Ch21}, and more comments will be given in Remark \ref{re:4.4}.

\end{remark}


\subsection{Asymptotic properties}\label{sec4.2}

We start with the consistency property for the group membership estimate when $R_0$ is pre-specified.

\renewcommand{\thetheorem}{4.\arabic{theorem}}\setcounter{theorem}{0}

\begin{theorem}\label{thm:4.1}

Suppose that Assumptions \ref{ass:1}--\ref{ass:4} are satisfied, $N=O(T^{\kappa/2})$ with $\kappa$ defined in Assumption \ref{ass:3}(i), and the group number $R_0$ is known a priori. Then we have
\begin{equation}\label{eq4.8}
{\sf P}\left(\big\{\widehat{\cal G}_r,\ r=1, \cdots,R_0\big\}=\big\{{\cal G}_r,\ r=1, \cdots,R_0\big\}\right)\rightarrow1,\ \ T\rightarrow\infty.
\end{equation}

\end{theorem}

\begin{remark}\label{re:4.3}

The consistency result (\ref{eq4.8}) is similar to Theorem 3.1 in \cite{VL17}, Theorem 1 in \cite{Ch19} and Theorem 4.1(a) in \cite{VL20}, all of which study nonparametric mean regression for panel data with a latent group structure. By the clustering algorithm, to achieve the consistency property, it is sufficient to show that $\max_{1\leq j,k\leq N}\left\vert \widehat\Delta(j,k)-\Delta(j,k)\right\vert$ is of order smaller than the minimum distance between true group-specific functional coefficients. This can be proved by using the uniform consistency result for the preliminary local linear quantile estimation (see Lemma A.2 in Appendix A) and Assumption \ref{ass:4}(ii). It is also worthwhile to point out that Theorem \ref{thm:4.1} holds when the cross-sectional dimension $N$ is either fixed or diverging. The condition $N=O(T^{\kappa/2})$ shows that there exists a trade-off between the possible divergence rate of $N$ and the moment condition on ${\mathbf X}_{it}$. In particular, $N$ is allowed to be larger than $T$. We do not need to impose any restriction on the cross-sectional dependence structure for panel observations in this theorem.

\end{remark}

The following theorem shows that the simple ratio criterion proposed in Section \ref{sec3.2} consistently estimates the group number $R_0$.

\begin{theorem}\label{thm:4.2} 

Suppose that Assumptions \ref{ass:1}--\ref{ass:5} are satisfied, and $N=O(T^{\kappa/2})$. Then 
\begin{equation}\label{eq4.9}
	{\sf P}\left(\widehat{R}=R_0\right)\rightarrow1,\ \ T\rightarrow\infty.
\end{equation}

\end{theorem}

Combining Theorems \ref{thm:4.1} and \ref{thm:4.2}, we readily have the following corollary on consistency of the group membership estimate when $R_0$ is unknown.

\renewcommand{\thecorollary}{4.\arabic{corollary}}\setcounter{corollary}{0}

\begin{corollary}\label{cor:4.1}

{\em Suppose that the assumptions in Theorem \ref{thm:4.2} are satisfied. Then
	\begin{equation}\label{eq4.10}
		{\sf P}\left(\widetilde{\boldsymbol{\mathcal G}}=\boldsymbol{{\cal G}}\right)\rightarrow1,\ \ T\rightarrow\infty,
	\end{equation}
	where $\widetilde{\boldsymbol{\mathcal G}}=\{\widetilde{\cal G}_1, \widetilde{\cal G}_2,\cdots, \widetilde{\cal G}_{\widehat{R}}\}$ and $\boldsymbol{\mathcal G}=\left\{{\cal G}_1,{\cal G}_2,\cdots,{\cal G}_{R}\right\}$.}
	
\end{corollary}

We finally turn to the asymptotic normal distribution theory for the post-grouping estimate $\widetilde{\boldsymbol\gamma}_j(z)$ of the group-specific functional coefficient. It follows from Corollary \ref{cor:4.1} that there exists $1\leq j_\ast\leq \widehat{R}$ such that ${\sf P}\left({\cal G}_j=\widetilde{\cal G}_{j_\ast}\right)\rightarrow1$. Without loss of generality, we let $j_\ast=j$ as in Section \ref{sec3.3}, and define $\mu_k=\int z^kK(z)dz$ and $\nu_k=\int z^kK^2(z)dz$ for $k=0,1,2,\cdots$.

\begin{theorem}\label{thm:4.3}

Suppose that Assumptions \ref{ass:1}--\ref{ass:7} are satisfied. For $z$, an interior point of $[0,1]$, and $j=1,\cdots,R_0$, we have
	\begin{equation}\label{eq4.11}
		\sqrt{N_jTh_1}\left[\widetilde{\boldsymbol\gamma}_j(z)-{\boldsymbol\gamma}_j(z)-{\mathbf B}_j(z)\right]\stackrel{d}\longrightarrow {\sf N}\left({\bf 0}_d,\ {\boldsymbol\Sigma}(z;{\cal G}_j)\right),
	\end{equation}
	where ${\mathbf B}_j(z)=\frac{1}{2}h_1^2{\boldsymbol\gamma}_j^{\prime\prime}(z)\mu_2$ with ${\boldsymbol\gamma}_j^{\prime\prime}(z)$ being the second-order derivative of ${\boldsymbol\gamma}_j(z)$, and ${\boldsymbol\Sigma}(z;{\cal G}_j)=\left[{\boldsymbol\Omega}(z;{\cal G}_j)\right]^{-1}\left[\nu_0{\boldsymbol\Lambda}(z;{\cal G}_j)\right]\left[{\boldsymbol\Omega}(z;{\cal G}_j)\right]^{-1}$ with	${\boldsymbol\Omega}(z;{\cal G}_j)$ defined in (\ref{eq4.6}) and ${\boldsymbol\Lambda}(z;{\cal G}_j)$ defined in Assumption \ref{ass:7}(iii).
	
\end{theorem}

\begin{remark}\label{re:4.4}

(i) Theorem \ref{thm:4.3} can be seen as an extension of the asymptotic normality given in Theorem 3.2 of \cite{KGM12} from linear quantile regression to functional-coefficient quantile regression. To obtain the root-($N_jTh_1$) convergence, we need to asymptotically remove the influence of the nuisance parameters $\alpha_i$, $i\in{\cal G}_j$, and impose a somehow restrictive condition on the divergence rate of $N_j$. Specifically, we assume that $N_j=o\left((Th_1)^{1/2}/(\log T)^{3/2}\right)$ in Assumption \ref{ass:6}(ii), analogous to the condition $N=o\left(T^{1/2}/(\log T)^{3/2}\right)$ in \cite{KGM12} if $Th_1$ is treated as the {\em effective} sample size for each subject in the kernel-based estimation. \cite{GGV20} show that the latter restriction may be relaxed to $N=o\left(T/(\log T)^{2}\right)$ in the context of linear panel quantile regression with fixed effects. This improvement is achieved by more precisely computing the orders of the remainder terms in the Bahadur representation of the quantile regression estimation. In particular, they show that the main remainder term can be approximated by a cross-sectional average of independent random elements under the cross-sectional independence restriction. Extension of this technique in the asymptotic proofs to our more general setting is non-trivial since we allow the panel observations to be cross-sectionally correlated. However, we conjecture that the restriction of $N_j=o\left((Th_1)^{1/2}/(\log T)^{3/2}\right)$ may be similarly relaxed by imposing some additional high-level conditions and handling the remainder terms of the Bahadur representation more carefully. This will be left in our future research.


(ii) Theorem \ref{thm:4.3} can be used to conduct point-wise statistical inference on the group-specific functional coefficients. For this, we have to use the bias and asymptotic variance matrix in (\ref{eq4.11}), both of which contain some unknown quantities. The asymptotic bias term may be removed by appropriate undersmoothing or bias correction. Appendix C.3 in the supplement introduces nonparametric methods to estimate these unknown quantities and subsequently construct the point-wise confidence interval.

\end{remark}

\section{Numerical studies}\label{sec5}
\renewcommand{\theequation}{5.\arabic{equation}}
\setcounter{equation}{0}



\subsection{Monte-Carolo simulation}\label{sec5.1}

We consider two data generating processes (DGP) to examine the small-sample performance of the proposed methodologies in various scenarios.

\medskip

\noindent{\bf DGP 1}.\ \ Data is generated via the following functional-coefficient quantile regression:
\begin{equation}\label{eq5.1}
Y_{it} = {\mathbf X}_{it}^{^\intercal} {\boldsymbol\beta}_{i}(Z_t)+\alpha_i+e_{it}, \quad i=1,\cdots,N, \quad t=1,\cdots,T,
\end{equation}
where $Z_t$, $t=1,\cdots T$, are independently drawn from the uniform distribution $U[0,1]$, ${\mathbf X}_{it}=(X_{it,1}, X_{it,2})^{^\intercal}$, $i=1,\cdots,N, t=1,\cdots, T$, are independently drawn from a bivariate normal distribution with zero means, unit variances, and  a correlation coefficient of $1/2$, $\alpha_i=\left(\bar{X}_{i,1}^2+\bar{X}_{i,2}^2\right)/5$ with $\bar{X}_{i,k}=\frac{1}{T}\sum_{t=1}^TX_{it,k}$, and the idiosyncratic errors $e_{it}$ are independently generated from one of the following distributions: ${\sf N}(0,1)$, $t(5)$, and $0.4\big[\chi^2(3)-3\big]$, and are independent of $Z_t$ and ${\mathbf X}_{it}$. The results for the ${\sf N}(0,1)$ errors can be used as the benchmark for assessing how the proposed method performs when the errors are heavy tailed (with $t(5)$ distribution) or asymmetrically distributed (with $0.4\big[\chi^2(3)-3\big]$ distribution, where the scaling factor $0.4$ is used to give a comparable error variance to ${\sf N}(0,1)$).

As in \cite{Ch19} and \cite{SWJ19}, the heterogenous functional coefficients ${\boldsymbol\beta}_{i}(\cdot)=\left[\beta_{i,1}(\cdot), \beta_{i,2}(\cdot)\right]^{^\intercal}$ satisfy the following group structure:
\[
\beta_{i,1}(z)=\left\{
\begin{array}{ll}
	\gamma_{1,1}(z)=3F(z;0.5,0.1)&\quad {\rm if}\ i\in{\cal G}_1, \\
	\gamma_{2,1}(z)=3[2z-6z^2+4z^3+F(z;0.7,0.05)]&\quad {\rm if}\ i\in{\cal G}_2, \\
	\gamma_{3,1}(z)=3[4z-8z^2+4z^3+F(z;0.6,0.05)] &\quad {\rm if}\ i\in{\cal G}_3,
\end{array}
\right.
\]
and
\[
\beta_{i,2}(z)=\left\{
\begin{array}{ll}
	\gamma_{1,2}(z)=3[2z-4z^2+2z^3+F(z;0.6,0.1)]&\quad {\rm if}\ i\in{\cal G}_1, \\
	\gamma_{2,2}(z)=3[z-3z^2+2z^3+F(z;0.7,0.04)]&\quad {\rm if}\ i\in{\cal G}_2, \\
	\gamma_{3,2}(z)=3[0.5z-0.5z^2+F(z;0.4,0.07)]&\quad {\rm if}\ i\in{\cal G}_3,
\end{array}
\right.
\]
where $F(z;\xi,\eta)=1/(1+\exp[-(z-\xi)/\eta])$, ${\cal G}_1=\left\{1,2,\cdots,N_1\right\}$, ${\cal G}_2=\left\{N_1+1,\cdots,N_1+N_2\right\}$, and ${\cal G}_3=\left\{N_1+N_2+1,\cdots,N_1+N_2+N_3\right\}$ with $N_1=\lfloor 0.3N\rfloor $, $N_2=\lfloor 0.3N\rfloor $ and $N_3=N-N_1-N_2$.

\medskip

\noindent{\bf DGP 2}.\ \ This DGP has the same model equation as (\ref{eq5.1}), but with quantile-dependent functional coefficients and group structure:
\[
\beta_{i,1}(z)=\beta_{\tau,i,1}(z)=\left\{
\begin{array}{ll}
	2\tau\gamma_{1,1}(z)+2(1-\tau)\gamma_{2,1}(z)&\quad {\rm if}\ i\in{\cal G}_1,\\
	2(1-\tau)\gamma_{1,1}(z)+2\tau\gamma_{2,1}(z)&\quad {\rm if}\ i\in{\cal G}_2,\\
	\gamma_{3,1}(z)&\quad {\rm if}\ i\in{\cal G}_3,
\end{array}
\right.
\]
and
\[
\beta_{i,2}(z)=\beta_{\tau,i,2}(z)=\left\{
\begin{array}{ll}
	2\tau\gamma_{1,2}(z)+2(1-\tau)\gamma_{2,2}(z)& \quad {\rm if}\ i\in{\cal G}_1, \\
	2(1-\tau)\gamma_{1,2}(z)+2\tau\gamma_{2,2}(z)&\quad {\rm if}\ i\in{\cal G}_2,\\
	\gamma_{3,2}(z)&\quad {\rm if}\ i\in{\cal G}_3,
\end{array}
\right.
\]
where ${\cal G}_k$ and $\gamma_{k,j}(z)$, $k=1,2,3$ and $j=1,2$, are the same as in DGP 1. Data for $\mathbf{X}_{it}$, $Z_t$, $\alpha_i$, and $e_{it}$ are generated in the same way as in DGP 1. Note that when $\tau=0.5$, ${\cal G}_1$ and ${\cal G}_2$ merge so there are only two groups, whereas when $\tau\neq0.5$, there are three groups.

\smallskip

The sample size is $N=50, 100$ and $T=50, 100$, and the number of replications is $M=200$. We consider three quantile levels, $\tau=0.25, 0.50$ and $0.75$, and use the Gaussian kernel, $K(u)=e^{-u^2/2}/\sqrt{2\pi}$, in the local linear estimation. The bandwidth is selected via the leave-one-out cross-validation method. To gauge the performance of the ratio criterion (\ref{eq3.3}) in estimating the number of groups, we set $\overline{R}=5$ and report the percentage of replications where each integer (between 1 and $\overline{R}$) is chosen. To understand the accuracy of the estimated groups (and their membership) from the agglomerative clustering algorithm in Section \ref{sec3.1}, we consider two measures: purity and the normalised mutual information (NMI) of the estimated groups $\widetilde{\boldsymbol{\mathcal G}}=\left\{\widetilde{{\cal G}}_1, \cdots, \widetilde{{\cal G}}_{\widehat{R}}\right\}$ with the true groups $\boldsymbol{{\mathcal G}}=\left\{{\cal G}_1, \cdots, {\cal G}_{R}\right\}$, both of which are classic criteria of clustering quality and are defined respectively as
$${\rm Purity}(\widetilde{\boldsymbol{\mathcal G}}, \boldsymbol{{\mathcal G}})=\frac{1}{N}\sum_{k=1}^{\widehat{R}}\max_{1\leq j\leq R_0}\left|\widetilde{{\cal G}}_k \cap {\cal G}_j\right|,\ \ \ \ \ \ {\rm NMI}(\widetilde{\boldsymbol{\mathcal G}}, \boldsymbol{\mathcal G})=2\cdot\frac{I(\widetilde{\boldsymbol{\mathcal G}}, \boldsymbol{\mathcal G})}{H(\widetilde{\boldsymbol{\mathcal G}})+H(\boldsymbol{\mathcal G})},$$
where $I(\widetilde{\boldsymbol{\mathcal G}}, \boldsymbol{\mathcal G})$ is the mutual information between $\widetilde{\boldsymbol{\mathcal G}}$ and $\boldsymbol{\mathcal G}$ defined as
$$I(\widetilde{\boldsymbol{\mathcal G}}, \boldsymbol{\mathcal G})=\sum_{k=1}^{\widehat{R}}\sum_{j=1}^{R_0}\frac{\left|\widetilde{{\cal G}}_k \cap {\cal G}_j \right|}{N}\cdot\log_2\left(\frac{N\left |\widetilde{{\cal G}}_k \cap {\cal G}_j\right|}{\left|\widetilde{{\cal G}}_k\right| \cdot\left|{\cal G}_j \right|}\right),$$
$H(\widetilde{\boldsymbol{\mathcal G}})$ is the entropy of $\widetilde{\boldsymbol{\mathcal G}}$ defined as
$$H(\widetilde{\boldsymbol{\mathcal G}})=-\sum_{k=1}^{\widehat{R}}\frac{\left|\widetilde{\cal G}_k\right|}{N}\log_2\left(\frac{\left|\widetilde{\cal G}_k\right|}{N}\right),$$ and $H(\boldsymbol{\mathcal G})$ is defined analogously. The closer the values of NMI and purity are to 1, the more accurate the estimated groups are to the true ones. We also look at the estimation accuracy of the functional coefficients for both the preliminary local linear quantile estimator defined in Section \ref{sec3.1} and the post-grouping local linear quantile estimator defined in Section \ref{sec3.3}. For this we compute the average root mean squared errors (RMSE) defined as
\begin{align}
	\mbox{RMSE}(\widehat{\boldsymbol\beta})=\frac{1}{N}\sum\limits_{i=1}^N\left[\frac{1}{T}\sum\limits_{t=1}^T\big\|\widehat{\boldsymbol\beta}_i(Z_t)-{\boldsymbol\beta}_i(Z_t)\big\|_2^2\right]^{1/2},\notag
\end{align}
for an estimate, $\widehat{\boldsymbol\beta}(\cdot)=\big(\widehat{\boldsymbol\beta}_1(\cdot),\cdots,\widehat{\boldsymbol\beta}_N(\cdot)\big)^{^\intercal}$, of the true functional coefficients ${\boldsymbol\beta}(\cdot)=\big({\boldsymbol\beta}_1(\cdot),\cdots,{\boldsymbol\beta}_N(\cdot)\big)^{^\intercal}$. As a benchmark, we also compute the RMSE of the oracle estimator, which assumes that the true group structure is known a priori and pools data belonging to each group to obtain group-specific estimates of the functional coefficients. The results for DGP 1 are reported in Table \ref{tab5.1} (for ${\sf N}(0,1)$ errors), Table \ref{tab5.2} (for $t(5)$ errors), and Table \ref{tab5.3} (for $0.4\big[\chi^2(3)-3\big]$ errors). Results for DGP 2 are reported in Tables \ref{tab5.4}--\ref{tab5.6}.

\smallskip

Table \ref{tab5.1} shows that for DGP 1, when the errors follow the ${\sf N}(0,1)$ distribution, for the smallest sample size considered (i.e., $N=50$ and $T=50$), the ratio criterion (\ref{eq3.3}) picks the correct number of groups in about 75\% of the replications at $\tau=0.50$ and about 60\% of the replications at $\tau=0.25$ and 0.75. These percentages increase markedly as $T$ increases and usually as $N$ increases, but to a lesser extent. When $T=100$, the correct $R_0$ is chosen in 100\% of the replications at $\tau=0.50$ and more than 85\% of the replications at $\tau=0.25$ and 0.75. This is consistent with the theoretical result in Theorem \ref{thm:4.2}. On the other hand, the NMI values are above 0.77 and purities above 0.91 at all quantiles when $T=50$, and they increase to above 0.95 when $T$ increases to $100$, verifying the consistency results in Theorem \ref{thm:4.1} and Corollary \ref{cor:4.1} for the agglomerative clustering algorithm. The lower block of Table \ref{tab5.1} shows that by pooling data belonging to the same group in the estimation, the post-grouping local linear estimator cuts the RMSE of the preliminary local linear estimator by more than 30\%, and its RMSE values are not far from those of the oracle estimator. When $T$ increases to $100$ and the groups are accurately estimated, the RMSEs of the post-grouping estimator are very close to the oracle estimator. Similar findings can be drawn from Tables \ref{tab5.2} and \ref{tab5.3}, where the errors in DGP 1 follow the heavier tailed $t(5)$ distribution and the asymmetric $0.4\big[\chi^2(3)-3\big]$ distribution, respectively. The results for $t(5)$ errors are in general worse than those of ${\sf N}(0,1)$ and $0.4[\chi^2(3)-3]$ errors, which may be due to the fact that the $t(5)$ distribution has a larger variance than the other two. For $0.4[\chi^2(3)-3]$ errors, the results are better than those of ${\sf N}(0,1)$ at $\tau=0.25$, worse than ${\sf N}(0,1)$ at $\tau=0.75$ and comparable at median $\tau=0.50$. This may be because at $\tau=0.25$, the $0.4\big[\chi^2(3)-3\big]$ distribution has more data points than the ${\sf N}(0,1)$ distribution and hence, the preliminary functional coefficients estimation and latent group estimation are more accurate. At $\tau=0.75$, the reverse is true.

\smallskip

For DGP 2, the performance of the proposed methods is similar, despite the fact that the functional coefficients and the group structure are now quantile dependent. In Tables \ref{tab5.4}--\ref{tab5.6}, we can observe the same evolving patterns when $T$ or $N$  increases as in Tables \ref{tab5.1}--\ref{tab5.3}. This demonstrates the robustness of our methods to the different settings considered.


\begin{table}[htp]\footnotesize
	\renewcommand\arraystretch{0.7}
	\caption{\label{tab5.1}Simulation results for DGP1 with ${\sf N}(0,1)$ idiosyncratic errors}%
	\vspace{0.3cm}
	\setlength{\tabcolsep}{1.2mm}{
	\begin{tabular}{r|cccccccccc}
	\hline
	\hline
	 &\multicolumn{10}{c}{\bf Percentage of replications each $\widehat{R}$ value is selected by the ratio criterion}\\
	&\multicolumn{10}{|c}{with true $R_0=3$ for $\tau=0.25,0.50,0.75$} \\ \hline
	{\multirow{2}{*}{\diagbox[width=6em]{$T$}{$N$}}}& \multicolumn{5}{c|}{$N=50$} & \multicolumn{5}{c}{$N=100$} \\
		& $\widehat{R}=1$ & $\widehat{R}=2$ & $\widehat{R}=3$ & $\widehat{R}=4$ & \multicolumn{1}{c|}{$\widehat R=5$} & $\widehat{R}=1$ & $\widehat{R}=2$ & $\widehat{R}=3$ & $\widehat{R}=4$ & $\widehat{R}=5$ \\ \hline
		~~~~~$\tau=0.25$ & 0.0\% & 2.5\% & 63.0\% & 27.5\% & \multicolumn{1}{c|}{7.0\%} & 0.0\% & 32.5\% & 65.0\% & 2.5\% & 0.0\% \\
		$50$~~$\tau=0.50$ & 0.0\% & 2.5\% & 74.5\% & 19.0\% & \multicolumn{1}{c|}{4.0\%} & 0.0\% & 2.0\% & 92.5\% & 4.5\% & 1.0\% \\
		~~~~~$\tau=0.75$ & 0.0\% & 42.5\% & 56.5\% & 1.0\% & \multicolumn{1}{c|}{0.0\%} & 0.0\% & 4.0\% & 89.0\% & 7.0\% & 0.0\% \\ \hline
		~~~~~$\tau=0.25$ & 0.0\% & 0.0\% & 96.5\% & 3.5\% & \multicolumn{1}{c|}{0.0\%} & 0.0\% & 0.0\% & 99.5\% & 0.5\% & 0.0\% \\
		$100$~$\tau=0.50$ & 0.0\% & 0.0\% & 100.0\% & 0.0\% & \multicolumn{1}{c|}{0.0\%} & 0.0\% & 0.0\% & 100.0\% & 0.0\% & 0.0\% \\
		~~~~~$\tau=0.75$ & 0.0\% & 15.0\% & 85.0\% & 0.0\% & \multicolumn{1}{c|}{0.0\%} & 0.0\% & 0.0\% & 100.0\% & 0.0\% & 0.0\% \\ 
	\end{tabular}}

	\setlength{\tabcolsep}{5mm}{
	\begin{tabular}{r|cccccccc}
		\hline
		\hline
		&\multicolumn{8}{c}{\bf Average NMI and purity (with standard deviation in parentheses)}\\
       \hline
       {\multirow{2}{*}{\diagbox[width=8em]{$T$}{$N$}}} & \multicolumn{4}{c|}{$N=50$} & \multicolumn{4}{c}{$N=100$} \\
       & \multicolumn{2}{c}{NMI} & \multicolumn{2}{c|}{Purity}  & \multicolumn{2}{c}{NMI}  & \multicolumn{2}{c}{Purity}  \\ \hline
       ~~~~~$\tau=0.25$ & \multicolumn{2}{c}{0.8363 (0.1305)} &  \multicolumn{2}{c|}{0.9174 (0.0868)} & \multicolumn{2}{c}{0.7807 (0.1466)} &  \multicolumn{2}{c}{0.9556 (0.0552)}   \\
       $50$~~$\tau=0.50$ & \multicolumn{2}{c}{0.8923 (0.1084)} &  \multicolumn{2}{c|}{0.9508 (0.0644)} & \multicolumn{2}{c}{0.8994 (0.0956)} & \multicolumn{2}{c}{0.9646 (0.0468)}   \\
       ~~~~~$\tau=0.75$ & \multicolumn{2}{c}{0.7766 (0.1416)} & \multicolumn{2}{c|}{0.9647 (0.0502)} & \multicolumn{2}{c}{0.8347 (0.1385)} & \multicolumn{2}{c}{0.9452 (0.0575)}   \\
      \hline
      ~~~~~$\tau=0.25$ &\multicolumn{2}{c}{0.9915 (0.0356)}  & \multicolumn{2}{c|}{0.9953 (0.0202)}  &  \multicolumn{2}{c}{0.9938 (0.0257)} & \multicolumn{2}{c}{0.9973 (0.0156)}  \\
      $100$~$\tau=0.50$ &\multicolumn{2}{c}{0.9986 (0.0099)} & \multicolumn{2}{c|}{0.9996 (0.0028)}  &  \multicolumn{2}{c}{0.9986 (0.0082)} & \multicolumn{2}{c}{0.9997 (0.0021)}  \\
      ~~~~~$\tau=0.75$ &\multicolumn{2}{c}{0.9522 (0.1021)} & \multicolumn{2}{c|}{0.9987 (0.0053)}  &  \multicolumn{2}{c}{0.9921 (0.0328)} & \multicolumn{2}{c}{0.9975 (0.0122)}  \\
     \end{tabular}}
 
	\setlength{\tabcolsep}{0.5mm}{
	\begin{tabular}{l|ccc|ccc}
	\hline
	\hline
	&\multicolumn{6}{|c}{\bf Average RMSE of ${\boldsymbol\beta}(\cdot)$ estimates (with standard deviation in parentheses) }\\
    \hline
    {\multirow{2}{*}{\diagbox[width=6em]{$T$}{$N$}}}  & \multicolumn{3}{c|}{$N=50$} & \multicolumn{3}{c}{$N=100$}  \\
    & Oracle & Preliminary & Post-grouping & Oracle & Preliminary & Post-grouping\\
    \hline
    ~~~~~~$\tau=0.25$ & {0.4211 (0.0213)} & {0.7085 (0.0280)} &{0.4770 (0.0653)} &{0.3877 (0.0181)} & {0.7105 (0.0235)} & {0.4910 (0.0906)}  \\
    $50$~~$\tau=0.50$ & {0.3783 (0.0195)} & {0.6608 (0.0282)} &{0.4190 (0.0602)} &{0.3693 (0.0170)} & {0.6646 (0.0204)} & {0.4025 (0.0497)}  \\
    ~~~~~~$\tau=0.75$ & {0.3866 (0.0191)} & {0.7038 (0.0287)} &{0.5057 (0.0924)} &{0.4003 (0.0176)} & {0.7114 (0.0214)} & {0.4598 (0.0727)}  \\ \hline
    ~~~~~~$\tau=0.25$ & {0.3028 (0.0124)}  & {0.5213 (0.0166)}  & {0.3056 (0.0205)} & {0.2742 (0.0091)} & {0.5160 (0.0129)}  & {0.2762 (0.0146)} \\
    $100$~$\tau=0.50$ & {0.2407 (0.0131)}  & {0.4805 (0.0170)}  & {0.2410 (0.0135)} & {0.2291 (0.0090)} & {0.4783 (0.0129)}  & {0.2294 (0.0090)} \\
    ~~~~~~$\tau=0.75$ & {0.2675 (0.0134)}  & {0.5171 (0.0170)}  & {0.3016 (0.0792)} & {0.2753 (0.0093)} & {0.5170 (0.0128)}  & {0.2780 (0.0186)} \\
    \hline
    \hline
    \end{tabular}}
\end{table}


\begin{table}[htp]\footnotesize
	\renewcommand\arraystretch{0.7}
	\caption{\label{tab5.2}Simulation results for DGP1 with $t(5)$ idiosyncratic errors}%
	\vspace{0.3cm}
	\setlength{\tabcolsep}{1.2mm}{
		\begin{tabular}{r|ccccc|ccccc}
			\hline
			\hline			
			&\multicolumn{10}{c}{\bf Percentage of replications each $\widehat{R}$ value is selected by the ratio criterion}\\
			&\multicolumn{10}{|c}{ with true $R_0=3$ for $\tau=0.25,0.50,0.75$}\\
			\hline
			{\multirow{2}{*}{\diagbox[width=6em]{$T$}{$N$}}} & \multicolumn{5}{c|}{$N=50$} & \multicolumn{5}{c}{$N=100$} \\
			& $\widehat{R}=1$ & $\widehat{R}=2$ & $\widehat{R}=3$& $\widehat{R}=4$ & $\widehat{R}=5$ &  $\widehat{R}=1$ & $\widehat{R}=2$ & $\widehat{R}=3$ & $\widehat{R}=4$ & $\widehat{R}=5$ \\\hline
			$\tau=0.25$ &  0.0\% & 7.5\% &67.0\% & 21.0\%& 4.5\%&   0.5\% & 4.5\% & 59.0\% & 28.0\% & 8.0\%  \\
			$50$\ \ $\tau=0.50$ &   0.0\% & 23.0\% &70.0\% & 6.5\%& 0.5\%&   0.0\% & 17.5\% & 78.0\% & 4.5\% & 0.0\%  \\
			$\tau=0.75$ &  0.0\% & 15.5\% &71.5\% & 10.0\%& 1.0\%&   0.0\% & 15.5\% & 71.5\% & 10.5\% & 2.5\%  \\
			\hline
			$\tau=0.25$ &  0.0\% & 0.5\%  & 85.5\%& 12.0\%&2.0\% &   0.0\% & 0.0\% &94.5\% &5.0\% &0.5\%  \\
			$100$\ \ $\tau=0.50$ &  0.0\% & 0.0\%  & 95.5\%& 4.5\%&0.0\% &   0.0\% & 0.0\% &100.0\% &0.0\% &0.0\%  \\
			$\tau=0.75$ &  0.0\% & 0.0\%  & 91.5\%& 6.0\%&2.5\% &   0.0\% & 0.0\% &98.5\% &1.5\% &0.0\%  \\ 
	\end{tabular}}

	\setlength{\tabcolsep}{5mm}{
		\begin{tabular}{r|cccccccc}
			\hline
			\hline			
			 &\multicolumn{8}{|c}{\bf Average NMI and purity (with standard deviation in parentheses)}\\
			\hline
           {\multirow{2}{*}{\diagbox[width=8em]{$T$}{$N$}}} & \multicolumn{4}{c|}{$N=50$} & \multicolumn{4}{c}{$N=100$} \\
			&  \multicolumn{2}{c}{NMI} & \multicolumn{2}{c|}{Purity}  & \multicolumn{2}{c}{NMI}  & \multicolumn{2}{c}{Purity}  \\ \hline
			$\tau=0.25$ & \multicolumn{2}{c}{0.7272 (0.1414)} & \multicolumn{2}{c|}{0.8885 (0.0897)} & \multicolumn{2}{c}{0.6954 (0.1689)} & \multicolumn{2}{c}{0.8770 (0.0943)}   \\
			$50$\ \ $\tau=0.50$ & \multicolumn{2}{c}{0.7940 (0.1641)} & \multicolumn{2}{c|}{0.9497 (0.0618)} & \multicolumn{2}{c}{0.8026 (0.1364)} & \multicolumn{2}{c}{0.9494 (0.0589)}   \\
			$\tau=0.75$ & \multicolumn{2}{c}{0.7292 (0.1555)} & \multicolumn{2}{c|}{0.9026 (0.0857)} & \multicolumn{2}{c}{0.6839 (0.1474)} & \multicolumn{2}{c}{0.9018 (0.0835)}   \\
			\hline
			$\tau=0.25$ &\multicolumn{2}{c}{0.9601 (0.0892)}  & \multicolumn{2}{c|}{0.9786 (0.0483)}  &  \multicolumn{2}{c}{0.9783 (0.0430)} & \multicolumn{2}{c}{0.9915 (0.0227)}  \\
			$100$\ \  $\tau=0.50$ &\multicolumn{2}{c}{0.9922 (0.0263)}  & \multicolumn{2}{c|}{0.9949 (0.0220)}  &  \multicolumn{2}{c}{0.9946 (0.0177)} & \multicolumn{2}{c}{0.9986 (0.0054)}  \\
			$\tau=0.75$ &\multicolumn{2}{c}{0.9690 (0.0617)}  & \multicolumn{2}{c|}{0.9850 (0.0382)}  &  \multicolumn{2}{c}{0.9768 (0.0505)} & \multicolumn{2}{c}{0.9926 (0.0218)}  \\
	\end{tabular}}

	\setlength{\tabcolsep}{0.5mm}{
	\begin{tabular}{l|ccc|ccc}
			\hline
			\hline
    &\multicolumn{6}{|c}{\bf Average RMSE of ${\boldsymbol\beta}(\cdot)$ estimates (with standard deviation in parentheses) }\\
   \hline
    {\multirow{2}{*}{\diagbox[width=6em]{$T$}{$N$}}}  & \multicolumn{3}{c|}{$N=50$} &  \multicolumn{3}{c}{$N=100$}  \\
   & Oracle & Preliminary & Post-grouping & Oracle & Preliminary & Post-grouping\\
   \hline
   ~~~~~$\tau=0.25$ & {0.4269 (0.0202)} & {0.7903 (0.0358)} &{0.5386 (0.0783)} &{0.4259 (0.0176)} & {0.7914 (0.0249)} & {0.5401 (0.0928)}  \\
   $50$~~$\tau=0.50$ & {0.3999 (0.0207)} & {0.7178 (0.0308)} &{0.4980 (0.0967)} &{0.3776 (0.0182)} & {0.7195 (0.0223)} & {0.4646 (0.0850)}  \\
   ~~~~~$\tau=0.75$ & {0.4187 (0.0221)} & {0.7935 (0.0381)} &{0.5340 (0.0864)} &{0.4095 (0.0174)} & {0.7949 (0.0266)} & {0.5462 (0.0922)}  \\ \hline
   ~~~~~~$\tau=0.25$ & {0.3206 (0.0139)}  & {0.5819 (0.0217)}  & {0.33421 (0.0471)} & {0.3117 (0.0104)} & {0.5804 (0.0161)}  & {0.3184 (0.0259)} \\
   $100$~$\tau=0.50$ & {0.2753 (0.0130)}  & {0.5185 (0.0189)}  & {0.2778 (0.0156)} & {0.2673 (0.0098)} & {0.5183 (0.0148)}  & {0.2686 (0.0110)} \\
   ~~~~~~$\tau=0.75$ & {0.3204 (0.0143)}  & {0.5825 (0.0206)}  & {0.3324 (0.0377)} & {0.2955 (0.0102)} & {0.5810 (0.0165)}  & {0.3037 (0.0305)} \\
   \hline
   \hline
	\end{tabular}}
\end{table}

\begin{table}[htp]\footnotesize
	\renewcommand\arraystretch{0.7}
	\caption{\label{tab5.3}Simulation results for DGP1 with $0.4*[\chi^2(3)-3]$ idiosyncratic errors}%
	\vspace{0.3cm}
	\setlength{\tabcolsep}{1.2mm}{
		\begin{tabular}{r|ccccc|ccccc}
			\hline
			\hline			
			&\multicolumn{10}{c}{\bf Percentage of replications each $\widehat{R}$ value is selected by the ratio criterion}\\
			&\multicolumn{10}{|c}{ with true $R_0=3$ for $\tau=0.25,0.50,0.75$}\\
			\hline
			{\multirow{2}{*}{\diagbox[width=6em]{$T$}{$N$}}} & \multicolumn{5}{c|}{$N=50$} & \multicolumn{5}{c}{$N=100$} \\
			& $\widehat{R}=1$ & $\widehat{R}=2$ & $\widehat{R}=3$& $\widehat{R}=4$ & $\widehat{R}=5$ &  $\widehat{R}=1$ & $\widehat{R}=2$ & $\widehat{R}=3$ & $\widehat{R}=4$ & $\widehat{R}=5$ \\\hline
			$\tau=0.25$ &  0.0\% & 0.0\% &95.5\% & 4.5\%& 0.0\%&   0.0\% & 0.5\% & 97.5\% & 2.0\% & 0.0\%  \\
			$50$\ \ $\tau=0.50$ &   0.0\% & 1.0\% &69.0\% & 24.5\%& 5.5\%&   0.0\% & 0.5\% & 96.0\% & 3.5\% & 0.0\%  \\
			$\tau=0.75$ &  0.0\% & 11.0\% &66.5\% & 18.5\%& 4.0\%&   0.0\% & 22.5\% & 70.5\% & 6.5\% & 0.5\%  \\
			\hline
			$\tau=0.25$ &  0.0\% & 0.5\%  & 99.5\%& 0.0\%&0.0\% &   0.0\% & 0.0\% &100.0\% &0.0\% &0.0\%  \\
			$100$\ \ $\tau=0.50$ &  0.0\% & 0.0\%  & 100.0\%& 0.0\%&0.0\% &   0.0\% & 0.0\% &100.0\% &0.0\% &0.0\%  \\
			$\tau=0.75$ &  0.0\% & 0.0\%  & 98.0\%& 2.0\%&0.0\% &   0.0\% & 0.0\% &98.0\% &1.5\% &0.5\%  \\ 
	\end{tabular}}

	\setlength{\tabcolsep}{5mm}{
		\begin{tabular}{r|cccccccc}
			\hline
			\hline			
			 &\multicolumn{8}{|c}{\bf Average NMI and purity (with standard deviation in parentheses)}\\
			\hline
           {\multirow{2}{*}{\diagbox[width=8em]{$T$}{$N$}}} & \multicolumn{4}{c|}{$N=50$} & \multicolumn{4}{c}{$N=100$} \\
			&  \multicolumn{2}{c}{NMI} & \multicolumn{2}{c|}{Purity}  & \multicolumn{2}{c}{NMI}  & \multicolumn{2}{c}{Purity}  \\ \hline
			$\tau=0.25$ & \multicolumn{2}{c}{0.9766 (0.0574)} & \multicolumn{2}{c|}{0.9927 (0.0182)} & \multicolumn{2}{c}{0.9748 (0.0691)} & \multicolumn{2}{c}{0.9945 (0.0187)}   \\
			$50$\ \ $\tau=0.50$ & \multicolumn{2}{c}{0.9382 (0.0789)} & \multicolumn{2}{c|}{0.9608 (0.0595)} & \multicolumn{2}{c}{0.9477 (0.0831)} & \multicolumn{2}{c}{0.9827 (0.0354)}   \\
			$\tau=0.75$ & \multicolumn{2}{c}{0.7418 (0.1690)} & \multicolumn{2}{c|}{0.9010 (0.0873)} & \multicolumn{2}{c}{0.7469 (0.1573)} & \multicolumn{2}{c}{0.9342 (0.0735)}   \\
			\hline
			$\tau=0.25$ &\multicolumn{2}{c}{0.9986 (0.0199)}  & \multicolumn{2}{c|}{1.0000 (0.0000)}  &  \multicolumn{2}{c}{1.0000 (0.0000)} & \multicolumn{2}{c}{1.0000 (0.0000)}  \\
			$100$\ \  $\tau=0.50$ &\multicolumn{2}{c}{1.0000 (0.0000)}  & \multicolumn{2}{c|}{1.0000 (0.0000)}  &  \multicolumn{2}{c}{1.0000 (0.0000)} & \multicolumn{2}{c}{1.0000 (0.0000)}  \\
			$\tau=0.75$ &\multicolumn{2}{c}{0.9811 (0.0502)}  & \multicolumn{2}{c|}{0.9933 (0.0195)}  &  \multicolumn{2}{c}{0.9755 (0.0554)} & \multicolumn{2}{c}{0.9904 (0.0299)}  \\
	\end{tabular}}

	\setlength{\tabcolsep}{0.5mm}{
	\begin{tabular}{l|ccc|ccc}
			\hline
			\hline
    &\multicolumn{6}{|c}{\bf Average RMSE of ${\boldsymbol\beta}(\cdot)$ estimates (with standard deviation in parentheses) }\\
   \hline
    {\multirow{2}{*}{\diagbox[width=6em]{$T$}{$N$}}}  & \multicolumn{3}{c|}{$N=50$} &  \multicolumn{3}{c}{$N=100$}  \\
   & Oracle & Preliminary & Post-grouping & Oracle & Preliminary & Post-grouping\\
   \hline
   ~~~~~$\tau=0.25$ & {0.3259 (0.0191)} & {0.5420 (0.0267)} &{0.3363 (0.0436)} &{0.3212 (0.0171)} & {0.5420 (0.0201)} & {0.3334 (0.0470)}  \\
   $50$~~$\tau=0.50$ & {0.3526 (0.0189)} & {0.6041 (0.0278)} &{0.3736 (0.0448)} &{0.3300 (0.0162)} & {0.6007 (0.0223)} & {0.3498 (0.0498)}  \\
   ~~~~~$\tau=0.75$ & {0.4172 (0.0220)} & {0.7685 (0.0380)} &{0.5251 (0.0935)} &{0.3944 (0.0158)} & {0.7638 (0.0238)} & {0.5039 (0.0944)}  \\ \hline
   ~~~~~~$\tau=0.25$ & {0.1953 (0.0089)}  & {0.3531 (0.0136)}  & {0.1964 (0.0204)} & {0.1911 (0.0072)} & {0.3534 (0.0113)}  & {0.1911 (0.0072)} \\
   $100$~$\tau=0.50$ & {0.2109 (0.0103)}  & {0.4241 (0.0175)}  & {0.2109 (0.0103)} & {0.2027 (0.0094)} & {0.4210 (0.0134)}  & {0.2027 (0.0094)} \\
   ~~~~~~$\tau=0.75$ & {0.3119 (0.0136)}  & {0.5704 (0.0192)}  & {0.3188 (0.0290)} & {0.3005 (0.0098)} & {0.5686 (0.0162)}  & {0.3093 (0.0302)} \\
   \hline
   \hline
	\end{tabular}}
\end{table}


\begin{table}[htp]\footnotesize
	\renewcommand\arraystretch{0.7}
	\caption{\label{tab5.4}Simulation results for DGP2 with ${\sf N}(0,1)$ idiosyncratic errors}%
	\vspace{0.3cm}
	
	\setlength{\tabcolsep}{1.2mm}{
		\begin{tabular}{r|ccccc|ccccc}
			\hline
			\hline			
    &\multicolumn{10}{c}{\bf Percentage of replications each $\widehat{R}$ value is selected by the ratio criterion}\\
   &\multicolumn{10}{|c}{ with true $R_0=3$ for $\tau=0.25,0.75$ and $R_0=2$ for $\tau=0.50$}\\
   \hline
   {\multirow{2}{*}{\diagbox[width=6em]{$T$}{$N$}}} & \multicolumn{5}{c|}{$N=50$} & \multicolumn{5}{c}{$N=100$} \\
   & $\widehat{R}=1$ & $\widehat{R}=2$ & $\widehat{R}=3$& $\widehat{R}=4$ & $\widehat{R}=5$ &  $\widehat{R}=1$ & $\widehat{R}=2$ & $\widehat{R}=3$ & $\widehat{R}=4$ & $\widehat{R}=5$ \\\hline
   $\tau=0.25$ &  0.0\% & 24.0\% &74.5\% & 1.5\%& 0.0\%&   0.0\% & 1.5\% & 89.5\% & 7.0\% & 2.0\%  \\
  $50$\ \ $\tau=0.50$ &   0.5\% & 99.5\% &0.0\% & 0.0\%& 0.0\%&   0.5\% & 98.5\% & 1.0\% & 0.0\% & 0.0\%  \\
  $\tau=0.75$  & 0.0\% & 4.5\% &86.5\% & 6.5\%& 2.5\%&   0.0\% & 1.5\% & 88.0\% & 8.0\% & 2.5\%  \\
  \hline
  $\tau=0.25$ &  0.0\% & 16.0\%  & 84.0\%& 0.0\%&0.0\% &   0.0\% & 0.0\% &100.0\% &0.0\% &0.0\%  \\
  $100$\ \ $\tau=0.50$ &  0.0\% & 100.0\%  & 0.0\%& 0.0\%&0.0\% &   0.0\% & 100.0\% &0.0\% &0.0\% &0.0\%  \\
  $\tau=0.75$ &  0.0\% & 11.0\%  & 89.0\%& 0.0\%&0.0\% &   0.0\% & 0.0\% &100.0\% &0.0\% &0.0\%  \\
  \end{tabular}}

  \setlength{\tabcolsep}{5mm}{
  \begin{tabular}{r|cccccccc}
  \hline
  &\multicolumn{8}{|c}{\bf Average NMI and purity (with standard deviation in parentheses)}\\ \hline
  {\multirow{2}{*}{\diagbox[width=8em]{$T$}{$N$}}} & \multicolumn{4}{c|}{$N=50$} & \multicolumn{4}{c}{$N=100$} \\
  &  \multicolumn{2}{c}{NMI} & \multicolumn{2}{c|}{Purity}  & \multicolumn{2}{c}{NMI}  & \multicolumn{2}{c}{Purity}  \\\hline
  $\tau=0.25$ & \multicolumn{2}{c}{0.8965 (0.1418)} & \multicolumn{2}{c|}{0.9868 (0.0312)} & \multicolumn{2}{c}{0.9501 (0.0792)} & \multicolumn{2}{c}{0.9805 (0.0373)}   \\
  $50$\ \ $\tau=0.50$ & \multicolumn{2}{c}{0.9773 (0.1215)} & \multicolumn{2}{c|}{0.9952 (0.0312)} & \multicolumn{2}{c}{0.9712 (0.1073)} & \multicolumn{2}{c}{0.9933 (0.0336)}   \\
  $\tau=0.75$ & \multicolumn{2}{c}{0.9407 (0.0860)} & \multicolumn{2}{c|}{0.9770 (0.0482)} & \multicolumn{2}{c}{0.9392 (0.1184)} & \multicolumn{2}{c}{0.9762 (0.0472)}   \\
 \hline
  $\tau=0.25$ &\multicolumn{2}{c}{0.9622 (0.0868)}  & \multicolumn{2}{c|}{1.0000 (0.0000)}  &  \multicolumn{2}{c}{1.0000 (0.0000)} & \multicolumn{2}{c}{1.0000 (0.0000)}  \\
  $100$\ \ $\tau=0.50$ &\multicolumn{2}{c}{1.0000 (0.0000)}  & \multicolumn{2}{c|}{1.0000 (0.0000)}  &  \multicolumn{2}{c}{1.0000 (0.0000)} & \multicolumn{2}{c}{1.0000 (0.0000)}  \\
  $\tau=0.75$ &\multicolumn{2}{c}{0.9740 (0.0740)}  & \multicolumn{2}{c|}{1.0000 (0.0000)}  &  \multicolumn{2}{c}{1.0000 (0.0000)} & \multicolumn{2}{c}{1.0000 (0.0000)}  \\			
 \end{tabular}}

 \setlength{\tabcolsep}{0.5mm}{
 \begin{tabular}{l|ccc|ccc}
 \hline
 \hline			
 &\multicolumn{6}{|c}{\bf Average RMSE of ${\boldsymbol\beta}(\cdot)$ estimates (with standard deviation in parentheses) }\\
 \hline
 {\multirow{2}{*}{\diagbox[width=6em]{$T$}{$N$}}}  & \multicolumn{3}{c|}{$N=50$} & \multicolumn{3}{c}{$N=100$}  \\
 & Oracle & Preliminary & Post-grouping & Oracle & Preliminary & Post-grouping\\
 \hline
 ~~~~~ $\tau=0.25$ & {0.4346 (0.0261)} & {0.7906 (0.0357)} &{0.5035 (0.1095)} &{0.4319 (0.0221)} & {0.7913 (0.0290)} & {0.4534 (0.0551)}  \\
 $50$~~$\tau=0.50$ & {0.3661 (0.0215)} & {0.7164 (0.0357)} &{0.3741 (0.0456)} &{0.3694 (0.0203)} & {0.7161 (0.0288)} & {0.3778 (0.0377)}  \\
 ~~~~~ $\tau=0.75$ & {0.4568 (0.0265)} & {0.7900 (0.0348)} &{0.4854 (0.0595)} &{0.4329 (0.0216)} & {0.7904 (0.0288} & {0.4589 (0.0759)}  \\ \hline
 ~~~~~~$\tau=0.25$ & {0.2746 (0.0154)}  & {0.5587 (0.0186)}  & {0.3152 (0.0951)} & {0.2618 (0.0123)} & {0.5577 (0.0137)}  & {0.2618 (0.0123)} \\
 $100$~$\tau=0.50$ & {0.2331 (0.0134)}  & {0.5122 (0.0176)}  & {0.2331 (0.0134)} & {0.2409 (0.0121)} & {0.5083 (0.0131)}  & {0.2409 (0.0121)} \\
 ~~~~~~ $\tau=0.75$ & {0.2725 (0.0146)}  & {0.5590 (0.0193)}  & {0.3003 (0.0816)} & {0.2617 (0.0119)} & {0.5575 (0.0139)}  & {0.2617 (0.0119)} \\
 \hline
 \hline
 \end{tabular}}
\end{table}

\begin{table}[htp]\footnotesize
\renewcommand\arraystretch{0.7}
\caption{\label{tab5.5}Simulation results for DGP2 with $t(5)$ idiosyncratic errors}%
	\vspace{0.3cm}
	
	\setlength{\tabcolsep}{1.2mm}{
		\begin{tabular}{r|ccccc|ccccc}
			\hline
			\hline	
			&\multicolumn{10}{c}{\bf Percentage of replications each $\widehat{R}$ value is selected by the ratio criterion}\\
			&\multicolumn{10}{|c}{ with true $R_0=3$ for $\tau=0.25,0.75$ and $R_0=2$ for $\tau=0.50$}\\
			\hline
			{\multirow{2}{*}{\diagbox[width=6em]{$T$}{$N$}}} & \multicolumn{5}{c|}{$N=50$} & \multicolumn{5}{c}{$N=100$} \\
			& $\widehat{R}=1$ & $\widehat{R}=2$ & $\widehat{R}=3$& $\widehat{R}=4$ & $\widehat{R}=5$ & $\widehat{R}=1$ & $\widehat{R}=2$ & $\widehat{R}=3$ & $\widehat{R}=4$ & $\widehat{R}=5$ \\\hline
			$\tau=0.25$  & 0.0\% & 24.0\% &65.0\% & 10.0\%& 1.0\%&  0.0\% & 20.0\% & 72.0\% & 6.0\% & 2.0\%  \\
			$50$\ \ $\tau=0.50$ &   0.0\% & 88.0\% &11.5\% & 0.5\%& 0.0\%&   0.5\% & 97.5\% & 1.5\% & 0.5\% & 0.0\%  \\
			$\tau=0.75$ &  0.0\% & 11.5\% &72.0\% & 12.0\%& 4.5\%&   0.0\% & 3.5\% & 67.0\% & 22.0\% & 7.5\%  \\
			\hline
			$\tau=0.25$ &  0.0\% & 0.5\%  & 99.5\%& 0.0\%&0.0\% &   0.0\% & 0.0\% &99.5\ &0.5\% &0.0\%  \\
			$100$\ \  $\tau=0.50$ &  0.0\% & 100.0\%  & 0.0\%& 0.0\%&0.0\% &   0.0\% & 100.0\% &0.0\% &0.0\% &0.0\%  \\
			$\tau=0.75$ &  0.0\% & 0.0\%  & 99.5\%& 0.5\%&0.0\% &  0.0\% & 0.0\% &100.0\% &0.0\% &0.0\%  \\
	\end{tabular}}

	\setlength{\tabcolsep}{5mm}{
		\begin{tabular}{r|cccccccc}	
			\hline
			\hline					
			&\multicolumn{8}{c}{\bf Average NMI and purity (with standard deviation in parentheses)}\\
			\hline
			{\multirow{2}{*}{\diagbox[width=8em]{$T$}{$N$}}} & \multicolumn{4}{c|}{$N=50$} & \multicolumn{4}{c}{$N=100$} \\
			&  \multicolumn{2}{c}{NMI} & \multicolumn{2}{c|}{Purity}  & \multicolumn{2}{c}{NMI}  & \multicolumn{2}{c}{Purity}  \\\hline
			$\tau=0.25$ & \multicolumn{2}{c}{0.8100 (0.1563)} & \multicolumn{2}{c|}{0.9510 (0.0673)} & \multicolumn{2}{c}{0.8308 (0.1568)} & \multicolumn{2}{c}{0.9595 (0.0568)}   \\
			$50$\ \ $\tau=0.50$ & \multicolumn{2}{c}{0.9392 (0.1161)} & \multicolumn{2}{c|}{0.9754 (0.0602)} & \multicolumn{2}{c}{0.9400 (0.1601)} & \multicolumn{2}{c}{0.9851 (0.0475)}   \\
			$\tau=0.75$ & \multicolumn{2}{c}{0.8152 (0.1570)} & \multicolumn{2}{c|}{0.9381 (0.0704)} & \multicolumn{2}{c}{0.8282 (0.1490)} & \multicolumn{2}{c}{0.9266 (0.0757)}   \\
			\hline
			$\tau=0.25$ &\multicolumn{2}{c}{0.9985 (0.0174)}  & \multicolumn{2}{c|}{0.9999 (0.0014)}  &  \multicolumn{2}{c}{0.9989 (0.0080)} & \multicolumn{2}{c}{0.9996 (0.0044)}  \\
			$100$\ \ $\tau=0.50$ &\multicolumn{2}{c}{1.0000 (0.0000)}  & \multicolumn{2}{c|}{1.0000 (0.0000)}  &  \multicolumn{2}{c}{1.0000 (0.0000)} & \multicolumn{2}{c}{1.0000 (0.0000)}  \\
			$\tau=0.75$ &\multicolumn{2}{c}{0.9964 (0.0223)}  & \multicolumn{2}{c|}{0.9992 (0.0039)}  &  \multicolumn{2}{c}{0.9986 (0.0074)} & \multicolumn{2}{c}{0.9997 (0.0018)}  \\			
	\end{tabular}}

	\setlength{\tabcolsep}{0.5mm}{
		\begin{tabular}{l|ccc|ccc}
			\hline
			\hline
			&\multicolumn{6}{c}{\bf Average RMSE of ${\boldsymbol\beta}(\cdot)$ estimates (with standard deviation in parentheses) }\\
			\hline
			{\multirow{2}{*}{\diagbox[width=6em]{$T$}{$N$}}}  & \multicolumn{3}{c|}{$N=50$} & \multicolumn{3}{c}{$N=100$}  \\
			& Oracle & Preliminary & Post-grouping & Oracle & Preliminary & Post-grouping\\
			\hline
			~~~~~$\tau=0.25$ & {0.5024 (0.0300)} & {0.8822 (0.0384)} &{0.6117 (0.1148)} &{0.4754 (0.0276)} & {0.8879 (0.0319)} & {0.5684 (0.1178)}  \\
			$50$~~$\tau=0.50$ & {0.4436 (0.0288)} & {0.7799 (0.0348)} &{0.4580 (0.0380)} &{0.4209 (0.0256)} & {0.7829 (0.0283)} & {0.4400 (0.0654)}  \\
			~~~~~$\tau=0.75$ & {0.5141 (0.0293)} & {0.8878 (0.0356)} &{0.6057 (0.1029)} &{0.4978 (0.0275)} & {0.8895 (0.0290)} & {0.5739 (0.1050)}  \\ \hline
			~~~~~~$\tau=0.25$ & {0.3134 (0.0177)}  & {0.6338 (0.0242)}  & {0.3146 (0.0228)} & {0.3157 (0.0127)} & {0.6346 (0.0173)}  & {0.3160 (0.0127)} \\
			$100$~$\tau=0.50$ & {0.2663 (0.0142)}  & {0.5569 (0.0208)}  & {0.2663 (0.0142)} & {0.2581 (0.0114)} & {0.5567 (0.0180)}  & {0.2581 (0.0114)} \\
			~~~~~~$\tau=0.75$ & {0.3300 (0.0147)}  & {0.6344 (0.0235)}  & {0.3323 (0.0266)} & {0.3171 (0.0131)} & {0.6351 (0.0176)}  & {0.3175 (0.0131)} \\
			\hline\hline
	\end{tabular}}
\end{table}


\begin{table}[htp]\footnotesize
	\renewcommand\arraystretch{0.7}
	\caption{\label{tab5.6}Simulation results for DGP1 with $0.4*[\chi^2(3)-3]$ idiosyncratic errors}%
	\vspace{0.3cm}
	\setlength{\tabcolsep}{1.2mm}{
		\begin{tabular}{r|ccccc|ccccc}
			\hline
			\hline			
			&\multicolumn{10}{c}{\bf Percentage of replications each $\widehat{R}$ value is selected by the ratio criterion}\\
			&\multicolumn{10}{|c}{ with true $R_0=3$ for $\tau=0.25,0.50,0.75$}\\
			\hline
			{\multirow{2}{*}{\diagbox[width=6em]{$T$}{$N$}}} & \multicolumn{5}{c|}{$N=50$} & \multicolumn{5}{c}{$N=100$} \\
			& $\widehat{R}=1$ & $\widehat{R}=2$ & $\widehat{R}=3$& $\widehat{R}=4$ & $\widehat{R}=5$ &  $\widehat{R}=1$ & $\widehat{R}=2$ & $\widehat{R}=3$ & $\widehat{R}=4$ & $\widehat{R}=5$ \\\hline
			$\tau=0.25$ &  0.0\% & 4.5\% &94.5\% & 1.0\%& 0.0\%&   0.0\% & 3.0\% & 96.5\% & 0.5\% & 0.0\%  \\
			$50$\ \ $\tau=0.50$ &   0.0\% & 94.0\% &5.0\% & 1.0\%& 5.5\%&   0.0\% & 99.0\% & 0.5\% & 0.5\% & 0.0\%  \\
			$\tau=0.75$ &  0.0\% & 9.5\% &79.5\% & 10.0\%& 1.0\%&   0.0\% & 16.5\% & 76.0\% & 6.5\% & 1.0\%  \\
			\hline
			$\tau=0.25$ &  0.0\% & 0.5\%  & 99.5\%& 0.0\%&0.0\% &   0.0\% & 0.0\% &100.0\% &0.0\% &0.0\%  \\
			$100$\ \ $\tau=0.50$ &  0.0\% & 100.0\%  & 0.0\%& 0.0\%&0.0\% &   0.0\% & 100.0\% &0.0\% &0.0\% &0.0\%  \\
			$\tau=0.75$ &  0.0\% & 0.0\%  & 99.5\%& 0.5\%&0.0\% &   0.0\% & 0.0\% &100.0\% &0.0\% &0.0\%  \\ 
	\end{tabular}}

	\setlength{\tabcolsep}{5mm}{
		\begin{tabular}{r|cccccccc}
			\hline
			\hline			
			 &\multicolumn{8}{|c}{\bf Average NMI and purity (with standard deviation in parentheses)}\\
			\hline
           {\multirow{2}{*}{\diagbox[width=8em]{$T$}{$N$}}} & \multicolumn{4}{c|}{$N=50$} & \multicolumn{4}{c}{$N=100$} \\
			&  \multicolumn{2}{c}{NMI} & \multicolumn{2}{c|}{Purity}  & \multicolumn{2}{c}{NMI}  & \multicolumn{2}{c}{Purity}  \\ \hline
			$\tau=0.25$ & \multicolumn{2}{c}{0.9803 (0.0631)} & \multicolumn{2}{c|}{0.9977 (0.0112)} & \multicolumn{2}{c}{0.9857 (0.0788)} & \multicolumn{2}{c}{0.9988 (0.0066)}   \\
			$50$\ \ $\tau=0.50$ & \multicolumn{2}{c}{0.9728 (0.1130)} & \multicolumn{2}{c|}{0.9901 (0.0416)} & \multicolumn{2}{c}{0.9906 (0.0737)} & \multicolumn{2}{c}{0.9991 (0.0054)}   \\
			$\tau=0.75$ & \multicolumn{2}{c}{0.8896 (0.1613)} & \multicolumn{2}{c|}{0.9715 (0.0431)} & \multicolumn{2}{c}{0.8576 (0.1657)} & \multicolumn{2}{c}{0.9682 (0.0503)}   \\
			\hline
			$\tau=0.25$ &\multicolumn{2}{c}{0.9988 (0.0167)}  & \multicolumn{2}{c|}{1.0000 (0.0000)}  &  \multicolumn{2}{c}{1.0000 (0.0000)} & \multicolumn{2}{c}{1.0000 (0.0000)}  \\
			$100$\ \  $\tau=0.50$ &\multicolumn{2}{c}{1.0000 (0.0000)}  & \multicolumn{2}{c|}{1.0000 (0.0000)}  &  \multicolumn{2}{c}{1.0000 (0.0000)} & \multicolumn{2}{c}{1.0000 (0.0000)}  \\
			$\tau=0.75$ &\multicolumn{2}{c}{0.9984 (0.0117)}  & \multicolumn{2}{c|}{0.9994 (0.0044)}  &  \multicolumn{2}{c}{0.9998 (0.0029)} & \multicolumn{2}{c}{1.0000 (0.0007)}  \\
	\end{tabular}}

	\setlength{\tabcolsep}{0.5mm}{
	\begin{tabular}{l|ccc|ccc}
			\hline
			\hline
    &\multicolumn{6}{|c}{\bf Average RMSE of ${\boldsymbol\beta}(\cdot)$ estimates (with standard deviation in parentheses) }\\
   \hline
    {\multirow{2}{*}{\diagbox[width=6em]{$T$}{$N$}}}  & \multicolumn{3}{c|}{$N=50$} &  \multicolumn{3}{c}{$N=100$}  \\
   & Oracle & Preliminary & Post-grouping & Oracle & Preliminary & Post-grouping\\
   \hline
   ~~~~~$\tau=0.25$ & {0.3157 (0.0202)} & {0.6084 (0.0425)} &{0.3320 (0.0615)} &{0.3282 (0.0192)} & {0.6088 (0.0323)} & {0.3373 (0.0598)}  \\
   $50$~~$\tau=0.50$ & {0.3231 (0.0215)} & {0.6547 (0.0423)} &{0.3324 (0.0598)} &{0.3025 (0.0175)} & {0.6562 (0.0334)} & {0.3064 (0.0433)}  \\
   ~~~~~$\tau=0.75$ & {0.4316 (0.0253)} & {0.8524 (0.0421)} &{0.4974 (0.1181)} &{0.4691 (0.0237)} & {0.8515 (0.0271)} & {0.5465 (0.1112)}  \\ \hline
   ~~~~~~$\tau=0.25$ & {0.2252 (0.0118)}  & {0.3942 (0.0174)}  & {0.2263 (0.0182)} & {0.2007 (0.0092)} & {0.3887 (0.0137)}  & {0.2007 (0.0092)} \\
   $100$~$\tau=0.50$ & {0.2268 (0.0123)}  & {0.4516 (0.0174)}  & {0.2268 (0.0123)} & {0.2347 (0.0110)} & {0.4513 (0.0126)}  & {0.2347 (0.0110)} \\
   ~~~~~~$\tau=0.75$ & {0.3063 (0.0162)}  & {0.6164 (0.0222)}  & {0.3070 (0.0162)} & {0.3067 (0.0133)} & {0.6152 (0.0157)}  & {0.3067 (0.0133)} \\
   \hline
   \hline
	\end{tabular}}
\end{table}



\renewcommand{\arraystretch}{1}

\subsection{Empirical analysis}\label{sec5.2}

To further illustrate the applicability and usefulness of our methods, we next consider a panel house price growth model for UK local authority districts (LADs). Similarly to \cite{CSZ22}, we use quarterly house price data over the period  Q1/1997 -- Q4/2016\footnote{Due to data unavailability, we consider LADs only in England and Wales.}, downloaded from the UK Office of National Statistics (ONS) website: \url{https://www.ons.gov.uk/}. Growth rates in population and the nominal per capita personal income are used as the explanatory variables. Their data at the individual LAD level are available at the annual rate on the ONS website\footnote{Four LADs from England and Wales - Aylesbury Vale, Gloucester, Norwich, and Powys, are excluded due to their outlying values. This gives a total of 335 LADs for the subsequent analysis.}, and we construct quarterly data for the two variables using the interpolation method in \cite{Den71} and \cite{DC06}. For the index variable, we use quarterly inflation rates, which are available at the country level at \url{https://www.bls.gov/cpi/data.htm}. This allows for interaction between inflation and the explanatory variables and for the effects of the explanatory variables on house prices to vary with inflation. Data for all the variables have been de-seasoned and de-trended\footnote{More data details can be found in \cite{CSZ22}.}.

Assume the following panel quantile regression:
\begin{equation}\label{eq5.2}
{\sf Q}_{\tau,i}(hp_{it}|pop_{it},inc_{i,t-1},inf_{t},\alpha_i)=\alpha_i+\beta_{i,1}(inf_t)pop_{it}+\beta_{i,2}(inf_t)inc_{i,t-1},
\end{equation}
for $i=1,\cdots, 335$ and $t=2,\cdots, 80$, where $hp_{it}$ is the house price growth rate (in \%) for the $i$-th LAD in the $t$-th quarter, $pop_{it}$ is the population growth (in \%), $inc_{i,t-1}$ is the growth in per capita personal income (in \%), $inf_t$ is the inflation rate (in \%), and $\alpha_i$ is the fixed effect. Here $inc$ is lagged by one time period due to the likely lagged effect of income growth on house price.

\begin{figure}
	\centering
	{\includegraphics[height=8.5cm]{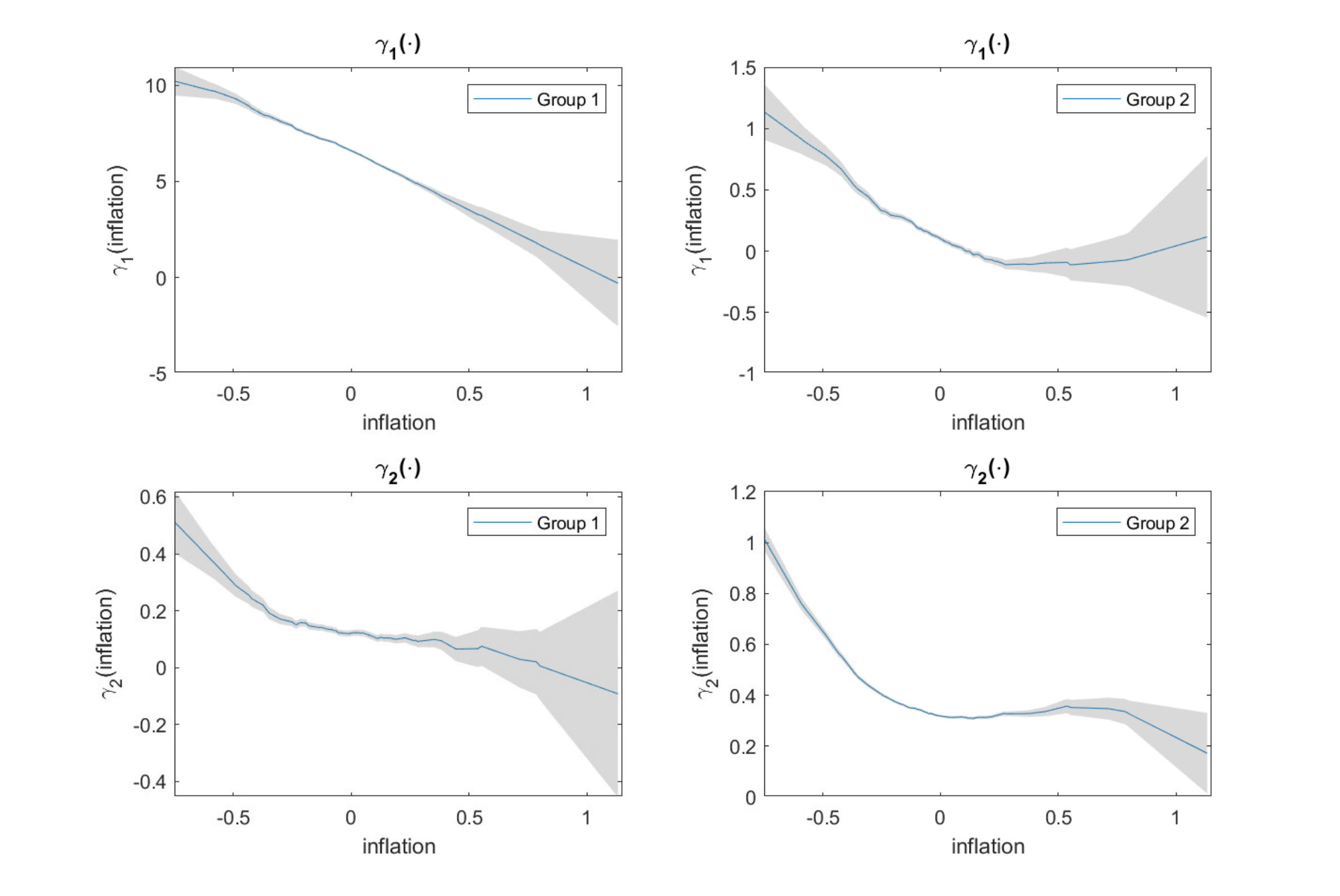}}
	\caption{\small \label{fig1}Post-grouping local linear estimates of the functional coefficients (with 95\% confidence intervals) for population growth ($\gamma_1(\cdot)$, top row) and income growth ($\gamma_2(\cdot)$, bottom row) at $\tau=0.25$: left - for Group 1; right - for Group 2.}
\end{figure}

While the model (\ref{eq5.2}) offers great flexibility by allowing the coefficients to vary across  LADs and quantiles as well as with inflation, there may exist some homogeneity groups of LADs at each quantile level, where the coefficients are homogeneous within each group (while still varying with inflation) but heterogeneous across groups. That is, at a given quantile level $\tau$, there may exist a partition of the index set $\{1,2,,\cdots,335\}$, denoted as $\{{\cal G}_1,{\cal G}_2,\cdots,{\cal G}_{R_0}\}$, such that
\[
{\boldsymbol\beta}_i(\cdot):=\left(\begin{array}{c}\beta_{i,1}(\cdot)\\\beta_{i,2}(\cdot)\end{array}\right)={\boldsymbol\gamma}_j(\cdot):=\left(\begin{array}{c}\gamma_{j,1}(\cdot)\\ \gamma_{j,2}(\cdot)\end{array}\right)\ \ {\rm for}\ i\in{\cal G}_j,\ \ \ \ i=1,\cdots,335,\ j=1,\cdots,R_0.\notag
\]
We consider 3 quantile levels, i.e., $\tau=0.25, 0.50, 0.75$, and at each quantile, we then use the methods proposed in Section \ref{sec3} to estimate the number of latent groups as well as the group membership. The results are summarised in Table \ref{tab5.7}. The post-grouping local linear estimates of the group-specific functional coefficients, together with their 95\% confidence intervals, are plotted in Figures \ref{fig1}--\ref{fig3} for $\tau=0.25, 0.50, 0.75$, respectively. Construction of the confidence intervals is introduced in Appendix C.3 of the online supplement. The estimated groups at each quantile level, projected onto a choropleth map, are shown in Figures \ref{fig4}--\ref{fig6}.

\smallskip

At both $\tau=0.25$ and $\tau=0.50$, two groups of LADs are identified, although the membership of the groups is not exactly the same\footnote{The membership of the two groups at $\tau=0.50$ is similar to that of the two groups at $\tau=0.25$ with a large number of overlapping member LADs (e.g. for Group 1, there are 45 overlapping LADs and for Group 2, there are 247 overlapping LADs).}. Furthermore, Figures \ref{fig1} and \ref{fig2} show that the coefficient functions for each group have similar patterns at these two quantiles, especially for that of population growth. The coefficients are mostly positive, consistent with the economic theory that growth in population and income leads to growth in demand for housing and hence, a rise in house price. At $\tau=0.25$, the effects of population and income growth in general decrease as inflation increases. This decreasing trend is more marked for the effect of population growth in Group 1 LADs (88\% of which are non-metropolitan districts or unitary authorities), but much less so in Group 2 LADs (which include 88\% of the London boroughs and 89\% of the metropolitan districts). For example, when inflation rate is -0.5\%, for each 1\% increase in population growth, house price growth is increased by around 8\%. But when inflation rate is 0.5\%, a 1\% increase in population growth leads only to a 2\% increase in house price growth. At $\tau=0.50$, we observe a similar trend. We can also find, from Figures \ref{fig1} and \ref{fig2}, that at $\tau=0.25$ and $\tau=0.50$, the effect of population growth on house price growth dominates that of income growth for Group 1 LADs, which are mainly non-metropolitan or unitary districts. At $\tau=0.75$, where five groups are identified, we see some different trends for different groups (see Figure \ref{fig3}). While most values of the coefficient functions are positive, there are negative values for some groups in some subintervals. For example, the coefficient function of population growth for Group 1 exhibits a downward sloping pattern with increasing inflation and is negative when inflation is above $-0.7\%$.  A closer examination of the member LADs reveals that three out of the four LADs of this group are urban districts with major or minor conurbation, where immigration is more likely to occur. If population growth is driven by migration inflows, its effect on house price growth might be ambiguous and sometimes even negative \citep{Sa15}. \cite{CSZ22} also find negative effects of population growth on house price growth in some LADs.

For comparison, we also conduct a grouping analysis based on the functional-coefficient mean regression for the same variables. The same ratio criterion and clustering algorithm are used for choosing the number of groups and estimating the group membership. The obtained results are similar to those from the quantile analysis at $\tau=0.50$: two groups are found with similar, although not exactly the same, membership to those from the quantile regression at $\tau=0.50$. Results from the mean regression are more susceptible to the influence of outliers. More detail about the mean regression grouping results can be found in Appendix D of the online supplement.

\begin{table}
	\renewcommand\arraystretch{0.6}
	\centering
	\caption{\label{tab5.7}Estimated latent groups for house price data from 335 UK LADs}%
	\vspace{0.2cm}
	\begin{tabular}{c|c|c}
		\hline
		\hline
		Quantile & Number of groups & Cardinality of each group \\ \hline
		$\tau=0.25$ & $\widehat{R}=2$ & 69, 266 \\
		$\tau=0.50$ & $\widehat{R}=2$ & 64, 271 \\
		$\tau=0.75$ & $\widehat{R}=5$ & 4, 59, 45, 204, 23 \\ 
		\hline
		\hline
	\end{tabular}
\end{table}

The above analysis reveals that at $\tau=0.25$ and $\tau=0.50$, there is more homogeneity  in the effects of population and income growth on house price growth across LADs, and in general these effects are positive and decrease as inflation increases. At $\tau=0.75$, more heterogeneity is observed, and for some identified groups the effects of population and income growth are negative for some values of inflation. It also appears that at $\tau=0.75$, the effect of income growth is larger than at lower quantiles. This empirical application demonstrates the benefit of the quantile grouping analysis: it can shed more light on the impact of population and income growth on house price growth across LADs than a mean regression based grouping analysis does. The results may be useful for policy makers in developing more targeted policies for specific groups of districts at specific quantiles.

\begin{figure}
	\centering
	{\includegraphics[height=8.5cm]{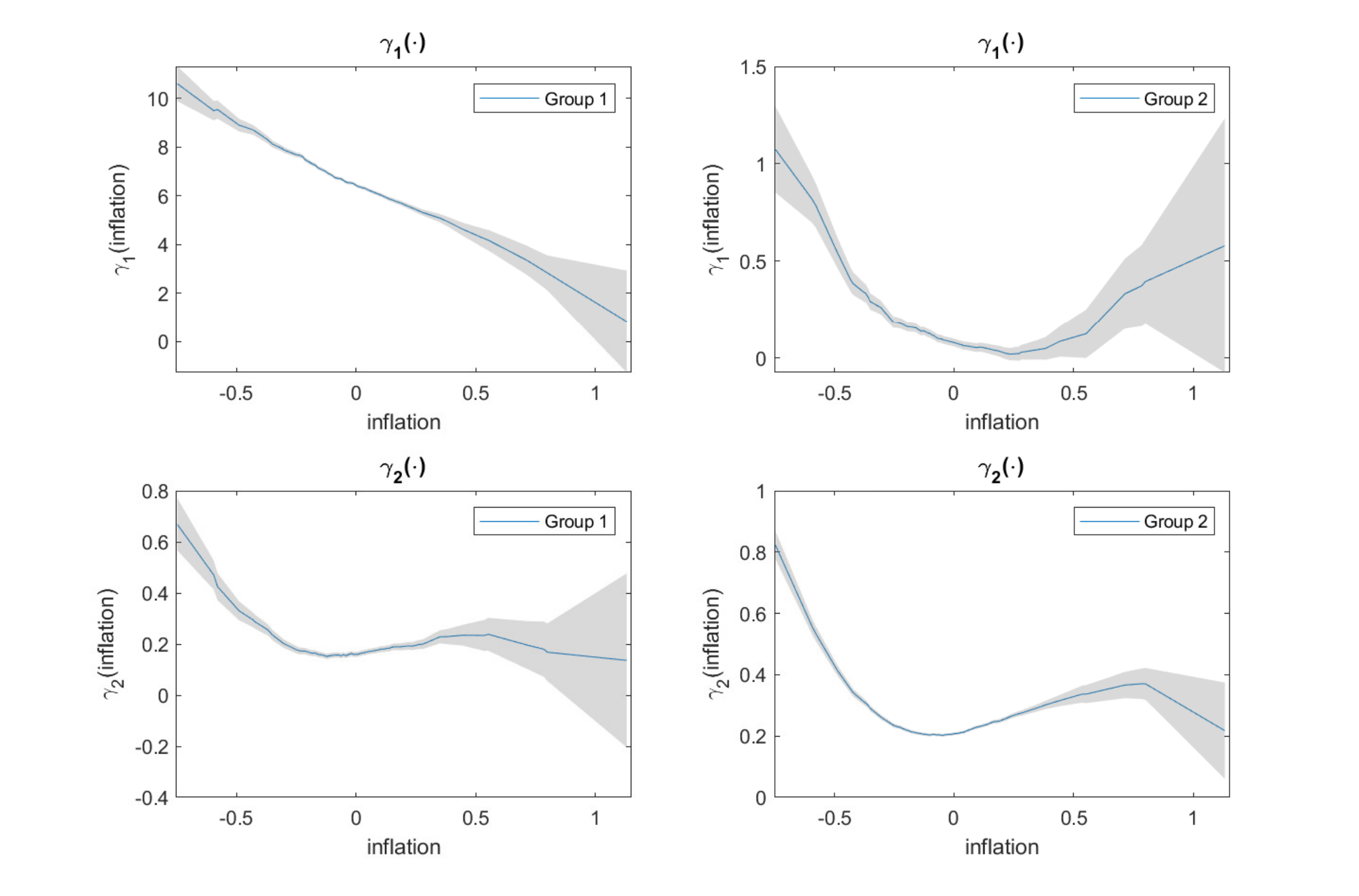}}
	\caption{\small \label{fig2}Post-grouping local linear estimates of the functional coefficients (with 95\% confidence intervals) for population growth ($\gamma_1(\cdot)$, top row) and income growth ($\gamma_2(\cdot)$, bottom row) at $\tau=0.50$: left - for Group 1; right - for Group 2.}
\end{figure}

\begin{figure}
	\begin{center}
	{\includegraphics[height=8.6cm]{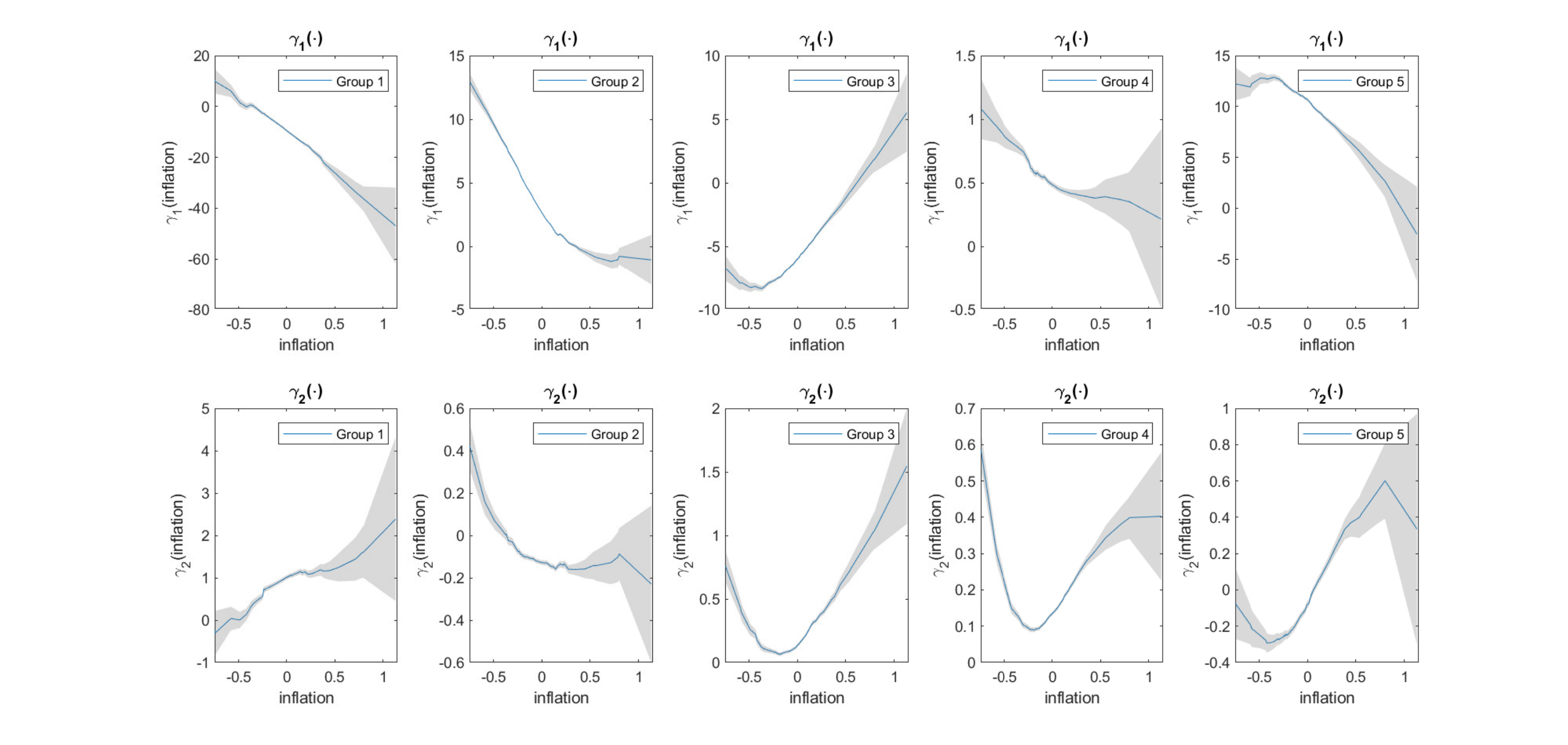}}
	\caption{\small \label{fig3}Post-grouping local linear estimates of the functional coefficients (with 95\% confidence intervals) for population growth ($\gamma_1(\cdot)$, top row) and income growth ($\gamma_2(\cdot)$, bottom row) at $\tau=0.75$: from left to right - Group 1,  Group 2,  Group 3, Group 4, Group 5.}
	\end{center}
\end{figure}

\begin{figure}
	\centering
	{\includegraphics[width=1\textwidth,height=9cm]{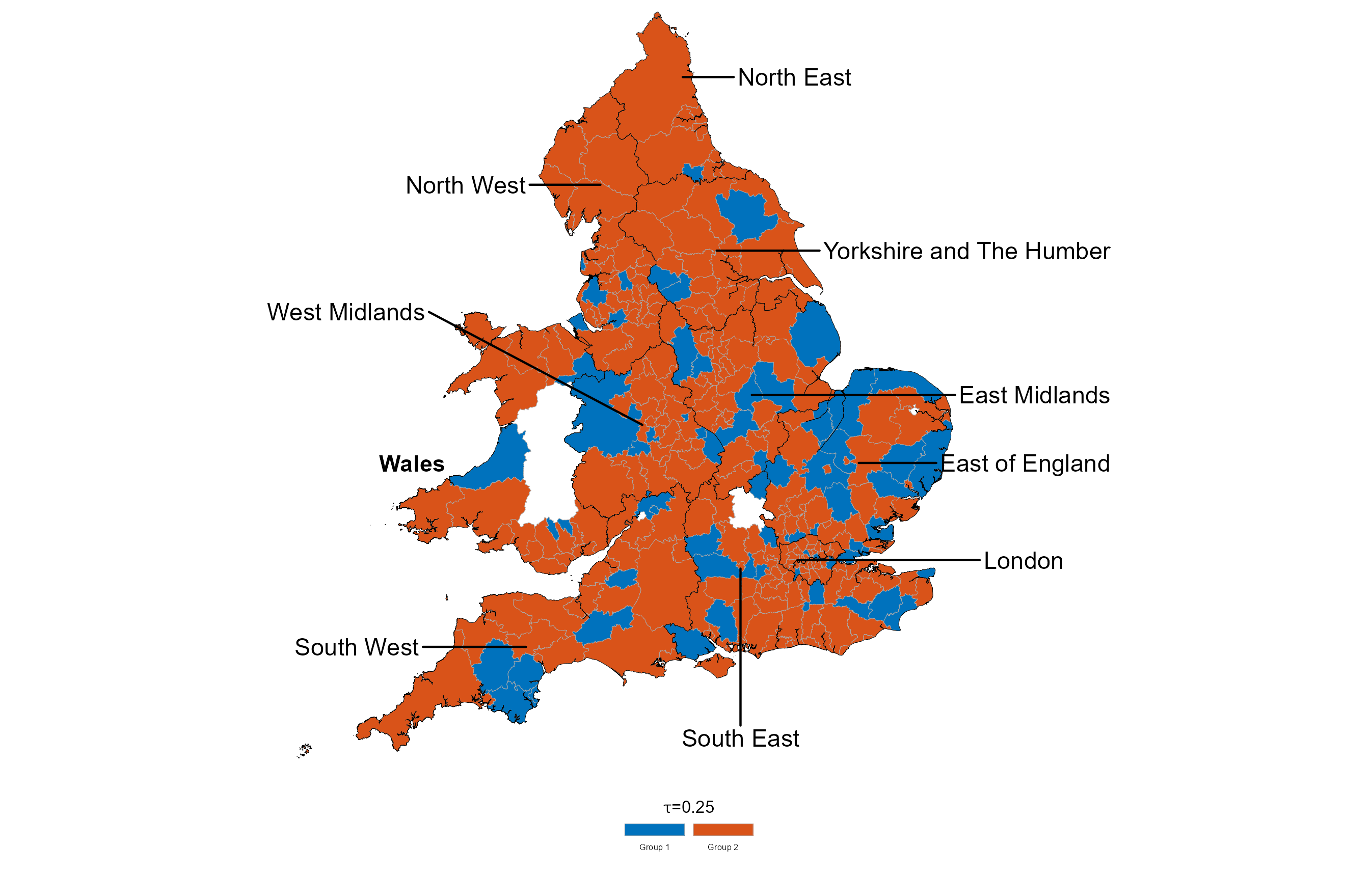}}
	\caption{\small \label{fig4}Spatial pattern of estimated groups for $\tau=0.25$: blue LADs - Group 1; red LADs - Group 2.}
\end{figure}

\begin{figure}
	\centering
	{\includegraphics[width=1\textwidth,height=9cm]{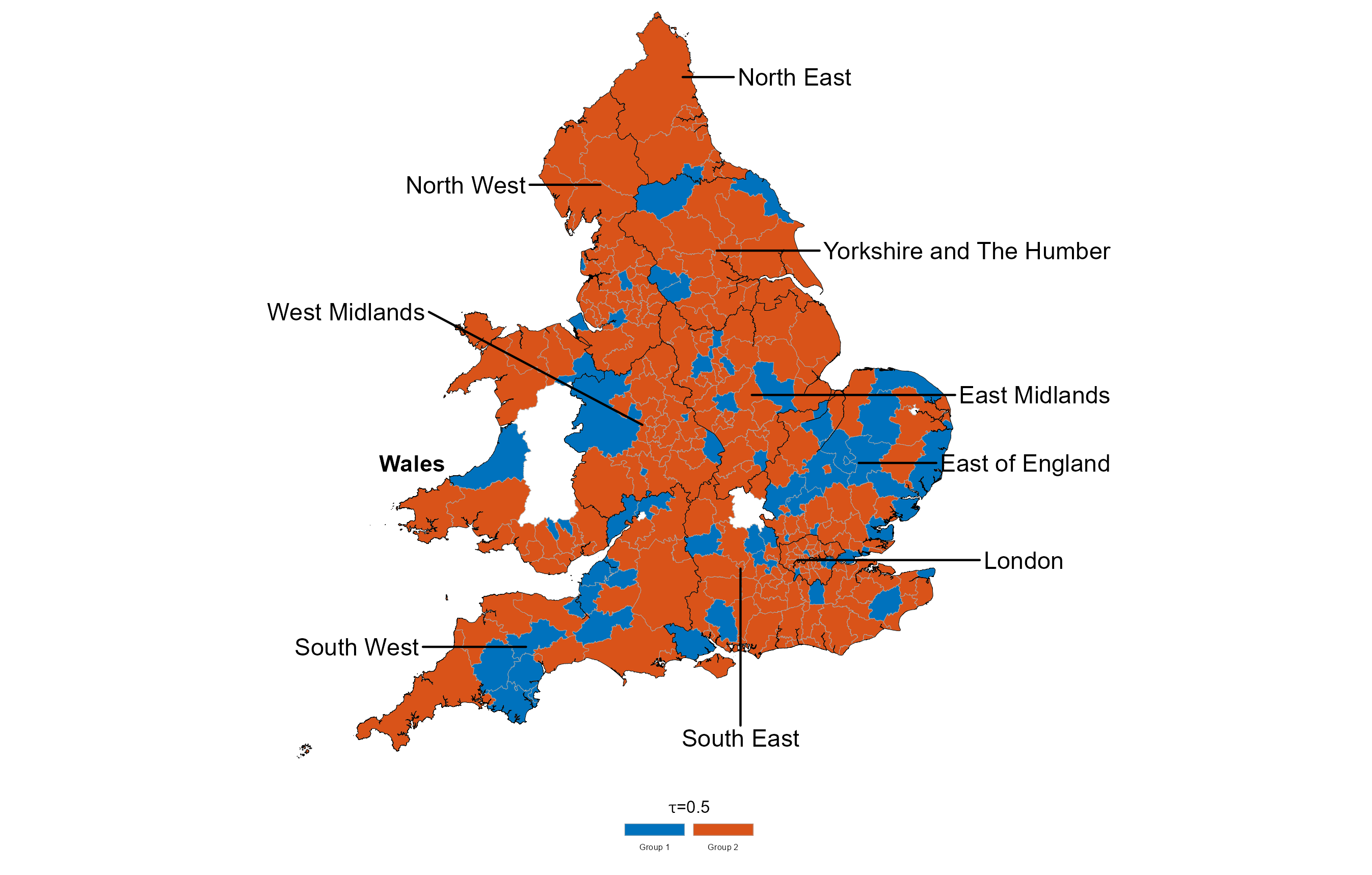}}
	\caption{\small \label{fig5}Spatial pattern of estimated groups for $\tau=0.50$: blue LADs - Group 1; red LADs - Group 2.}
\end{figure}

\begin{figure}
	\centering
	{\includegraphics[width=1\textwidth,height=9cm]{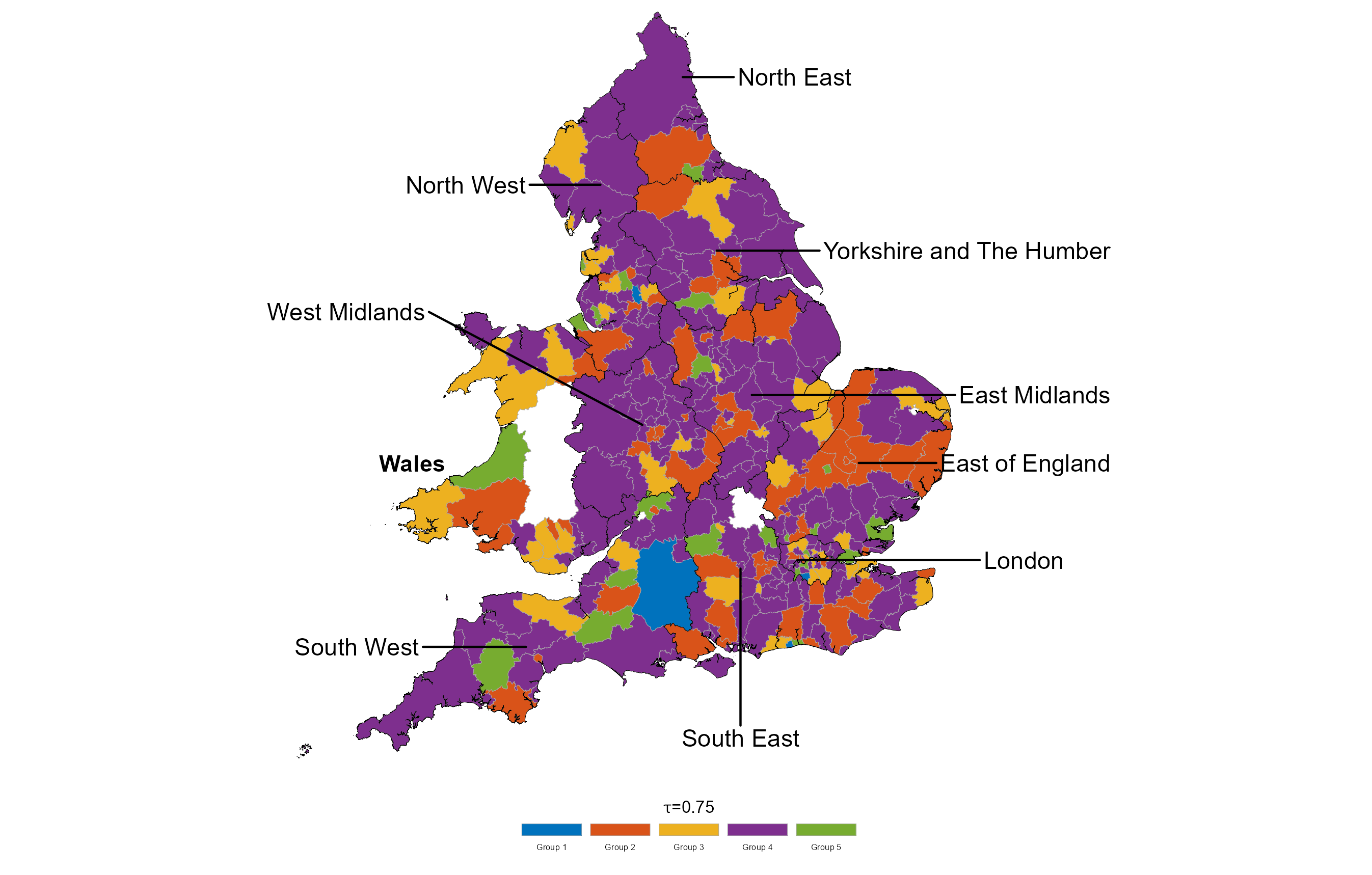}}
	\caption{\small \label{fig6}Spatial pattern of estimated groups for $\tau=0.75$: blue LADs - Group 1; red LADs - Group 2; yellow LADs - Group 3; purple LADs - Group 4; green LADs - Group 5.}
\end{figure}


\section{Conclusion}\label{sec6}

In this paper, we propose a general functional-coefficient quantile regression model for large panel data and assume a latent group structure on the heterogenous functional coefficients. An estimation methodology which combines preliminary functional coefficient estimates (ignoring the latent group structure), an agglomerative clustering algorithm and a simple ratio criterion is introduced to consistently estimate the group number and membership. Furthermore, a post-grouping local linear quantile regression method is used to estimate the group-specific functional coefficients, aiming to achieve faster convergence rates than the preliminary local linear estimator. To asymptotically remove the influence of nuisance parameters and derive an asymptotic normal distribution theory comparable to that in the literature such as \cite{KGM12}, we impose a relatively restrictive condition on the divergence rate of the group size, but allow weak cross-sectional and temporal dependence for large panel observations. The simulation studies show that the proposed methods have reliable finite-sample performance. The empirical application to the UK house price data reveals that the latent structures vary over different quantile levels with more heterogeneity observed at the upper quartile.

\section*{Acknowledgements}

The authors would like to thank an Associate Editor and two reviewers for the constructive comments, which helped to substantially improve the paper. The authors also thank Dr Chaowen Zheng for collecting the empirical data and preparing Figures \ref{fig4}-\ref{fig6}. The first author is partially supported by Natural Science Foundation of Zhejiang Province (No. LY22A010006), National Social Science Foundation of China (No. 17BTJ027) and the Fundamental Research Funds for Provincial Universities of Zhejiang. The second author is partially supported by the Economic and Social Research Council in the UK (No. ES/T01573X/1). The third author is partially supported by the National Natural Science Foundation of China (No. 72033002). The usual disclaimer applies.

\bigskip
\begin{center}
{\large\bf SUPPLEMENT}
\end{center}

The supplemental document contains proofs of the main asymptotic theorems, technical lemmas with proofs, some extensions of the developed method and theory, and additional empirical result.


{\footnotesize\bibliographystyle{Chicago}

\bibliography{MYREF}}

\begin{thebibliography}{}

\bibitem[\protect\citeauthoryear{Ahn and Horenstein}{Ahn and Horenstein}{2013}]{AH13}
Ahn, S.~C. and A.~R. Horenstein (2013). \newblock Eigenvalue ratio test for the number of factors.
\newblock {\em Econometrica\/}~{\bf 81}(3), 1203--1227.

\bibitem[\protect\citeauthoryear{Ando and Bai}{Ando and Bai}{2017}]{AB17} Ando, T. and J.~Bai (2017). \newblock Clustering huge number of financial time series: A panel data approach with high-dimensional predictors and factor structures. \newblock {\em Journal of the American Statistical Association\/}~{\bf
  112\/}(519), 1182--1198.

\bibitem[\protect\citeauthoryear{Belloni, Chernozhukov, Chetverikov, and Fern{\'a}ndez-Val}{Belloni et~al.}{2019}]{BCCF19} Belloni, A., V.~Chernozhukov, D.~Chetverikov, and I.~Fern{\'a}ndez-Val (2019).
\newblock Conditional quantile processes based on series or many regressors. \newblock {\em Journal of Econometrics\/}~{\bf 213}(1), 4--29.

\bibitem[\protect\citeauthoryear{Bonhomme and Manresa}{Bonhomme and Manresa}{2015}]{BM15}
Bonhomme, S. and E.~Manresa (2015). \newblock Grouped patterns of heterogeneity in panel data.
\newblock {\em Econometrica\/}~{\bf 83\/}(3), 1147--1184.

\bibitem[\protect\citeauthoryear{Bosq}{Bosq}{1998}]{Bo98}
Bosq, D. (1998).
\newblock {\em Nonparametric Statistics for Stochastic Processes: Estimation
  and Prediction (2nd Edition)}.
\newblock Lecture Notes in Statistics 110, Springer-Verlag, Berlin.

\bibitem[\protect\citeauthoryear{Cai}{Cai}{2002}]{Ca02} Cai, Z. (2002). \newblock Regression quantiles for time series. \newblock {\em Econometric Theory\/}~{\bf 18\/}(1), 169--192.

\bibitem[\protect\citeauthoryear{Cai, Chen, and Fang}{Cai et~al.}{2018}]{CCF18} Cai, Z., L.~Chen, and Y.~Fang (2018). \newblock A semiparametric quantile panel data model with an application to estimating the growth effect of fdi. \newblock {\em Journal of Econometrics\/}~{\bf 206\/}(2), 531--553.

\bibitem[\protect\citeauthoryear{Cai and Xu}{Cai and Xu}{2008}]{CX08} Cai, Z. and X.~Xu (2008).
\newblock Nonparametric quantile estimations for dynamic smooth coefficient models. \newblock {\em Journal of the American Statistical Association\/}~{\bf 103\/}(484), 1595--1608.

\bibitem[\protect\citeauthoryear{Canay}{Canay}{2011}]{Ca11} Canay, I.~A. (2011). \newblock A simple approach to quantile regression for panel data. \newblock {\em The Econometrics Journal\/}~{\bf 14\/}(3), 368--386.

\bibitem[\protect\citeauthoryear{Chen}{Chen}{2019}]{Ch19} Chen, J. (2019). \newblock Estimating latent group structure in time-varying coefficient panel data models. \newblock {\em The Econometrics Journal\/}~{\bf 22\/}(3), 223--240.

\bibitem[\protect\citeauthoryear{Chen, Li, Wei, and Zhang}{Chen et~al.}{2021}]{CLWZ21} Chen, J., D.~Li, L.~Wei, and W.~Zhang (2021). \newblock Nonparametric homogeneity pursuit in functional-coefficient models. \newblock {\em Journal of Nonparametric Statistics\/}~{\bf 33\/}(3-4), 387--416.

\bibitem[\protect\citeauthoryear{Chen, Shin, and Zheng}{Chen et~al.}{2022}]{CSZ22} Chen, J., Y.~Shin, and C.~Zheng (2022). \newblock Estimation and inference in heterogeneous spatial panels with a
  multifactor error structure. \newblock {\em Journal of Econometrics\/}~{\bf 229\/}(1), 55--79.

\bibitem[\protect\citeauthoryear{Chen}{Chen}{2021}]{Ch21} Chen, L. (2021). \newblock Nonparametric quantile regressions for panel data models with large $T$. \newblock Working paper available at
  \url{https://arxiv.org/abs/1911.01824}.
  
  
\bibitem[\protect\citeauthoryear{Cheng and Peng}{Cheng and Peng}{2002}]{CP02}
Cheng, M. and L.~Peng (2002).
\newblock Regression modeling for nonparametric estimation of distribution and quantile functions.
\newblock {\em Statistica Sinica\/}~{\bf 12\/}(4), 1043--1060.

\bibitem[\protect\citeauthoryear{Chernozhukov and Hansen}{Chernozhukov and  Hansen}{2006}]{CH06} Chernozhukov, V. and C.~Hansen (2006). \newblock Instrumental quantile regression inference for structural and treatment effect models. \newblock {\em Journal of Econometrics\/}~{\bf 132\/}(2), 491--525.

\bibitem[\protect\citeauthoryear{Chetverikov, Larsen and Palmer}{Chetverikov et~al.}{2016}]{CLP16} Chetverikov, D., B. Larsen and C. Palmer (2016).
\newblock IV quantile regression for group-level treatments, with an application to the distributional effects of trade.\newblock {\em Econometrica\/}~{\bf 84\/}(2), 809--833.


\bibitem[\protect\citeauthoryear{Dagum and Cholette}{Dagum and Cholette}{2006}]{DC06}
Dagum, E.~B. and P.~A. Cholette (2006). \newblock {\em Benchmarking, Temporal Distribution, and Reconciliation Methods for Time Series}. \newblock Lecture Notes in Statistics, Springer.

\bibitem[\protect\citeauthoryear{Denton}{Denton}{1971}]{Den71} Denton, F.~T. (1971).
\newblock Adjustment of monthly or quarterly series to annual totals: an approach based on quadratic minimization. \newblock {\em Journal of the American Statistical Association\/}~{\bf 66}(333), 99--102.

\bibitem[\protect\citeauthoryear{Everitt, Landau, Leese, and Stahl}{Everitt et~al.}{2011}]{ELLS11}
Everitt, B.~S., S.~Landau, M.~Leese, and D.~Stahl (2011). \newblock {\em Cluster Analysis (5th Edition)}.
\newblock Wiley Series in Probability and Statistics, Wiley.


\bibitem[\protect\citeauthoryear{Fan and Yao}{Fan and Yao}{2003}]{FY03}
Fan, J. and Q.~Yao (2003).
\newblock {\em Nonlinear Time Series: Nonparametric and Parametric Methods}.
\newblock Springer.

\bibitem[\protect\citeauthoryear{Galvao, Gu, and Volgushev}{Galvao et~al.}{2020}]{GGV20}
Galvao, A.~F., J.~Gu, and S.~Volgushev (2020). \newblock On the unbiased asymptotic normality of quantile regression with fixed effects. \newblock {\em Journal of Econometrics\/}~{\bf 218}(1), 178--215.

\bibitem[\protect\citeauthoryear{Galvao and Kato}{Galvao and Kato}{2016}]{GK16} Galvao, A.~F. and K.~Kato (2016). \newblock Smoothed quantile regression for panel data. \newblock {\em Journal of Econometrics\/}~{\bf 193\/}(1), 92--112.

\bibitem[\protect\citeauthoryear{Galvao, Lamarche, and Lima}{Galvao et~al.}{2013}]{GLL13}
Galvao, A.~F., C.~Lamarche, and L.~R. Lima (2013). \newblock Estimation of censored quantile regression for panel data with fixed effects.\newblock {\em Journal of the American Statistical Association\/}~{\bf
  108\/}(503), 1075--1089.

\bibitem[\protect\citeauthoryear{Gu and Volgushev}{Gu and Volgushev}{2019}]{GV19} Gu, J. and S.~Volgushev (2019). \newblock Panel data quantile regression with grouped fixed effects. \newblock {\em Journal of Econometrics\/}~{\bf 213\/}(1), 68--91.

\bibitem[\protect\citeauthoryear{Hahn and Moon}{Hahn and Moon}{2010}]{HM10} Hahn, J. and H.~R. Moon (2010). \newblock Panel data models with finite number of multiple equilibria. \newblock {\em Econometric Theory\/}~{\bf 26\/}(3), 863--881.

\bibitem[\protect\citeauthoryear{Honda}{Honda}{2004}]{Ho04} Honda, T. (2004). \newblock Quantile regression in varying coefficient models. \newblock {\em Journal of Statistical Planning and Inference\/}~{\bf 121\/}(1), 113--125.

\bibitem[\protect\citeauthoryear{Kai, Li, and Zou}{Kai et~al.}{2011}]{KLZ11}Kai, B., R.~Li, and H.~Zou (2011).
\newblock New efficient estimation and variable selection methods for semiparametric varying-coefficient partially linear models. \newblock {\em The Annals of Statistics\/}~{\bf 39\/}(1), 305--332.

\bibitem[\protect\citeauthoryear{Kato}{Kato}{2012}]{K12} Kato, K. (2012). \newblock Estimation in functional linear quantile regression. \newblock {\em The Annals of Statistics\/}~{\bf 40\/}(6), 3108--3136.

\bibitem[\protect\citeauthoryear{Kato, Galvao~Jr, and Montes-Rojas}{Kato et~al.}{2012}]{KGM12}
Kato, K., A.~F. Galvao~Jr, and G.~V. Montes-Rojas (2012). \newblock Asymptotics for panel quantile regression models with individual effects. \newblock {\em Journal of Econometrics\/}~{\bf 170\/}(1), 76--91.

\bibitem[\protect\citeauthoryear{Ke, Li, and Zhang}{Ke et~al.}{2016}]{KLZ16} Ke, Y., J.~Li, and W.~Zhang (2016). \newblock Structure identification in panel data analysis. \newblock {\em The Annals of Statistics\/}~{\bf 44\/}(3), 1193--1233.

\bibitem[\protect\citeauthoryear{Kim}{Kim}{2007}]{Ki07} Kim, M.-O. (2007). \newblock Quantile regression with varying coefficients. \newblock {\em The Annals of Statistics\/}~{\bf 35\/}(1), 92--108.

\bibitem[\protect\citeauthoryear{Koenker}{Koenker}{2004}]{Ko04} Koenker, R. (2004). \newblock Quantile regression for longitudinal data. \newblock {\em Journal of Multivariate Analysis\/}~{\bf 91\/}(1), 74--89.

\bibitem[\protect\citeauthoryear{Koenker}{Koenker}{2005}]{Ko05} Koenker, R. (2005). \newblock {\em Quantile regression}. \newblock Cambridge University Press.

\bibitem[\protect\citeauthoryear{Koenker and Bassett}{Koenker and Bassett}{1978}]{KB78}
Koenker, R. and G.~Bassett (1978). \newblock Quantile regression. \newblock {\em Econometrica\/}~{\bf 46}(1), 33--50.

\bibitem[\protect\citeauthoryear{Koenker, Chernozhukov, He, and Peng}{Koenker et~al.}{2017}]{KCHP17}
Koenker, R., V.~Chernozhukov, X.~He, and L.~Peng (2017). \newblock {\em Handbook of Quantile Regression}. \newblock Chapman and Hall/CRC.

\bibitem[\protect\citeauthoryear{Koenker and Xiao}{Koenker and Xiao}{2006}]{KX06} Koenker, R. and Z.~Xiao (2006). \newblock Quantile autoregression. \newblock {\em Journal of the American Statistical Association\/}~{\bf 101\/}(475), 980--990.

\bibitem[\protect\citeauthoryear{Lam and Yao}{Lam and Yao}{2012}]{LY12} Lam, C. and Q.~Yao (2012).
\newblock Factor modeling for high-dimensional time series: inference for the number of factors.
\newblock {\em The Annals of Statistics\/}~{\bf 40}(2), 694--726.

\bibitem[\protect\citeauthoryear{Li, Chen, and Gao}{Li et~al.}{2011}]{LCG11} Li, D., J.~Chen, and J.~Gao (2011). \newblock Non-parametric time-varying coefficient panel data models with fixed effects.
\newblock {\em The Econometrics Journal\/}~{\bf 14\/}(3), 387--408.

\bibitem[\protect\citeauthoryear{Li, Li, and Li}{Li et~al.}{2021}]{LLL21} Li, D., Q.~Li, and Z.~Li (2021).
\newblock Nonparametric quantile regression estimation with mixed discrete and continuous data.
\newblock {\em Journal of Business \& Economic Statistics\/}~{\bf 39\/}(3), 741--756.

\bibitem[\protect\citeauthoryear{Li, Robinson, and Shang}{Li  et~al.}{2020}]{LRS20} Li, D., P.~M. Robinson, and H.~L. Shang (2020). \newblock Long-range dependent curve time series. \newblock {\em Journal of the American Statistical Association\/}~{\bf 115\/}(530), 957--971.

\bibitem[\protect\citeauthoryear{Li, Lin, and Racine}{Li et~al.}{2013}]{LLR13} Li, Q., J.~Lin, and J.~S. Racine (2013). \newblock Optimal bandwidth selection for nonparametric conditional distribution and quantile functions. \newblock {\em Journal of Business \& Economic Statistics\/}~{\bf 31\/}(1), 57--65.

\bibitem[\protect\citeauthoryear{Lian, Qiao, and Zhang}{Lian  et~al.}{2021}]{LQZ21} Lian, H., X.~Qiao, and W.~Zhang (2021). \newblock Homogeneity pursuit in single index models based panel data analysis.
\newblock {\em Journal of Business \& Economic Statistics\/}~{\bf 39\/}(2), 386--401.

\bibitem[\protect\citeauthoryear{Liu, Shang, Zhang, and Zhou}{Liu et~al.}{2020}]{LSZZ20} Liu, R., Z.~Shang, Y.~Zhang, and Q.~Zhou (2020). \newblock Identification and estimation in panel models with overspecified number of groups. \newblock {\em Journal of Econometrics\/}~{\bf 215\/}(2), 574--590.

\bibitem[\protect\citeauthoryear{Phillips and Sul}{Phillips and Sul}{2007}]{PS07} Phillips, P.C.B. and D.~Sul (2007). \newblock Transition modeling and econometric convergence tests. \newblock {\em Econometrica\/}~{\bf 75\/}(6), 1771--1855.

\bibitem[\protect\citeauthoryear{Phillips and Wang}{Phillips and Wang}{2022}]{PW22} Phillips, P.C.B. and Y.~Wang (2022). \newblock Functional coefficient panel modeling with communal smoothing covariates. \newblock {\em Journal of Econometrics\/}~{\bf 227\/}(2), 371--407.

\bibitem[\protect\citeauthoryear{Rencher and Christensen}{Rencher and Christensen}{2012}]{RC12} Rencher, A.~C. and W.~F. Christensen (2012). \newblock {\em Methods of Multivariate Analysis (3rd Edition)}.\newblock Wiley Series in Probability and Statistics, Wiley.

\bibitem[\protect\citeauthoryear{Ruppert and Carroll}{Ruppert and
  Carroll}{1980}]{RC80}
Ruppert, D. and R.~J. Carroll (1980).
\newblock Trimmed least squares estimation in the linear model.
\newblock {\em Journal of the American Statistical Association\/}~{\bf
  75\/}(372), 828--838.


\bibitem[\protect\citeauthoryear{S{\'a}}{S{\'a}}{2015}]{Sa15} S{\'a}, F. (2015). \newblock Immigration and house prices in the UK. \newblock {\em The Economic Journal\/}~{\bf 125\/}(587), 1393--1424.

\bibitem[\protect\citeauthoryear{Shao and Yu}{Shao and Yu}{1996}]{SY96}
Shao, Q.-M. and H.~Yu (1996).
\newblock Weak convergence for weighted empirical processes of dependent
  sequences.
\newblock {\em The Annals of Probability\/} {\bf 24}(4), 2098--2127.


\bibitem[\protect\citeauthoryear{Su and Hoshino}{Su and Hoshino}{2016}]{SH16} Su, L. and T.~Hoshino (2016). \newblock Sieve instrumental variable quantile regression estimation of functional coefficient models.
\newblock {\em Journal of Econometrics\/}~{\bf 191\/}(1), 231--254.

\bibitem[\protect\citeauthoryear{Su, Shi, and Phillips}{Su et~al.}{2016}]{SSP16} Su, L., Z.~Shi, and P.C.B. Phillips (2016). \newblock Identifying latent structures in panel data. \newblock {\em Econometrica\/}~{\bf 84\/}(6), 2215--2264.

\bibitem[\protect\citeauthoryear{Su, Wang, and Jin}{Su et~al.}{2019}]{SWJ19} Su, L., X.~Wang, and S.~Jin (2019). \newblock Sieve estimation of time-varying panel data models with latent structures. \newblock {\em Journal of Business \& Economic Statistics\/}~{\bf 37\/}(2), 334--349.

\bibitem[\protect\citeauthoryear{Su and White}{Su and White}{2012}]{SW12}
Su, L. and H.~L. White (2012).
\newblock Conditional independence specification testing for dependent
  processes with local polynomial quantile regression.
\newblock {\em Advances in Econometrics\/}~{\bf 29}, 355--434.

\bibitem[\protect\citeauthoryear{Tang, Song, Wang, and Zhu}{Tang et~al.}{2013}]{TSWZ13} Tang, Y., X.~Song, H.~J. Wang, and Z.~Zhu (2013). \newblock Variable selection in high-dimensional quantile varying coefficient models. \newblock {\em Journal of Multivariate Analysis\/}~{\bf 122}, 115--132.

\bibitem[\protect\citeauthoryear{Vogt and Linton}{Vogt and Linton}{2017}]{VL17} Vogt, M. and O.~Linton (2017). \newblock Classification of non-parametric regression functions in longitudinal data models. \newblock {\em Journal of the Royal Statistical Society: Series B (Statistical Methodology)\/}~{\bf 79\/}(1), 5--27.

\bibitem[\protect\citeauthoryear{Vogt and Linton}{Vogt and Linton}{2020}]{VL20} Vogt, M. and O.~Linton (2020). \newblock Multiscale clustering of nonparametric regression curves. \newblock {\em Journal of Econometrics\/}~{\bf 216\/}(1), 305--325.

\bibitem[\protect\citeauthoryear{Wang, Zhu, and Zhou}{Wang et~al.}{2009}]{WZZ09} Wang, H.~J., Z.~Zhu, and J.~Zhou (2009). \newblock Quantile regression in partially linear varying coefficient models.
\newblock {\em The Annals of Statistics\/}~{\bf 37\/}(6B), 3841--3866.

\bibitem[\protect\citeauthoryear{Wang and Su}{Wang and Su}{2021}]{WS20} Wang, W. and L.~Su (2021).
\newblock Identifying latent group structures in nonlinear panels. \newblock {\em Journal of Econometrics\/}~{\bf 220\/}(2), 272--295.

\bibitem[\protect\citeauthoryear{Xia}{Xia}{1998}]{X98}
Xia, Y. (1998).
\newblock Bias-corrected confidence bands in nonparametric regression.
\newblock {\em Journal of the Royal Statistical Society Series B (Statistical Methodology)}~{\bf 60}(4), 797--811.


\bibitem[\protect\citeauthoryear{Yan and Li}{Yan and Li}{2018}]{YL18} Yan, K.~X. and Q.~Li (2018).
\newblock Nonparametric estimation of a conditional quantile function in a fixed effects panel data model.
\newblock {\em Journal of Risk and Financial Management\/}~{\bf 11\/}(3), 44.


\bibitem[\protect\citeauthoryear{Yokoyama}{Yokoyama}{1980}]{Yo80}
Yokoyama, R. (1980).
\newblock Moment bounds for stationary mixing sequences.
\newblock {\em Zeitschrift f{\"u}r Wahrscheinlichkeitstheorie und Verwandte
  Gebiete\/}~{\bf 52\/}(1), 45--57.

\bibitem[\protect\citeauthoryear{Yu and Jones}{Yu and Jones}{1998}]{YJ98} Yu, K. and M.~Jones (1998).
\newblock Local linear quantile regression. \newblock {\em Journal of the American Statistical Association\/}~{\bf 93\/}(441), 228--237.

\bibitem[\protect\citeauthoryear{Yu and Lu}{Yu and Lu}{2004}]{YL04} Yu, K. and Z.~Lu (2004). \newblock Local linear additive quantile regression. \newblock {\em Scandinavian Journal of Statistics\/}~{\bf 31\/}(3), 333--346.

\bibitem[\protect\citeauthoryear{Zhang, Wang, and Zhu}{Zhang et~al.}{2019a}]{ZWZ19} Zhang, Y., H.~J. Wang, and Z.~Zhu (2019a). \newblock Quantile-regression-based clustering for panel data. \newblock {\em Journal of Econometrics\/}~{\bf 213\/}(1), 54--67.

\bibitem[\protect\citeauthoryear{Zhang, Wang, and Zhu}{Zhang  et~al.}{2019b}]{ZZW19} Zhang, Y., H.~J. Wang, and Z.~Zhu (2019b). \newblock Robust subgroup identification. \newblock {\em Statistica Sinica\/}~{\bf 29\/}(4), 1873--1889.

\end{thebibliography}

\newpage


\if0\blind
{
  \begin{center}
  {\Large\bf Supplement to ``Functional-Coefficient Quantile Regression for Panel Data with Latent Group Structure"}
  \end{center}
\date{}
  \maketitle
} \fi

\if1\blind
{
  \bigskip
  \bigskip
  \bigskip
  \begin{center}
    {\Large\bf Supplement to ``Functional-Coefficient Quantile Regression for Panel Data with Latent Group Structure"}
\end{center}
  \medskip
} \fi

\spacingset{1.68} 

In this supplement, we prove the main asymptotic theorems in Appendix A, prove some technical lemmas in Appendix B, apply the modified methodology to identify the latent group structure (uniformly over quantile levels) in linear panel quantile regression in Appendix C.1, discuss the methodology and theory for the time-varying coefficient panel quantile regression with the latent structure in Appendix C.2, construct the point-wise confidence intervals in Appendix C.3, and report the empirical result of functional-coefficient mean regression based grouping analysis in Appendix D.

\section*{Appendix A:\ Proofs of the main asymptotic theorems}\label{app:A}
\renewcommand{\theequation} {A.\arabic{equation}}
\setcounter{equation}{0}

In this appendix, we give the detailed proofs of the main theoretical results. We start with some key lemmas whose proofs are available in Appendix B. Let
\[
{\mathbf W}_{iT,0}(z)=\frac{1}{Th}\sum_{t=1}^T\eta_{it}(z)\left(
\begin{array}{c}
	1\\
	{\mathbf X}_{it}
\end{array}\right)
K_h(Z_t-z),
\]
where $\eta_{it}(z)$ is defined as in (4.7). Lemma A.1 below gives a uniform Bahadur representation for the preliminary local linear quantile estimates, which is similar to (A.1) in \cite{CX08} and (7.3) in \cite{KLZ11}.

\renewcommand{\thelemma}{A.\arabic{lemma}}\setcounter{lemma}{0}

\begin{lemma}\label{le:A.1}

 Suppose that Assumptions 1--3, 4(i) and (4.3) are satisfied. Then we have
\begin{equation}\label{eqA.1}
	\max_{1\leq i\leq N}\sup_{0\leq z\leq 1}\left\Vert \left[
	\begin{array}{c}
		\widehat{\alpha}_i(z)\\
		\widehat{\boldsymbol\beta}_i(z)
	\end{array}\right]
	-\left[
	\begin{array}{c}
		\alpha_i(z)\\
		{\boldsymbol\beta}_i(z)
	\end{array}\right]-{\boldsymbol\Omega}_{i}^{-1}(z){\mathbf W}_{iT,0}(z)\right\Vert=O_P\left(\xi_{NT}^{3/2}+h\xi_{NT}\right),
\end{equation}
where $\xi_{NT}=\left(\frac{\log N\vee T)}{Th}\right)^{1/2}$ and ${\boldsymbol\Omega}_{i}(z)$ is defined in (4.1).

\end{lemma}

With the uniform Bahadur representation in (\ref{eqA.1}), we readily have the following lemma, giving the uniform convergence rate for $\widehat{\boldsymbol\beta}_i(z)$.

\begin{lemma}\label{le:A.2}

Suppose that the assumptions in Lemma \ref{le:A.1} are satisfied. Then we have
\begin{equation}\label{eqA.2}
	\max_{1\leq i\leq N}\sup_{0\leq z\leq 1}\left\Vert \widehat{\boldsymbol\beta}_i(z)-{\boldsymbol\beta}_i(z)\right\Vert=O_P\left(\xi_{NT}\right).
\end{equation}

\end{lemma}

\medskip

\noindent{\bf Proof of Theorem 4.1}.\ \ The proof is similar to the proof of Theorem 3.1 in \cite{Ch19}. By Lemma \ref{le:A.2}, we readily have that
\begin{equation}\label{eqA.3}
	\max_{1\leq i\leq N}\max_{1\leq t\leq T}\left\Vert \widehat{\boldsymbol\beta}_i(Z_t)-{\boldsymbol\beta}_i(Z_t)\right\Vert=O_P\left(\xi_{NT}\right).
\end{equation}
Letting
\[
\overline\Delta(j,k)=\frac{1}{T}\sum_{t=1}^T\left\Vert {\boldsymbol\beta}_j(Z_t)-{\boldsymbol\beta}_k(Z_t)\right\Vert,
\]
and using (\ref{eqA.3}) and the triangle inequality, we may show that
\begin{eqnarray}
	\max_{1\leq j,k\leq N}\left\vert \widehat\Delta(j,k)-\overline\Delta(j,k)\right\vert&=&\max_{1\leq j\neq k\leq N}\left\vert \widehat\Delta(j,k)-\overline\Delta(j,k)\right\vert\notag\\
	&\leq&2\max_{1\leq i\leq N}\max_{1\leq t\leq T}\left\Vert \widehat{\boldsymbol\beta}_i(Z_t)-{\boldsymbol\beta}_i(Z_t)\right\Vert\notag\\
	&=&O_P\left(\xi_{NT}\right).\label{eqA.4}
\end{eqnarray}
For each index pair $(j,k)$, $\left\{\left\Vert {\boldsymbol\beta}_j(Z_t)-{\boldsymbol\beta}_k(Z_t)\right\Vert: t\geq1\right\}$ is a stationary sequence of bounded and $\alpha$-mixing dependent random variables with $\Delta(j,k)$ being the mean value. Then, by Assumption 1 and the Bernstein inequality for the $\alpha$-mixing process \citep[e.g., Theorem 1.3(2) in][]{Bo98}, we may show that
\begin{equation}\label{eqA.5}
	\max_{1\leq j,k\leq N}\left\vert \overline\Delta(j,k)-\Delta(j,k)\right\vert=O_P\left(\sqrt{(\log N)/T}\right).
\end{equation}
By (\ref{eqA.4}) and (\ref{eqA.5}), noting that $(\log N)/T=o\left(\xi_{NT}^2\right)$ and using (4.4) in Assumption 4(ii), we readily have that
\begin{equation}\label{eqA.6}
	\max_{1\leq j,k\leq N}\left\vert \widehat\Delta(j,k)-\Delta(j,k)\right\vert=O_P\left(\xi_{NT}\right)=o_P(\zeta_{NT}),
\end{equation}
which, together with the computational property for the clustering algorithm, completes the proof of Theorem 4.1.\hfill$\blacksquare$

\medskip

\noindent{\bf Proof of Theorem 4.2}.\ \ Let ${\cal E}(R_0)$ denote the event that
\[\big\{\widehat{\cal G}_r,\ r=1, \cdots,R_0\big\}=\big\{{\cal G}_r,\ r=1, \cdots,R_0\big\}.\]
By Theorem 4.1, we readily have that ${\sf P}({\cal E}(R_0))\rightarrow1$ as $T\rightarrow\infty$. Hence, to prove Theorem 4.2, it is sufficient to show that
\begin{equation}\label{eqA.7}
	{\sf P}\left(\widehat{R}=R_0\ |\ {\cal E}(R_0)\right)\rightarrow1.
\end{equation}

For the case $R_0+1\leq R\leq \overline{R}$, conditional on the event ${\cal E}(R_0)$, the panel quantile functional-coefficient model is over-fitted as some of ${\cal G}_r$, $r=1,\cdots,R_0$, are further split into smaller groups. Without loss of generality, we next only consider the case of $R=R_0+1$ and conditional on the event ${\cal E}(R_0)$, we assume that the final group ${\cal G}_{R_0}$ is split into ${\cal G}_{R_0,\star}$ and ${\cal G}_{R_0+1,\star}$. In this case, ${\boldsymbol\beta}_j(\cdot)$ have the same functional pattern when $j$ belongs to one of the $R_0+1$ estimated groups. By (2.4), Assumption 5(iii) and Lemma \ref{le:A.2}, we may show that
\begin{eqnarray}
	{\sf D}(R_0+1)&=& \frac{1}{T(R_0+1)}\left[\sum_{r=1}^{R_0-1}\frac{1}{\left\vert{\cal G}_{r}\right\vert}\sum_{j\in{\cal G}_{r}}\sum_{t=1}^T\left\Vert{\boldsymbol\beta}_j(Z_t)-{\boldsymbol\gamma}_{r}(Z_t)\right\Vert\right.+\notag\\
	&&\left.\sum_{r=R_0}^{R_0+1}\frac{1}{\left\vert{\cal G}_{r,\star}\right\vert}\sum_{j\in{\cal G}_{r,\star}}\sum_{t=1}^T\left\Vert{\boldsymbol\beta}_j(Z_t)-{\boldsymbol\gamma}_{R_0}(Z_t)\right\Vert\right]+O_P(\xi_{NT})\notag\\
	&=&O_P(\xi_{NT})=o_P(\omega_{NT})\notag
\end{eqnarray}
conditional on ${\cal E}(R_0)$. Analogously, we can also prove that ${\sf D}(R)=o_P(\omega_{NT})$ for $R=R_0, R_0+2,\cdots,\overline{R}$. Consequently, with probability approaching one, we may set ${\sf D}(R)\equiv0$ as ${\sf D}(R)$ is smaller than $\omega_{NT}$. Then, we have
\begin{equation}\label{eqA.8}
	{\sf P}\left(\frac{{\sf D}(R)}{{\sf D}(R-1)}=\frac{0}{0}\equiv1,\ \ R=R_0+1,\cdots,\overline{R}\ \big|\ {\cal E}(R_0)\right)\rightarrow1.
\end{equation}

For the case $1\leq R\leq R_0-1$, conditional on the event ${\cal E}(R_0)$, the functional-coefficient panel model is under-fitted and at least two groups are falsely merged. Without loss of generality, we only consider the case of $R=R_0-1$ and conditional on ${\cal E}(R_0)$, we assume that the final two groups ${\cal G}_{R_0-1}$ and ${\cal G}_{R_0}$ are merged. In this case, we define ${\boldsymbol\gamma}_{r|R_0-1}(z)={\boldsymbol\gamma}_r(z)$ for $r=1,\cdots,R_0-2$, and
\[{\boldsymbol\gamma}_{R_0-1|R_0-1}(z)=\frac{1}{\left\vert {\cal G}_{R_0-1}\bigcup{\cal G}_{R_0}\right\vert}\left[\left\vert {\cal G}_{R_0-1}\right\vert{\boldsymbol\gamma}_{R_0-1}(z)+\left\vert{\cal G}_{R_0}\right\vert {\boldsymbol\gamma}_{R_0}(z)\right].\]
Then, by Assumption 5(i)(ii) and Lemma \ref{le:A.2}, we can prove that
\begin{eqnarray}
	{\sf D}(R_0-1)&=& \frac{1}{T(R_0-1)}\left[\sum_{r=1}^{R_0-2}\frac{1}{\left\vert{\cal G}_{r}\right\vert}\sum_{j\in{\cal G}_{r}}\sum_{t=1}^T\left\Vert{\boldsymbol\beta}_j(Z_t)-{\boldsymbol\gamma}_{r|R_0-1}(Z_t)\right\Vert\right.+\notag\\
	&&\left.\frac{1}{\left\vert {\cal G}_{R_0-1}\bigcup{\cal G}_{R_0}\right\vert}\sum_{j\in{\cal G}_{R_0-1}\bigcup{\cal G}_{R_0}}\sum_{t=1}^T\left\Vert{\boldsymbol\beta}_j(Z_t)-{\boldsymbol\gamma}_{R_0-1|R_0-1}(Z_t)\right\Vert\right]+O_P(\xi_{NT})\notag\\
	&=&\frac{1}{T(R_0-1)\left\vert {\cal G}_{R_0-1}\bigcup{\cal G}_{R_0}\right\vert}\sum_{j\in{\cal G}_{R_0-1}\bigcup{\cal G}_{R_0}}\sum_{t=1}^T\left\Vert{\boldsymbol\beta}_j(Z_t)-{\boldsymbol\gamma}_{R_0-1|R_0-1}(Z_t)\right\Vert \notag\\
	&&+O_P(\xi_{NT})\notag\\
	&\geq&\underline{C}_1+o_P(1)\notag
\end{eqnarray}
conditional on ${\cal E}(R_0)$, where $\underline{C}_1$ is a positive constant. Similarly, we can also show that ${\sf D}(R)\geq \underline{C}_1$ for $R=1,\cdots,R_0-2$, and consequently, there exists a positive constant $\underline{C}_2$ such that
\begin{equation}\label{eqA.9}
	{\sf P}\left(\frac{{\sf D}(R)}{{\sf D}(R-1)}\geq \underline{C}_2,\ \ R=1,\cdots,R_0-1\ \big|\ {\cal E}(R_0)\right)\rightarrow1,
\end{equation}
setting ${\sf D}(1)/{\sf D}(0)=1$, and
\begin{equation}\label{eqA.10}
	{\sf P}\left(\frac{{\sf D}(R_0)}{{\sf D}(R_0-1)}=0\ \big|\ {\cal E}(R_0)\right)\rightarrow1.
\end{equation}

In virtue of (\ref{eqA.8})--(\ref{eqA.10}), we complete the proof of (\ref{eqA.7}).\hfill$\blacksquare$

\medskip

Before proving Theorem 4.3, we first give the following Bahadur representation for the post-grouping local linear estimate $\widetilde{\boldsymbol\gamma}_j(z)$. Recall that $e_{it}=Y_{it}-{\mathbf X}_{it}^{^\intercal}{\boldsymbol\gamma}_j(Z_t)-\alpha_i$ and $b_{it}(z)={\mathbf X}_{it}^{^\intercal}\left[{\boldsymbol\gamma}_j(Z_t)-{\boldsymbol\gamma}_j(z)-{\boldsymbol\gamma}_j^\prime(z)(Z_t-z)\right]$ for $i\in{\cal G}_j$.

\begin{lemma}\label{le:A.3}

Suppose that Assumptions 1--6 and 7(i)(ii) are satisfied. For $z$ which is an interior point of $[0,1]$, and $j=1,\cdots,R_0$, we have
	\begin{equation}\label{eqA.11}
		\widetilde{\boldsymbol\gamma}_j(z)-{\boldsymbol\gamma}_j(z)={\boldsymbol\Omega}^{-1}(z;{\cal G}_j){\mathbf W}_T(z;{\cal G}_j)+o_P\left(h_1^2+\frac{1}{\sqrt{N_jTh_1}}\right),
	\end{equation}
	where ${\boldsymbol\Omega}(z; {\cal G}_j)$ is defined in (4.6), and
	\begin{eqnarray}
		{\mathbf W}_T(z;{\cal G}_j)&=&\frac{1}{N_jTh_1}\sum_{i\in{\cal G}_j}\left[\sum_{t=1}^T\eta_{it}{\mathbf X}_{it}K_{h_1}(Z_t-z)-\frac{{\boldsymbol\Omega}_{i}^{\gamma\alpha}(z)}{\omega_i^\alpha(z)}\sum_{t=1}^T\eta_{it}K_{h_1}(Z_t-z)\right]\notag\\
		&=&\frac{1}{N_jTh_1}\sum_{t=1}^T{\boldsymbol\Gamma}_{t1}({\cal G}_j)K_{h_1}(Z_t-z)\notag
	\end{eqnarray}
	with ${\boldsymbol\Gamma}_{t1}({\cal G}_j)$ defined in (4.7).
	
	\end{lemma}

\noindent{\bf Proof of Theorem 4.3}.\ \ Using the Bahadur representation (\ref{eqA.11}) in Lemma \ref{le:A.3}, it is sufficient to show that
\begin{equation}\label{eqA.12}
	\sqrt{N_jTh_1}\left\{{\mathbf W}_T(z;{\cal G}_j)-{\sf E}\left[{\mathbf W}_T(z;{\cal G}_j)\right]\right\}\stackrel{d}\longrightarrow {\sf N}\left({\bf 0},\ \nu_0{\boldsymbol\Lambda}(z;{\cal G}_j)\right)
\end{equation}
and
\begin{equation}\label{eqA.13}
	{\boldsymbol\Omega}^{-1}(z;{\cal G}_j){\sf E}\left[{\mathbf W}_T(z; {\cal G}_j)\right]=\frac{1}{2}h_1^2{\boldsymbol\gamma}_j^{\prime\prime}(z)\mu_2+o(h_1^2)={\mathbf B}_j(z)+o(h_1^2).
\end{equation}
By the smoothness conditions in Assumptions 2 and 3 and the standard Taylor expansion of $F_{ie}(\cdot|{\mathbf X}_{it},Z_t)$ (the conditional distribution function of $e_{it}$ given ${\mathbf X}_{it}$ and $Z_t$) and ${\boldsymbol\beta}_i(\cdot)$, we may prove (\ref{eqA.13}). To prove (\ref{eqA.12}), we note that
\begin{align}
&\sqrt{N_jTh_1}\left\{{\mathbf W}_T(z;{\cal G}_j)-{\sf E}\left[{\mathbf W}_T(z; {\cal G}_j)\right]\right\} \nonumber \\
&=\frac{1}{\sqrt{N_jTh_1}}\sum_{t=1}^T \left\{{\boldsymbol\Gamma}_{t1}({\cal G}_j)K_{h_1}(Z_t-z)-{\sf E}\left[{\boldsymbol\Gamma}_{t1}({\cal G}_j)K_{h_1}(Z_t-z)\right]\right\}. \nonumber
\end{align}
By Assumption 7(ii)(iii) and the standard central limit theorem for the kernel-based regression estimate \citep[e.g., Theorem 2.22 in][]{FY03}, we can prove (\ref{eqA.12}).\hfill$\blacksquare$


\renewcommand{\theequation}{B.\arabic{equation}}
\setcounter{equation}{0}

\section*{Appendix B:\ \ Proofs of the technical lemmas}

\noindent In this appendix, we prove the technical lemmas stated in Appendix A. Throughout the proofs, $C$ denotes a generic positive constant whose value is independent of $(N,T)$ and may change from line to line. The main idea to prove the uniform Bahadur representation in Lemma \ref{le:A.1} is similar to that in the proofs of Theorem 3.1 in \cite{SW12} and Proposition S.1 in \cite{LLL21}. Some further technical results are needed to prove Lemma \ref{le:A.1}.

\renewcommand{\thelemma}{B.\arabic{lemma}}\setcounter{lemma}{0}

\begin{lemma}\label{le:B.1}

Suppose that Assumption 1 is satisfied and
	\[\max_{1\leq i\leq N}{\sf E}\left[\Vert {\mathbf X}_{it}\Vert^{\kappa+\epsilon}\right]<\infty,\ \ N=O(T^{\kappa/2}),\]
	where $\kappa$ and $\epsilon$ are defined as in Assumption 3(i). Then we have
\begin{equation}\label{eqC.1}
	\max_{1\leq i\leq N}\frac{1}{T}\sum_{t=1}^T\Vert {\mathbf X}_{it}\Vert=O_P(1)\ \ as\ T\rightarrow\infty.
\end{equation}

\end{lemma}

\noindent{\bf Proof of Lemma \ref{le:B.1}}.\ \ Note that
\begin{eqnarray}
	\max_{1\leq i\leq N}\frac{1}{T}\sum_{t=1}^T\Vert {\mathbf X}_{it}\Vert&\leq&\max_{1\leq i\leq N}\left\vert\frac{1}{T}\sum_{t=1}^T\left(\Vert {\mathbf X}_{it}\Vert-{\sf E}\left[\Vert {\mathbf X}_{it}\Vert\right]\right)\right\vert+\max_{1\leq i\leq N}\left\vert\frac{1}{T}\sum_{t=1}^T{\sf E}\left[\Vert {\mathbf X}_{it}\Vert\right]\right\vert\notag\\
	&=&\max_{1\leq i\leq N}\left\vert\frac{1}{T}\sum_{t=1}^T\left(\Vert {\mathbf X}_{it}\Vert-{\sf E}\left[\Vert {\mathbf X}_{it}\Vert\right]\right)\right\vert+O(1).\label{eqC.2}
\end{eqnarray}
By the Bonferroni and Markov inequalities, for any $\delta>0$
\begin{eqnarray}
	&&{\sf P}\left(\max_{1\leq i\leq N}\left\vert\frac{1}{T}\sum_{t=1}^T\left(\Vert {\mathbf X}_{it}\Vert-{\sf E}\left[\Vert {\mathbf X}_{it}\Vert\right]\right)\right\vert>\delta\right)\notag\\
	&\leq&\sum_{i=1}^N{\sf P}\left(\left\vert\frac{1}{T}\sum_{t=1}^T\left(\Vert {\mathbf X}_{it}\Vert-{\sf E}\left[\Vert {\mathbf X}_{it}\Vert\right]\right)\right\vert>\delta\right)\notag\\
	&\leq& (\delta T)^{-\kappa-\epsilon/2}\sum_{i=1}^N{\sf E}\left[\left|\sum_{t=1}^T\left(\Vert {\mathbf X}_{it}\Vert-{\sf E}\left[\Vert {\mathbf X}_{it}\Vert\right]\right)\right|^{\kappa+\epsilon/2}\right].\label{eqC.3}
\end{eqnarray}
Furthermore, by Assumption 1 and Theorem 1 in \cite{Yo80}, we may show that
\begin{equation}\label{eqC.4}
	\sum_{i=1}^N{\sf E}\left[\left|\sum_{t=1}^T\left(\Vert {\mathbf X}_{it}\Vert-{\sf E}\left[\Vert {\mathbf X}_{it}\Vert\right]\right)\right|^{\kappa+\epsilon/2}\right]=O(NT^{\kappa/2+\epsilon/4}).
\end{equation}
By (\ref{eqC.3}), (\ref{eqC.4}) and the condition $N=O(T^{\kappa/2})$, we have
\[
{\sf P}\left(\max_{1\leq i\leq N}\left\vert\frac{1}{T}\sum_{t=1}^T\left(\Vert {\mathbf X}_{it}\Vert-{\sf E}\left[\Vert {\mathbf X}_{it}\Vert\right]\right)\right\vert>\delta\right)\rightarrow0\ \ {\rm as}\ T\rightarrow\infty,
\]
which, together with (\ref{eqC.2}), completes the proof of (\ref{eqC.1}).\hfill$\blacksquare$

\begin{lemma}\label{le:B.2}

 Suppose that the assumptions in Lemma \ref{le:B.1} and Assumption 2(i) are satisfied. Letting $r_\ast\rightarrow0$,
	\begin{equation}\label{eqC.5}
		\max_{1\leq i\leq N}\frac{1}{T}\sum_{t=1}^TI\left(|e_{it}|\leq r_\ast\Vert {\mathbf X}_{it}\Vert \right)\Vert {\mathbf X}_{it}\Vert=O_P\left(r_\ast^{1/(\kappa+\epsilon-1)}T^{1/(2\kappa-2)}\right),
	\end{equation}
	where $\kappa$ and $\epsilon$ are defined in Assumption 3(i) and $e_{it}$ is defined in Section 4.1.
	
\end{lemma}

\noindent{\bf Proof of Lemma \ref{le:B.2}}.\ The proof is similar to the proof of Lemma \ref{le:B.1}. By Assumption 2(i), 
\[\frac{1}{T}\sum_{t=1}^T{\sf E}\left[I\left(|e_{it}|\leq r_\ast\Vert {\mathbf X}_{it}\Vert \right)\Vert {\mathbf X}_{it}\Vert\right]\leq \frac{Cr_\ast}{T}\sum_{t=1}^T{\sf E}\left[\Vert {\mathbf X}_{it}\Vert^2\right]=O(r_\ast)=o\left(r_\ast^{1/(\kappa+\epsilon-1)}T^{1/(2\kappa-2)}\right)\]
uniformly over $i$. Letting $V_{it}=I\left(|e_{it}|\leq r_\ast\Vert {\mathbf X}_{it}\Vert \right)\Vert {\mathbf X}_{it}\Vert$, to prove (\ref{eqC.5}), we only need to show
\begin{equation}\label{eqC.6}
	\max_{1\leq i\leq N}\left|\frac{1}{T}\sum_{t=1}^T \left(V_{it}-{\sf E}[V_{it}]\right)\right|=O_P\left(r_\ast^{1/(\kappa+\epsilon-1)}T^{1/(2\kappa-2)}\right).
\end{equation}
By the Bonferroni and Markov inequalities, Theorem 4.1 in \cite{SY96} with $p=\kappa-1$ and $r=\kappa+\epsilon-1$ and Assumption 2(i), we have
\begin{eqnarray}
	&&{\sf P}\left(\max_{1\leq i\leq N}\left|\frac{1}{T}\sum_{t=1}^T \left(V_{it}-{\sf E}[V_{it}]\right)\right|>Mr_\ast^{1/(\kappa+\epsilon-1)}T^{1/(2\kappa-2)}\right)\notag\\
	&\leq&\sum_{i=1}^N{\sf P}\left(\left|\frac{1}{T}\sum_{t=1}^T \left(V_{it}-{\sf E}[V_{it}]\right)\right|>Mr_\ast^{1/(\kappa+\epsilon-1)}T^{1/(2\kappa-2)}\right)\notag\\
	&\leq&M^{-(\kappa-1)}r_\ast^{-(\kappa-1)/(\kappa+\epsilon-1)}T^{-(2\kappa-1)/2}\sum_{i=1}^N{\sf E}\left[\left|\sum_{t=1}^T \left(V_{it}-{\sf E}[V_{it}]\right)\right|^{\kappa-1}\right]\notag\\
	&\leq&CM^{-(\kappa-1)}r_\ast^{-(\kappa-1)/(\kappa+\epsilon-1)}T^{-(2\kappa-1)/2}\sum_{i=1}^NT^{(\kappa-1)/2}\left({\sf E} \left[\vert V_{it}\vert^{\kappa+\epsilon-1}\right]\right)^{(\kappa-1)/(\kappa+\epsilon-1)}\notag\\
	&\leq&CM^{-(\kappa-1)}r_\ast^{-(\kappa-1)/(\kappa+\epsilon-1)}T^{-(2\kappa-1)/2}\sum_{i=1}^NT^{(\kappa-1)/2}\left(r_\ast {\sf E} \left[\Vert {\mathbf X}_{it}\Vert^{\kappa+\epsilon}\right]\right)^{(\kappa-1)/(\kappa+\epsilon-1)}\notag\\
	&\leq&CM^{-(\kappa-1)}NT^{-(2\kappa-1)/2+(\kappa-1)/2}\leq CM^{-(\kappa-1)},\notag
\end{eqnarray}
which converges to zero by letting $M>0$ be sufficiently large. This completes the proof of (\ref{eqC.6}).\hfill$\blacksquare$

\begin{lemma}\label{le:B.3}

Suppose that the assumptions in Lemma \ref{le:A.1} are satisfied. Define ${\mathbf W}_{iT}(z)=\left[{\mathbf W}_{iT,0}^{^\intercal}(z),{\mathbf W}_{iT,1}^{^\intercal}(z)\right]^{^\intercal}$, where ${\mathbf W}_{iT,0}(z)$ is defined as in Lemma \ref{le:A.1}, and
	\[
	{\mathbf W}_{iT,1}(z)=\frac{1}{Th}\sum_{t=1}^T\eta_{it}(z)\left(
	\begin{array}{c}
		1\\
		{\mathbf X}_{it}
	\end{array}\right)
	\left(\frac{Z_t-z}{h}\right)K_h(Z_t-z).
	\]
	Then, we have
\begin{equation}\label{eqC.7}
	\max_{1\leq i\leq N}\sup_{0\leq z\leq 1}\left\Vert {\mathbf W}_{iT}(z)\right\Vert=O_P\left(h^2+\xi_{NT}\right)=O_P\left(\xi_{NT}\right),
\end{equation}
where $\xi_{NT}=\left[\frac{\log (T\vee N)}{Th}\right]^{1/2}$. 

\end{lemma}

\noindent{\bf Proof of Lemma \ref{le:B.3}}.\ \ Let
\[\overline{Z}_t(z)=\left(1, \frac{Z_t-z}{h}\right)^{^\intercal},\ \ \overline{\mathbf X}_{it}=\left(
\begin{array}{c}
	1\\
	{\mathbf X}_{it}
\end{array}\right),
\]
and re-write
\[{\mathbf W}_{iT}(z)=\frac{1}{Th}\sum_{t=1}^T\eta_{it}(z)\left[\overline{Z}_t(z)\otimes \overline{\mathbf X}_{it}\right]K_h(Z_t-z),\]
where $\otimes$ denotes the Kronecker product. By the smoothness conditions in Assumptions 2 and 3, the bandwidth condition in Assumption 4(ii) and the standard Taylor expansion of $F_{ie}(\cdot|{\mathbf X}_{it},Z_t)$ (the conditional distribution function of $e_{it}$ given ${\mathbf X}_{it}$ and $Z_t$) and ${\boldsymbol\beta}_i(\cdot)$, we may prove that
\begin{eqnarray}
	\left\Vert{\sf E}\left[{\mathbf W}_{iT}(z)\right]\right\Vert&=&\left\Vert{\sf E}\left[\frac{1}{Th}\sum_{t=1}^T\eta_{it}(z)\left(\overline{Z}_t(z)\otimes \overline{\mathbf X}_{it}\right)K_h(Z_t-z)\right]\right\Vert\notag\\
	&=&\left\Vert\frac{1}{h}{\sf E}\left[\eta_{it}(z)\left(\overline{Z}_t(z)\otimes \overline{\mathbf X}_{it}\right)K_h(Z_t-z)\right]\right\Vert\notag\\
	&=&\frac{1}{h}\left\Vert{\sf E}\left[\left(\tau-F_{ie}(-b_{it}(z)|{\mathbf X}_{it},Z_t)\right)\left(\overline{Z}_t(z)\otimes \overline{\mathbf X}_{it}\right)K_h(Z_t-z)\right]\right\Vert\notag\\
	&=&O(h^2)=o\left(\xi_{NT}\right)\label{eqC.8}
\end{eqnarray}
uniformly over $0\leq z\leq 1$ and $1\leq i\leq N$. By (\ref{eqC.8}), we only need to prove that
\begin{equation}\label{eqC.9}
	\max_{1\leq i\leq N}\sup_{0\leq z\leq 1}\left\Vert {\mathbf W}_{iT,k}(z)-{\sf E}\left[{\mathbf W}_{iT,k}(z)\right]\right\Vert=O_P\left(\xi_{NT}\right),\ \ k=0,1.
\end{equation}
To save the space, we only consider $k=0$ in the following proof.

The technique of proving (\ref{eqC.9}) is standard in proofs of the uniform consistency results for kernel-based quantities. Consider covering the closed interval $[0,1]$ by some disjoint sub-intervals ${\cal Z}_s$ with equal length, $s=1,\cdots, J$. Let the center point of ${\cal Z}_s$ be $z_s$, and set the length of ${\cal Z}_s$ as $r=\left(h\xi_{NT}\right)^{\kappa+\epsilon-1}T^{-(\kappa+\epsilon-1)/(2\kappa-2)}$. Observe that
\begin{eqnarray}
	&&\max_{1\leq i\leq N}\max_{1\leq s\leq J}\sup_{z\in{\cal Z}_s}\left\Vert {\mathbf W}_{iT,0}(z)-{\mathbf W}_{iT,0}(z_s)\right\Vert\notag\\
	&=&\max_{1\leq i\leq N}\max_{1\leq s\leq J}\sup_{z\in{\cal Z}_s}\left\Vert \frac{1}{Th}\sum_{t=1}^T\eta_{it}(z)\overline{\mathbf X}_{it}K_h(Z_t-z)-\frac{1}{Th}\sum_{t=1}^T\eta_{it}(z)\overline{\mathbf X}_{it}K_h(Z_t-z_s)\right\Vert\notag\\
	&\leq&\max_{1\leq i\leq N}\max_{1\leq s\leq J}\sup_{z\in{\cal Z}_s}\left\Vert \frac{1}{Th}\sum_{t=1}^T\eta_{it}(z)\overline{\mathbf X}_{it}\left[K_h(Z_t-z)-K_h(Z_t-z_s)\right]\right\Vert+\notag\\
	&&\max_{1\leq i\leq N}\max_{1\leq s\leq J}\sup_{z\in{\cal Z}_s}\left\Vert \frac{1}{Th}\sum_{t=1}^T\left[\eta_{it}(z)-\eta_{it}(z_s)\right]\overline{\mathbf X}_{it} K_h(Z_t-z_s)\right\Vert.\label{eqC.10}
\end{eqnarray}

Noting that $\eta_{it}(z)$ is bounded and using the smoothness condition on $K(\cdot)$ in Assumption 4(i), we may show that
\begin{eqnarray}
&&\max_{1\leq i\leq N}\max_{1\leq s\leq J}\sup_{z\in{\cal Z}_s}\left\Vert \frac{1}{Th}\sum_{t=1}^T\eta_{it}(z)\overline{\mathbf X}_{it}\left[K_h(Z_t-z)-K_h(Z_t-z_s)\right]\right\Vert \notag\\
&\leq & C\cdot \frac{r}{h^2}\cdot \max_{1\leq i\leq N}\frac{1}{T}\sum_{t=1}^T(\Vert {\mathbf X}_{it}\Vert\vee 1). \notag
\end{eqnarray}
Then, by Lemma \ref{le:B.1} and the choice of $r$, we readily have
\begin{equation}\label{eqC.11}
	\max_{1\leq i\leq N}\max_{1\leq s\leq J}\sup_{z\in{\cal Z}_s}\left\Vert \frac{1}{Th}\sum_{t=1}^T\eta_{it}(z)\overline{\mathbf X}_{it}\left[K_h(Z_t-z)-K_h(Z_t-z_s)\right]\right\Vert=o_P\left(\xi_{NT}\right).
\end{equation}

On the other hand, with Assumption 3(ii), we have
\[\left\vert\eta_{it}(z)-\eta_{it}(z_s)\right\vert=I\left(|e_{it}|\leq 3c_\beta r\Vert {\mathbf X}_{it}\Vert \right)\]
for $z\in{\cal Z}_s$, where $c_\beta$ is defined in Assumption 3(ii). Consequently, by Lemma \ref{le:B.2}, we have
\begin{eqnarray}
	&&\max_{1\leq i\leq N}\max_{1\leq s\leq J}\sup_{z\in{\cal Z}_s}\left\Vert \frac{1}{Th}\sum_{t=1}^T\left[\eta_{it}(z)-\eta_{it}(z_s)\right]\overline{\mathbf X}_{it} K_h(Z_t-z_s)\right\Vert\notag\\
	&\leq& \frac{C}{h}\max_{1\leq i\leq N}\frac{1}{T}\sum_{t=1}^TI\left(|e_{it}|\leq 3c_\beta r\Vert {\mathbf X}_{it}\Vert \right)\Vert {\mathbf X}_{it}\Vert\notag\\
	&=&O_P\left(h^{-1}r^{1/(\kappa+\epsilon-1)}T^{1/(2\kappa-2)}\right)=O_P(\xi_{NT}).\label{eqC.12}
\end{eqnarray}

With (\ref{eqC.10})--(\ref{eqC.12}), we have
\begin{equation}\label{eqC.13}
	\max_{1\leq i\leq N}\max_{1\leq s\leq J}\sup_{z\in{\cal Z}_s}\left\Vert {\mathbf W}_{iT,0}(z)-{\mathbf W}_{iT,0}(z_s)\right\Vert=O_P\left(\xi_{NT}\right).
\end{equation}
Similarly, we may also show that
\begin{equation}\label{eqC.14}
	\max_{1\leq i\leq N}\max_{1\leq s\leq J}\sup_{z\in{\cal Z}_s}\left\Vert {\sf E}\left[{\mathbf W}_{iT,0}(z)\right]-{\sf E}\left[{\mathbf W}_{iT,0}(z_s)\right]\right\Vert=O_P\left(\xi_{NT}\right).
\end{equation}
By (\ref{eqC.13}) and (\ref{eqC.14}), in order to complete the proof of (\ref{eqC.9}), we only have to show that
\begin{equation}\label{eqC.15}
	\max_{1\leq i\leq N}\max_{1\leq s\leq J}\left\Vert {\mathbf W}_{iT,0}(z_s)-{\sf E}\left[{\mathbf W}_{iT,0}(z_s)\right]\right\Vert=O_P\left(\xi_{NT}\right).
\end{equation}

Let
\[\overline{\mathbf X}_{it}^\ast=\overline{\mathbf X}_{it}I\left(\Vert {\mathbf X}_{it}\Vert\leq (NT)^{1/\kappa}\right),\ \ \overline{\mathbf X}_{it}^\diamond=\overline{\mathbf X}_{it}-\overline{\mathbf X}_{it}^\ast,\]
and define
\[{\mathbf W}_{iT,0}^\ast(z)=\frac{1}{Th}\sum_{t=1}^T\eta_{it}(z)\overline{\mathbf X}_{it}^\ast K_h(Z_t-z),\ \ {\mathbf W}_{iT,0}^\diamond(z)=\frac{1}{Th}\sum_{t=1}^T\eta_{it}(z)\overline{\mathbf X}_{it}^\diamond K_h(Z_t-z).\]
To prove (\ref{eqC.15}), it is sufficient to show
\begin{equation}\label{eqC.16}
	\max_{1\leq i\leq N}\max_{1\leq s\leq J}\left\Vert {\mathbf W}_{iT,0}^\ast(z_s)-{\sf E}\left[{\mathbf W}_{iT,0}^\ast(z_s)\right]\right\Vert=O_P\left(\xi_{NT}\right),
\end{equation}
and
\begin{equation}\label{eqC.17}
	\max_{1\leq i\leq N}\max_{1\leq s\leq J}\left\Vert {\mathbf W}_{iT,0}^\diamond(z_s)-{\sf E}\left[{\mathbf W}_{iT,0}^\diamond(z_s)\right]\right\Vert=O_P\left(\xi_{NT}\right).
\end{equation}

As
\[\max_{1\leq i\leq N}\max_{1\leq s\leq J}\left\Vert{\sf E}\left[{\mathbf W}_{iT,0}^\diamond(z_s)\right]\right\Vert=o(\xi_{NT}),\]
and
\begin{eqnarray}
	&&{\sf P}\left(\max_{1\leq i\leq N}\max_{1\leq s\leq J}\left\Vert {\mathbf W}_{iT,0}^\diamond(z_s)\right\Vert>\delta\xi_{NT}\right)\notag\\
	&\leq&{\sf P}\left(\max_{1\leq i\leq N}\max_{1\leq t\leq T}\left\Vert {\mathbf X}_{it}\right\Vert>(NT)^{1/\kappa}\right)\notag\\
	&\leq&\sum_{i=1}^N\sum_{t=1}^T{\sf P}\left(\left\Vert{\mathbf X}_{it}\right\Vert>(NT)^{1/\kappa}\right)\notag\\
	&\leq&(NT)^{-\epsilon}\max_{1\leq i\leq N} {\sf E}\left[\Vert {\mathbf X}_{it}\Vert^{\kappa+\epsilon}\right]=o(1)\notag
\end{eqnarray}
for any $\delta>0$, we can then easily prove (\ref{eqC.17}).

Using the Bonferroni and Bernstein inequalities for the $\alpha$-mixing process \citep[e.g., Theorem 1.3(2) in][]{Bo98}, we can show that
\begin{eqnarray}
	&&{\sf P}\left(\max_{1\leq i\leq N}\max_{1\leq s\leq J}\left\Vert {\mathbf W}_{iT,0}^\ast(z_s)-{\sf E}\left[{\mathbf W}_{iT,0}^\ast(z_s)\right]\right\Vert>M\xi_{NT}\right)\notag\\
	&\leq&\sum_{i=1}^N\sum_{s=1}^J{\sf P}\left(\left\Vert {\mathbf W}_{iT,0}^\ast(z_s)-{\sf E}\left[{\mathbf W}_{iT,0}^\ast(z_s)\right]\right\Vert>M\xi_{NT}\right)\nonumber\\
	&\leq&CNJ\left(1+N^{3/(2\kappa)}T^{(3+2\kappa)/(2\kappa)}\xi_{NT}^{1/2}h^{-1/2}\right)\exp\left\{-\overline{M} \log (N\vee T))\right\},\notag
\end{eqnarray}
where $\overline{M}$ is a sufficiently large positive constant satisfying that
\[NJ\left(1+N^{3/(2\kappa)}T^{(3+2\kappa)/(2\kappa)}\xi_{NT}^{1/2}h^{-1/2}\right)\exp\left\{-\overline{M} \log (N\vee T))\right\}=o(1),\]
which is feasible as $J$ diverges to infinity at a polynomial rate of $NT$ (from the definition of $r$). Hence, we complete the proof of (\ref{eqC.16}), which, together with (\ref{eqC.17}), leads to (\ref{eqC.15}). With (\ref{eqC.13})--(\ref{eqC.15}), we can prove (\ref{eqC.9}), completing the proof of Lemma \ref{le:B.3}.\hfill$\blacksquare$

\medskip

Define 
\begin{eqnarray}
	{\mathbf D}_i(z)&:=&{\mathbf D}_i(a_1,a_2, {\mathbf b}_1,{\mathbf b}_2;z) \notag \\
	&=&(Th)^{1/2}\left[a_1-\alpha_i, \left({\mathbf b}_1-{\boldsymbol\beta}_i(z)\right)^{^\intercal}, ha_2, h\left({\mathbf b}_2-{\boldsymbol\beta}_i^\prime(z)\right)^{^\intercal}\right]^{^\intercal}, \notag
\end{eqnarray}
\[\eta_{it}(z,{\mathbf D}_i(z))=\tau-I\left(e_{it}+b_{it}(z)-(Th)^{-1/2}\left[\overline{Z}_t(z)\otimes\overline{\mathbf X}_{it}\right]^{^\intercal}{\mathbf D}_i(z)\leq0\right),
\]
and ${\cal D}_i=\left\{{\cal D}_i(z): 0\leq z\leq 1\right\}$ with
\[{\cal D}_i(z)=\left\{\left(a_1,a_2, {\mathbf b}_1^{^\intercal},{\mathbf b}_2^{^\intercal}\right)\in{\cal R}^{2(d+1)}:\ \Vert {\mathbf D}_i(a_1,a_2, {\mathbf b}_1,{\mathbf b}_2;z)\Vert\leq \xi_{NT}(c_1)\right\},\]
where $b_{it}(z)={\mathbf X}_{it}^{^\intercal}\left[{\boldsymbol\beta}_i(Z_t)-{\boldsymbol\beta}_i(z)-{\boldsymbol\beta}_i^\prime(z)(Z_t-z)\right]$ defined as in Section 4.1, and $\xi_{NT}(c_1)=c_1(Th)^{1/2}\xi_{NT}$ with $c_1$ being a positive constant. Let ${\mathbf W}_{iT}(z,{\mathbf D}_i(z))$ be defined as ${\mathbf W}_{iT}(z)$ in Lemma \ref{le:B.3} but with $\eta_{it}(z)$ replaced by $\eta_{it}(z,{\mathbf D}_i(z))$, and
\[{\mathbf W}_{iT}^\star\left(z,{\mathbf D}_i(z)\right)={\mathbf W}_{iT}(z,{\mathbf D}_i(z))-{\mathbf W}_{iT}(z).\]

\begin{lemma}\label{le:B.4} 

Suppose that the assumptions in Lemma \ref{le:A.1} are satisfied. Then, we have
	\begin{equation}\label{eqC.18}
		\max_{1\leq i\leq N}\sup_{0\leq z\leq 1}\sup_{\Vert {\mathbf D}_i(z)\Vert\leq \xi_{NT}(c_1)}\left\Vert {\mathbf W}_{iT}^\star\left(z, {\mathbf D}_i(z)\right)-{\sf E}\left[{\mathbf W}_{iT}^\star\left(z, {\mathbf D}_i(z)\right)\right]\right\Vert=O_P\left(\xi_{NT}^{3/2}\right),
	\end{equation}
	where $\xi_{NT}$ is defined as in Lemma \ref{le:B.3}.
	
\end{lemma}

\noindent{\bf Proof of Lemma \ref{le:B.4}}.\ \ The main idea in this proof is very similar to that in the proof of Lemma \ref{le:B.3} above. Hence, we next only sketch the proof. Consider covering the sets ${\cal Z}=[0,1]$ and ${\cal D}_i$ by ${\cal Z}_{s_1}^\star$ (with equal length), $s_1=1,\cdots,J_1$, and ${\cal D}_{i,s_2}^\star$ (with equal size), $s_2=1,\cdots,J_2$, respectively. Denote the center points of ${\cal Z}_{s_1}^\star$ and ${\cal D}_{i,s_2}^\star$ by $z_{s_1}^\star$ and $\left(a_{i,s_2,1},a_{i,s_2,2}, {\mathbf b}_{i,s_2,1}^{^\intercal},{\mathbf b}_{i,s_2,2}^{^\intercal}\right)^{^\intercal}$, respectively. Following the proofs of (\ref{eqC.13}) and (\ref{eqC.14}), we may choose the sizes of ${\cal Z}_{s_1}^\star$ and ${\cal D}_{i,s_2}^\star$ to be sufficiently small so that both $J_1$ and $J_2$ diverge at a polynomial rate of $NT$ and
\begin{eqnarray}
	&&\max_{1\leq i\leq N}\sup_{0\leq z\leq 1}\sup_{\Vert {\mathbf D}_i(z)\Vert\leq \xi_{NT}(c_1)}\left\Vert {\mathbf W}_{iT}^\star\left(z, {\mathbf D}_i(z)\right)-{\sf E}\left[{\mathbf W}_{iT}^\star\left(z, {\mathbf D}_i(z)\right)\right]\right\Vert\label{eqC.19}\\
	&=&\max_{1\leq i\leq N}\max_{1\leq s_1\leq J_1}\max_{s_2\in{\cal N}(z_{s_1}^\star)}\left\Vert {\mathbf W}_{iT}^\star\left(z_{s_1}^\star, {\mathbf D}_{i,s_2}(z_{s_1}^\star)\right)-{\sf E}\left[{\mathbf W}_{iT}^\star\left(z_{s_1}^\star, {\mathbf D}_{i,s_2}(z_{s_1}^\star)\right)\right]\right\Vert+O_P\left(\xi_{NT}^{3/2}\right),\notag
\end{eqnarray}
where
\[{\cal N}(z_{s_1}^\star)=\left\{1\leq s_2\leq J_2:\Vert {\mathbf D}_i(a_{i,s_2,1},a_{i,s_2,2}, {\mathbf b}_{i,s_2,1},{\mathbf b}_{i,s_2,2}; z_{s_1}^\star)\Vert \leq \xi_{NT}(c_1)\right\}\]
and
$${\mathbf D}_{i,s_2}(z_{s_1}^\star)=(Th)^{1/2}\left[a_{i,s_2,1}-\alpha_i, \left({\mathbf b}_{i,s_2,1}-{\boldsymbol\beta}_i(z_{s_1}^\star)\right)^{^\intercal}, ha_{i,s_2,2}, h\left({\mathbf b}_{i,s_2,2}-{\boldsymbol\beta}_i^\prime(z_{s_1}^\star)\right)^{^\intercal}\right]^{^\intercal}.$$
Then, using the truncation technique and applying the Bernstein inequality as in the proof (\ref{eqC.15}), we may prove that
\begin{equation}\label{eqC.20}
	\max_{1\leq i\leq N}\max_{1\leq s_1\leq J_1}\max_{s_2\in{\cal N}(z_{s_1}^\star)}\left\Vert {\mathbf W}_{iT}^\star\left(z_{s_1}^\star, {\mathbf D}_{i,s_2}(z_{s_1}^\star)\right)-{\sf E}\left[{\mathbf W}_{iT}^\star\left(z_{s_1}^\star, {\mathbf D}_{i,s_2}(z_{s_1}^\star)\right)\right]\right\Vert=O_P\left(\xi_{NT}^{3/2}\right),
\end{equation}
which, together with (\ref{eqC.19}), completes the proof of (\ref{eqC.18}).\hfill$\blacksquare$

\medskip

\noindent{\bf Proof of Lemma \ref{le:A.1}}.\ Note that
\begin{eqnarray}
	{\sf E}\left[{\mathbf W}_{iT}^\star\left(z, {\mathbf D}_i(z)\right)\right]&=& {\sf E}\left[{\mathbf W}_{iT}\left(z,{\mathbf D}_i(z)\right)-{\mathbf W}_{iT}(z)\right]\nonumber\\
	&=& \frac{1}{Th}\sum_{t=1}^T{\sf E}\left[\left(\eta_{it}(z, {\mathbf D}_i(z))-\eta_{it}(z)\right)\left(\overline{Z}_t(z)\otimes \overline{\mathbf X}_{it}\right)K_h(Z_t-z)\right].\nonumber
\end{eqnarray}
Letting $\overline{\boldsymbol\Omega}_i(z)={\sf diag}(1,\mu_2)\otimes{\boldsymbol\Omega}_i(z)$ and following the proof of (\ref{eqC.8}), we may show that
\begin{equation}
	\left\Vert {\sf E}\left[{\mathbf W}_{iT}^\star\left(z, {\mathbf D}_i(z)\right)\right]+(Th)^{-1/2}\overline{\boldsymbol\Omega}_i(z){\mathbf D}_i(z)\right\Vert=O\left(\xi_{NT}^2+h\xi_{NT}\right)\label{eqC.21}
\end{equation}
uniformly over $1\leq i\leq N$, $z\in[0,1]$ and $\Vert{\mathbf D}_i(z)\Vert\leq \xi_{NT}(c_1)$, where $\xi_{NT}(c_1)$ is defined in Lemma \ref{le:B.4}. By (\ref{eqC.21}) and Lemma \ref{le:B.4}, we readily have that
\begin{equation}\label{eqC.22}
	\left\Vert {\mathbf W}_{iT}(z,{\mathbf D}_i(z))-{\mathbf W}_{iT}(z)+(Th)^{-1/2}\overline{\boldsymbol\Omega}_i(z){\mathbf D}_{i}(u)\right\Vert=O_P\left(\xi_{NT}^{3/2}+h\xi_{NT}\right)
\end{equation}
uniformly over $1\leq i\leq N$, $z\in[0,1]$ and $\Vert{\mathbf D}_i(z)\Vert\leq \xi_{NT}(c_1)$. Define
\[\widehat{\mathbf D}_i(z)=(Th)^{1/2}\left[\widehat{\alpha}_i(z)-\alpha_i, \left(\widehat{\boldsymbol\beta}_i(z)-{\boldsymbol\beta}_i(z)\right)^{^\intercal}, h\widehat{\alpha}_i^\prime(z), h\left(\widehat{\boldsymbol\beta}_i^\prime(z)-{\boldsymbol\beta}_i^\prime(z)\right)^{^\intercal}\right]^{^\intercal}.\]
By (\ref{eqC.22}) and following the standard arguments in the proofs of Lemma A2 in \cite{RC80} and Lemma A.5 in \cite{SW12}, to prove Lemma \ref{le:A.1}, it is sufficient to show that, with probability approaching one,
\begin{equation}\label{eqC.23}
	\max_{1\leq i\leq N}\sup_{0\leq z\leq 1}\left\Vert\widehat {\mathbf D}_i(z)\right\Vert<\xi_{NT}(c_1),
\end{equation}
choosing $c_1>0$ to be sufficiently large.

Let ${\cal E}_{1}$ denote the event that
\[\inf_{\Vert{\mathbf D}_i(z)\Vert=\xi_{NT}(c_1)}\left\{-{\mathbf D}_{i}^{^\intercal}(z){\mathbf W}_{iT}(z,{\mathbf D}_i(z))\right\}<c_1c_2 (Th)^{1/2}\xi_{NT}^2\]
uniformly over $i=1,\cdots,N$ and $z\in[0,1]$, and let ${\cal E}_{2}$ denote the event that
\[
\inf_{\Vert{\mathbf D}_i(z)\Vert=\xi_{NT}(c_1)}\left\{{\mathbf D}_{i}^{^\intercal}(z)\overline{\boldsymbol\Omega}_i(z){\mathbf D}_{i}(z)-(Th)^{1/2}{\mathbf D}_{i}^{^\intercal}(z){\mathbf W}_{iT}(z)\right\}\geq 2c_1c_2(Th) \xi_{NT}^2
\]
uniformly over $i=1,\cdots,N$ and $z\in[0,1]$, where $c_2$ is a positive constant. By Assumption 3(i), there exists a positive constant $\underline{c}$ such that the smallest eigenvalue of $\overline{\boldsymbol\Omega}_i(z)$ is strictly larger than $\underline{c}$ uniformly over $i=1,\cdots,N$ and $z\in[0,1]$. Hence, we have
\begin{eqnarray}
	&&\inf_{\Vert{\mathbf D}_i(z)\Vert=\xi_{NT}(c_1)}\left\{{\mathbf D}_{i}^{^\intercal}(z)\overline{\boldsymbol\Omega}_i(z){\mathbf D}_{i}(z)-(Th)^{1/2}{\mathbf D}_{i}^{^\intercal}(z){\mathbf W}_{iT}(z)\right\}\notag\\
	&\geq&\inf_{\Vert{\mathbf D}_i(z)\Vert=\xi_{NT}(c_1)}{\mathbf D}_{i}^{^\intercal}(z)\overline{\boldsymbol\Omega}_i(z){\mathbf D}_{i}(z)- (Th)^{1/2} \xi_{NT}(c_1) \Vert {\mathbf W}_{iT}(z)\Vert \notag\\
	&\geq& \underline{c}c_1^2(Th) \xi_{NT}^2-c_1(Th)\xi_{NT}\Vert  {\mathbf W}_{iT}(z)\Vert\label{eqC.24}
\end{eqnarray}
uniformly over $i=1,\cdots,N$ and $z\in[0,1]$. Letting ${\cal E}_{2}^c$ be the complement of ${\cal E}_{2}$, if ${\cal E}_{2}^c$ holds, by (\ref{eqC.24}), we must have
\[
c_1(Th)\xi_{NT}\Vert {\mathbf W}_{iT}(z)\Vert\geq\left(\underline{c}c_1^2-2c_1c_2\right)(Th) \xi_{NT}^2\]
uniformly over $i=1,\cdots,N$ and $z\in[0,1]$, which is equivalent to
\[ \max_{1\leq i\leq N}\sup_{0\leq z\leq 1}\Vert {\mathbf W}_{iT}(z)\Vert\geq\left(\underline{c}c_1-2c_2\right) \xi_{NT}.\]
From Lemma \ref{le:B.3} and choosing $c_1>0$ to be sufficiently large, the latter holds with probability convergent to zero, which indicates that
\begin{equation}\label{eqC.25}
	{\sf P}({\cal E}_{2}^c)\rightarrow0\ \ {\rm as}\ T\rightarrow\infty.
\end{equation}
We next prove that
\begin{equation}\label{eqC.26}
	{\sf P}({\cal E}_{1})\rightarrow0\ \ {\rm as}\ T\rightarrow\infty.
\end{equation}
By (\ref{eqC.25}), it is sufficient to show that
\begin{equation}\label{eqC.27}
	{\sf P}({\cal E}_{1}\cap {\cal E}_{2})\rightarrow0\ \ {\rm as}\ T\rightarrow\infty.
\end{equation}
It is easy to show that, when the events ${\cal E}_{1}$ and  ${\cal E}_{2}$ hold jointly, we must have that
\begin{eqnarray}
&&\max_{1\leq i\leq N}\sup_{0\leq z\leq 1}\sup_{\Vert{\mathbf D}_i(z)\Vert=\xi_{NT}(c_1)}\left\{(Th)^{1/2} {\mathbf D}_{i}^{^\intercal}(z)\left[{\mathbf W}_{iT}(z,{\mathbf D}_i(z))-{\mathbf W}_{iT}(z)\right]+{\mathbf D}_{i}^{^\intercal}(z)\overline{\boldsymbol\Omega}_i(z){\mathbf D}_{i}(z)\right\}\notag \\
&\geq &c_1c_2(Th)\xi_{NT}^2. \notag
\end{eqnarray}
Then, using (\ref{eqC.22}) and noting that $(Th)^{1/2}\left(\xi_{NT}^{3/2}+h\xi_{NT}\right)=o\left(Th \xi_{NT}^2\right)$, we prove (\ref{eqC.27}).

We next consider a more general case: $\Vert{\mathbf D}_i(z)\Vert\geq\xi_{NT}(c_1)$. As in the proof of Theorem 1 in \cite{CX08}, we can verify that, for each fixed $i=1,\cdots,N$ and $z\in[0,1]$, $-{\mathbf D}_{i}^{^\intercal}(z){\mathbf W}_{iT}(z,\lambda{\mathbf D}_i(z))$ is a non-decreasing function of $\lambda$. For the case of $\Vert{\mathbf D}_i(z)\Vert\geq\xi_{NT}(c_1)$, we let
\[\lambda^\star=\frac{\Vert{\mathbf D}_i(z)\Vert}{\xi_{NT}(c_1)}\ \ {\rm and}\ \ {\mathbf D}_i^\star(z)=\frac{{\mathbf D}_i(z)}{\lambda^\star}.\]
Then, it is easy to show that $\lambda^\star\geq1$ and $\Vert{\mathbf D}_i^\star(z)\Vert=\xi_{NT}(c_1)$, indicating that
\begin{equation}\label{eqC.28}
	-{\mathbf D}_{i}^{\star^\intercal}(z){\mathbf W}_{iT}(z,{\mathbf D}_i(z))=-{\mathbf D}_{i}^{\star^\intercal}(z){\mathbf W}_{iT}(z,\lambda^\star{\mathbf D}_i^\star(z))\geq -{\mathbf D}_{i}^{\star^\intercal}(z){\mathbf W}_{iT}\left(z,{\mathbf D}_i^\star(z)\right).
\end{equation}
By (\ref{eqC.28}) and the Cauchy-Schwarz inequality, we have
\[
\Vert {\mathbf W}_{iT}(z,{\mathbf D}_i(z)) \Vert\geq \frac{-{\mathbf D}_{i}^{\star^\intercal}(z){\mathbf W}_{iT}(z, {\mathbf D}_\star(z))}{\xi_{NT}(c_1)}.
\]
Then, using (\ref{eqC.26}) and choosing $c_2>0$ to be sufficiently large, the event that
\[\inf_{\Vert{\mathbf D}_i(z)\Vert\geq\xi_{NT}(c_1)}\left\Vert(Th)^{1/2} {\mathbf W}_{iT}\left(z,{\mathbf D}_i(z)\right) \right\Vert<\xi_{NT}(c_2)=c_2(Th)^{1/2}\xi_{NT}\]
uniformly over $i=1,\cdots,N$ and $z\in[0,1]$, holds with probability convergent to zero. Let ${\cal E}_{3}$ denote the event that
\[\left\Vert (Th)^{1/2}{\mathbf W}_{iT}\left(z,\widehat{\mathbf D}_i(z)\right)\right\Vert< \xi_{NT}(c_2)\]
uniformly over $i=1,\cdots,N$ and $z\in[0,1]$, and let ${\cal E}_{4}$ denote the event that $
\left\Vert\widehat {\mathbf D}_i(z)\right\Vert\geq\xi_{NT}(c_1)$ uniformly over $i=1,\cdots,N$ and $z\in[0,1]$. From the above argument, we have
\begin{equation}\label{eqC.30}
	{\sf P}({\cal E}_{3}\cap{\cal E}_{4})\rightarrow0\ \ {\rm as}\ \ T\rightarrow\infty.
\end{equation}
On the other hand, from Lemma A.5 in \cite{SW12}, we readily have ${\sf P}({\cal E}_{3}^c)\rightarrow0$ as $T\rightarrow\infty$, which together with (\ref{eqC.30}), proves (\ref{eqC.23}) holds with probability approaching one. We have completed the proof of Lemma \ref{le:A.1}. \hfill$\blacksquare$

\medskip

\noindent{\bf Proof of Lemma \ref{le:A.2}}.\ \ By Lemma \ref{le:A.1} and the triangle inequality, we readily have that
\begin{eqnarray}
	\max_{1\leq i\leq N}\sup_{0\leq z\leq 1}\left\Vert \widehat{\boldsymbol\beta}_i(z)-{\boldsymbol\beta}_i(z)\right\Vert&\leq&\max_{1\leq i\leq N}\sup_{0\leq z\leq 1}\left\Vert \left({\mathbf 0}_d, {\mathbf I}_d\right){\boldsymbol\Omega}_{i}^{-1}(z){\mathbf W}_{iT,0}(z)\right\Vert+O_P\left(\xi_{NT}^{3/2}+h\xi_{NT}\right)\notag\\
	&=&\max_{1\leq i\leq N}\sup_{0\leq z\leq 1}\left\Vert \left({\mathbf 0}_d, {\mathbf I}_d\right){\boldsymbol\Omega}_{i}^{-1}(z){\mathbf W}_{iT,0}(z)\right\Vert+o_P\left(\xi_{NT}\right),\label{eqC.30}
\end{eqnarray}
where ${\mathbf 0}_d$ is a $d$-dimensional null vector and ${\mathbf I}_d$ is a $d\times d$ identity matrix. From Assumption 3(i), the maximum eigenvalue of ${\boldsymbol\Omega}_{i}^{-1}(z)$ is bounded uniformly over $i=1,\cdots,N$ and $z\in [0,1]$. Then, by (\ref{eqC.30}) and Lemma \ref{le:B.3}, we may complete the proof of Lemma \ref{le:A.2}.\hfill$\blacksquare$

\medskip

The main idea for proving Lemma \ref{le:A.3} is analogous to that used in the proofs of Theorem 3.2 in \cite{KGM12} and Lemma 3 in \cite{Ch21}. We start with Lemma \ref{le:B.5} which gives the consistency property for the post-grouping local linear quantile estimates.

\begin{lemma}\label{le:B.5}

Suppose that the assumptions of Lemma \ref{le:A.3} are satisfied. Then, we have
	\begin{equation}\label{eqC.31}
		\left\Vert\widetilde{\boldsymbol\gamma}_j(z)-{\boldsymbol\gamma}_j(z)\right\Vert_1+h_1\left\Vert\widetilde{\boldsymbol\gamma}_j^\prime(z)-{\boldsymbol\gamma}_j^\prime(z)\right\Vert_1=o_P(1),\ \ j=1,\cdots,R_0,
	\end{equation}
	and
	\begin{equation}\label{eqC.32}
		\max_{i\in{\cal G}_j^\tau}\big(\left\vert \widetilde{\alpha}_i(z)-\alpha_i\right\vert+h_1\left\vert \widetilde{\alpha}_i^\prime(z)\right\vert\big)=o_P(1),\ \ j=1,\cdots,R_0,
	\end{equation}
	where $\Vert \cdot\Vert_1$ denotes the $L_1$-norm for a vector.

\end{lemma}

\noindent{\bf Proof of Lemma \ref{le:B.5}}.\ \ For any small $\epsilon>0$ and $i\in{\cal G}_j$, $j=1,\cdots,R_0$, we define
	\begin{eqnarray}
		{\cal N}_i(\epsilon)=\left\{{\mathbf B}=\left(a_1,{\mathbf b}_1^{^\intercal},h_1a_2,h_1{\mathbf b}_2^{^\intercal}\right)^{^\intercal}\in{\cal R}^{2d+2}: \notag \right.\\   
		\left. \left\Vert {\mathbf b}_1-{\boldsymbol\gamma}_j(z)\right\Vert_1+h_1\left\Vert {\mathbf b}_2-{\boldsymbol\gamma}_j^\prime(z)\right\Vert_1+\left\vert a_1-\alpha_i\right\vert+h_1\left\vert a_2\right\vert\leq\epsilon\right\}. \notag
\end{eqnarray}
Let ${\cal N}_i^c(\epsilon)$ and $\partial {\cal N}_i(\epsilon)$ be the complement and boundary of ${\cal N}_i(\epsilon)$, respectively. Write
\[
\widetilde{\mathbf B}_{i}:=\widetilde{\mathbf B}_{i}(z)=\left[\widetilde\alpha_i(z), \widetilde{\boldsymbol\gamma}_j^{^\intercal}(z), h_1\widetilde\alpha_i^\prime(z), h_1\widetilde{\boldsymbol\gamma}_j^{\prime^{\intercal}}(z)\right]^{^\intercal},\ \ {\mathbf B}_{i,0}:={\mathbf B}_{i,0}(z)=\left[\alpha_i, {\boldsymbol\gamma}_j^{^\intercal}(z), 0, h_1{\boldsymbol\gamma}_j^{\prime^{\intercal}}(z)\right]^{^\intercal},
\]
and
\[{\rm Q}_{T,i}({\mathbf B})=\frac{1}{Th_1}\sum_{t=1}^T\rho_\tau\left(Y_{it}-{\mathbf X}_{it}^{^\intercal}{\mathbf b}_1-a_{i1}-(Z_t-z){\mathbf X}_{it}^{^\intercal}{\mathbf b}_2-(Z_t-z)a_{i2}\right)K_{h_1}(Z_t-z).\]
As in Section 3.3, it is sensible to prove (\ref{eqC.31}) and (\ref{eqC.32}) conditional on $\widetilde{\cal G}_j={\cal G}_j$.

If $\left\Vert \widetilde{\boldsymbol\gamma}_j(z)-{\boldsymbol\gamma}_j(z)\right\Vert_1+h_1\left\Vert \widetilde{\boldsymbol\gamma}_j^\prime(z)-{\boldsymbol\gamma}_j^\prime(z)\right\Vert_1>\epsilon$, we must have that $\widetilde{\mathbf B}_{i}\in {\cal N}_i^c(\epsilon)$ for $i\in{\cal G}_j$. Defining $\overline{\mathbf B}_{i}:=\overline{\mathbf B}_{i}(z)=\theta_i \widetilde{\mathbf B}_{i}(z)+(1-\theta_i){\mathbf B}_{i,0}(z)$ with
\[\theta_i=\frac{\epsilon}{\left\Vert \widetilde{\boldsymbol\gamma}_j(z)-{\boldsymbol\gamma}_j(z)\right\Vert_1+h_1\left\Vert \widetilde{\boldsymbol\gamma}_j^\prime(z)-{\boldsymbol\gamma}_j^\prime(z)\right\Vert_1+\left\vert \widetilde{\alpha}_i(z)-\alpha_i\right\vert+h_1\left\vert \widetilde{\alpha}_i(z)\right\vert}<1,\]
we can show that $\overline{\mathbf B}_{i}\in \partial{\cal N}_i(\epsilon)$. From the definitions of the elements in $\widetilde{\mathbf B}_{i}$, there exists at least one $i_0\in{\cal G}_j$ such that
\begin{equation}\label{eqC.33}
	{\rm Q}_{T,i_0}\big(\widetilde{\mathbf B}_{i_0}\big)\leq {\rm Q}_{T,i_0}\big({\mathbf B}_{i_0,0}\big).
\end{equation}
Meanwhile, by the convexity of ${\rm Q}_{T,i}(\cdot)$, we have
\begin{equation}\label{eqC.34}
	{\rm Q}_{T,i_0}\big(\overline{\mathbf B}_{i_0}\big)\leq \theta_{i_0}{\rm Q}_{T,i_0}\big(\widetilde{\mathbf B}_{i_0}\big)+(1-\theta_{i_0}){\rm Q}_{T,i_0}\big({\mathbf B}_{i_0,0}\big).
\end{equation}
By (\ref{eqC.33}) and (\ref{eqC.34}), we have
\begin{equation}\label{eqC.35}
	{\sf E}\left[\Delta_{T,i_0}\big(\overline{\mathbf B}_{i_0}\big)\right]\leq {\sf E}\left[\Delta_{T,i_0}\big(\overline{\mathbf B}_{i_0}\big)\right]-\Delta_{T,i_0}\big(\overline{\mathbf B}_{i_0}\big),
\end{equation}
where $\Delta_{T,i}({\mathbf B})={\rm Q}_{T,i}({\mathbf B})-{\rm Q}_{T,i}\big({\mathbf B}_{i,0}\big)$. Define
\[{\mathbf W}_{T,i}:={\mathbf W}_{T,i}(z)=\frac{1}{Th_1}\sum_{t=1}^T\eta_{it}(z)\left(\overline{Z}_t(z)\otimes \overline{\mathbf X}_{it}\right)K_{h_1}(Z_t-z),\]
where $\eta_{it}(z)$, $\overline{Z}_t(z)$ and $\overline{\mathbf X}_{it}$ are defined as in the proof of Lemma \ref{le:B.3}. Using the Taylor expansion for $\rho_\tau(\cdot)$ and Assumption 2(i), we have
\[{\sf E}\left[\Delta_{T,i_0}\big(\overline{\mathbf B}_{i_0}\big)\right]=\left(\overline{\mathbf B}_{i_0}-{\mathbf B}_{i_0,0}\right)^{^\intercal}{\sf E}\left[{\mathbf W}_{T,i_0}\right]+\frac{1}{2}\left(\overline{\mathbf B}_{i_0}-{\mathbf B}_{i_0,0}\right)^{^\intercal}{\boldsymbol\Omega}_{i_0}\left(\overline{\mathbf B}_{i_0}-{\mathbf B}_{i_0,0}\right)\left(1+o(1)\right),\]
where ${\boldsymbol\Omega}_i:={\boldsymbol\Omega}_i(z)$ is defined in (4.1). As ${\sf E}\left[{\mathbf W}_{T,i_0}\right]=O(h_1^2)=o(1)$, the smallest eigenvalue of ${\boldsymbol\Omega}_{i_0}$ is larger than zero, and $\overline{\mathbf B}_{i_0}\in\partial {\cal B}_{i_0}(\epsilon)$, we may show that
\begin{equation}\label{eqC.36}
	{\sf E}\left[\Delta_{T,i_0}\big(\overline{\mathbf B}_{i_0}\big)\right]\geq c_3\epsilon^2,
\end{equation}
where $c_3$ is a positive constant. By (\ref{eqC.35}) and (\ref{eqC.36}), in order to prove (\ref{eqC.31}), it is sufficient to show that
\begin{equation}\label{eqC.37}
	\sum_{i\in{\cal G}_j}{\sf P}\left(\sup_{{\mathbf B}\in{\cal N}_i(\epsilon)}\left\vert\Delta_{T,i}({\mathbf B})-{\sf E}\left[\Delta_{T,i}({\mathbf B})\right]\right\vert>c_3\epsilon^2\right)\rightarrow0.
\end{equation}
Following the proof of Lemma B.3, we may prove (\ref{eqC.37}). Details are thus omitted here. Then, we prove that
\begin{equation}\label{eqC.38}
	{\sf P}\left(\left\Vert \widetilde{\boldsymbol\gamma}_j(z)-{\boldsymbol\gamma}_j(z)\right\Vert_1+h_1\left\Vert \widetilde{\boldsymbol\gamma}_j^\prime(z)-{\boldsymbol\gamma}_j^\prime(z)\right\Vert_1>\epsilon\right)\rightarrow0,
\end{equation}
completing the proof of (\ref{eqC.31}).

We next turn to the proof of (\ref{eqC.32}). By the definitions of $\widetilde{\alpha}_i(z)$ and $\widetilde{\alpha}_i^\prime(z)$, for any $i\in{\cal G}_j$, ${\rm Q}_{T,i}\big(\widetilde{\mathbf B}_i\big)\leq {\rm Q}_{T,i}\big(\widetilde{\mathbf B}_{i,0}\big)$, where $\widetilde{\mathbf B}_{i,0}$ is defined similarly to $\widetilde{\mathbf B}_i$ but with $\widetilde{\alpha}_i(z)$ and $h_1\widetilde{\alpha}_i^\prime(z)$ replaced by $\alpha_i$ and $0$, respectively. If there exists $i\in{\cal G}_j$ such that $\left\vert \widetilde{\alpha}_{i}(z)-\alpha_i\right\vert+h_1\left\vert \widetilde{\alpha}_i(z)\right\vert>\epsilon$, we must have $\widetilde{\mathbf B}_i\in{\cal N}_i^c(\epsilon)$. The above arguments, together with the convexity of ${\rm Q}_{T,i}(\cdot)$, lead to
\begin{equation}\label{eqC.39}
	\frac{{\rm Q}_{T,i}\big(\overline{\mathbf B}_i\big)-{\rm Q}_{T,i}\big({\mathbf B}_{i,0}\big)}{\theta_i}\leq {\rm Q}_{T,i}\big(\widetilde{\mathbf B}_i\big)-{\rm Q}_{T,i}\big({\mathbf B}_{i,0}\big)\leq {\rm Q}_{T,i}\big(\widetilde{\mathbf B}_{i,0}\big)-{\rm Q}_{T,i}\big({\mathbf B}_{i,0}\big).
\end{equation}
By (\ref{eqC.39}), we have
\begin{equation}\label{eqC.40}
	{\sf E}\left[\Delta_{T,i}\big(\overline{\mathbf B}_{i}\big)\right]\leq \left\{{\sf E}\left[\Delta_{T,i}\big(\overline{\mathbf B}_{i}\big)\right]-\Delta_{T,i}\big(\overline{\mathbf B}_{i}\big)\right\}+\theta_i\left[{\rm Q}_{T,i}\big(\widetilde{\mathbf B}_{i,0}\big)-{\rm Q}_{T,i}\big({\mathbf B}_{i,0}\big)\right],
\end{equation}
By (\ref{eqC.36}), (\ref{eqC.37}) and (\ref{eqC.40}) and noting that $\theta_i\leq 1$, it is sufficient to show that
\begin{equation}\label{eqC.41}
	\max_{i\in{\cal G}_j}\left\vert {\rm Q}_{T,i}\big(\widetilde{\mathbf B}_{i,0}\big)-{\rm Q}_{T,i}\big({\mathbf B}_{i,0}\big)\right\vert=o_P(1).
\end{equation}
Noting that $\vert \rho_\tau(z_1)-\rho_\tau(z_2)\vert\leq 2\vert z_1-z_2\vert$ and using (\ref{eqC.31}), we have
\begin{eqnarray}
	&&\max_{i\in{\cal G}_j}\left\vert {\rm Q}_{T,i}\big(\widetilde{\mathbf B}_{i,0}\big)-{\rm Q}_{T,i}\big({\mathbf B}_{i,0}\big)\right\vert\notag\\
	&\leq&C\left[\left\Vert \widetilde{\boldsymbol\gamma}_j(z)-{\boldsymbol\gamma}_j(z)\right\Vert 
	+h_1\left\Vert \widetilde{\boldsymbol\gamma}_j^\prime(z)-{\boldsymbol\gamma}_j^\prime(z)\right\Vert\right] \notag\\ 
	&& \times \max_{i\in{\cal G}_j}\frac{1}{Th_1}\sum_{t=1}^T\Vert {\mathbf X}_{it}\Vert \left(1+\left\vert\frac{Z_t-z}{h_1}\right\vert\right)K_{h_1}(Z_t-z)\notag\\
	&\leq&o_P(1)\cdot\max_{i\in{\cal G}_j}\frac{1}{Th_1}\sum_{t=1}^T\Vert {\mathbf X}_{it}\Vert \left(1+\left\vert\frac{Z_t-z}{h_1}\right\vert\right)K_{h_1}(Z_t-z).\label{eqC.42}
\end{eqnarray}
Following the proof of Lemma \ref{le:B.3}, we may show that
\begin{equation}\label{eqC.43}
	\max_{i\in{\cal G}_j}\frac{1}{Th_1}\sum_{t=1}^T\Vert {\mathbf X}_{it}\Vert \left(1+\left\vert\frac{Z_t-z}{h_1}\right\vert\right)K_{h_1}(Z_t-z)=O_P(1).
\end{equation}
With (\ref{eqC.42}) and (\ref{eqC.43}), we prove (\ref{eqC.41}), thus completing the proof of (\ref{eqC.32}).\hfill$\blacksquare$

\medskip

\noindent{\bf Proof of Lemma \ref{le:A.3}}.\ \ Let
\[\eta_{it}(z,{\mathbf B})=\tau-I\left(e_{it}+b_{it}(z)-\left[\overline{Z}_t(z)\otimes \overline{\mathbf X}_{it}\right]({\mathbf B}-{\mathbf B}_{i,0})\leq 0\right),\]
which is similar to the definition of $\eta_{it}(z,{\mathbf D}_i(z))$ given before presenting Lemma \ref{le:B.4}. Define
\begin{eqnarray}
	{\mathbf W}_{T,i}^\gamma({\mathbf B})&=&\frac{1}{Th_1}\sum_{t=1}^T\eta_{it}(z,{\mathbf B})\left[\overline{Z}_t(z)\otimes {\mathbf X}_{it}\right]K_{h_1}(Z_t-z),\notag\\
	{\mathbf W}_{T,i}^\alpha({\mathbf B})&=&\frac{1}{Th_1}\sum_{t=1}^T\eta_{it}(z,{\mathbf B})\overline{Z}_t(z)K_{h_1}(Z_t-z),\notag
\end{eqnarray}
$\overline{\mathbf W}_{T,i}^\gamma({\mathbf B})={\sf E}\left[{\mathbf W}_{T,i}^\gamma({\mathbf B})\right]$ and $\overline{\mathbf W}_{T,i}^\alpha({\mathbf B})={\sf E}\left[{\mathbf W}_{T,i}^\alpha({\mathbf B})\right]$. Let
\begin{eqnarray}
	\overline{\mathbf W}_{T,i}^{\gamma\gamma}&=&\frac{\partial \overline{\mathbf W}_{T,i}^\gamma({\mathbf B})}{\partial \gamma}|_{{\mathbf B}={\mathbf B}_{i,0}},\ \ \overline{\mathbf W}_{T,i}^{\alpha\alpha}=\frac{\partial \overline{\mathbf W}_{T,i}^\gamma({\mathbf B})}{\partial \alpha}|_{{\mathbf B}={\mathbf B}_{i,0}},\notag\\
	\overline{\mathbf W}_{T,i}^{\gamma\alpha}&=&\frac{\partial \overline{\mathbf W}_{T,i}^\gamma({\mathbf B})}{\partial \alpha}|_{{\mathbf B}={\mathbf B}_{i,0}},\ \ \overline{\mathbf W}_{T,i}^{\alpha\gamma}=\frac{\partial \overline{\mathbf W}_{T,i}^\gamma({\mathbf B})}{\partial \gamma}|_{{\mathbf B}={\mathbf B}_{i,0}}.\notag
\end{eqnarray}
As in \cite{KGM12} and \cite{Ch21}, we may divide the proof into several steps. As in the proof of Lemma \ref{le:B.5}, we prove the Bahadur representation conditional on $\widetilde{\cal G}_j={\cal G}_j$.

\smallskip

\noindent{\sc Step 1}:\ \ Compute $\overline{\mathbf W}_{T,i}^{\gamma\gamma}$, $\overline{\mathbf W}_{T,i}^{\alpha\alpha}$, $\overline{\mathbf W}_{T,i}^{\gamma\alpha}$ and $\overline{\mathbf W}_{T,i}^{\alpha\gamma}$.

With standard calculations, we may show that
\begin{eqnarray}
	\overline{\mathbf W}_{T,i}^{\gamma\gamma}&=&{\boldsymbol\Omega}(K)\otimes {\boldsymbol\Omega}_{i}^\gamma(z)+O(h_1),\ \ \ \ \ \ \overline{\mathbf W}_{T,i}^{\alpha\alpha}=\omega_i^\alpha(z){\boldsymbol\Omega}(K)+O(h_1),\label{eqC.44}\\
	\overline{\mathbf W}_{T,i}^{\gamma\alpha}&=&{\boldsymbol\Omega}(K)\otimes {\boldsymbol\Omega}_{i}^{\gamma\alpha}(z)+O(h_1),\ \ \overline{\mathbf W}_{T,i}^{\alpha\gamma}={\boldsymbol\Omega}(K)\otimes {\boldsymbol\Omega}_{i}^{\alpha\gamma}(z)+O(h_1),\label{eqC.45}
\end{eqnarray}
where ${\boldsymbol\Omega}(K)={\sf diag}(1,\mu_2)$, $\omega_i^{\alpha}(z)$, ${\boldsymbol\Omega}_{i}^{\gamma}(z)$, ${\boldsymbol\Omega}_{i}^{\gamma\alpha}(z)$ and ${\boldsymbol\Omega}_{i}^{\alpha\gamma}(z)$ are defined in (4.5) of the main document.

\smallskip

\noindent{\sc Step 2}:\ \ Expand $\overline{\mathbf W}_{T,i}^\gamma(\widetilde{\mathbf B}_i)$ and $\overline{\mathbf W}_{T,i}^\alpha(\widetilde{\mathbf B}_i)$.

From Lemma \ref{le:B.5}, there exist non-increasing sequences $\left\{\delta_{NT}^\alpha\right\}$ and $\left\{\delta_{NT}^\gamma\right\}$ such that $\delta_{NT}^\alpha\rightarrow0$, $\delta_{NT}^\gamma\rightarrow0$,
\[\left\Vert\widetilde{\boldsymbol\gamma}_j(z)-{\boldsymbol\gamma}_j(z)\right\Vert+h_1\left\Vert\widetilde{\boldsymbol\gamma}_j^\prime(z)-{\boldsymbol\gamma}_j^\prime(z)\right\Vert=O_P\left(\delta_{NT}^\gamma\right)\]
and
\[\max_{i\in{\cal G}_j}\big(\left\vert \widetilde{\alpha}_i(z)-\alpha_i\right\vert+h_1\left\vert \widetilde{\alpha}_i^\prime(z)\right\vert\big)=O_P\left(\delta_{NT}^\alpha\right)\]
for $j=1,\cdots, R_\tau$. Let $\delta_{NT}=\delta_{NT}^\gamma\vee\delta_{NT}^\alpha$. By the Taylor expansion for $\overline{\mathbf W}_{T,i}^\gamma(\widetilde{\mathbf B}_i)$ and $\overline{\mathbf W}_{T,i}^\alpha(\widetilde{\mathbf B}_i)$, we have
\begin{eqnarray}
	\overline{\mathbf W}_{T,i}^\gamma(\widetilde{\mathbf B}_i)&=&\overline{\mathbf W}_{T,i}^\gamma({\mathbf B}_{i,0})+\overline{\mathbf W}_{T,i}^{\gamma\gamma}\left(
	\begin{array}{c}
		\widetilde{\boldsymbol\gamma}_j(z)-{\boldsymbol\gamma}_j(z)\\
		h_1\widetilde{\boldsymbol\gamma}_j^\prime(z)-h_1{\boldsymbol\gamma}_j^\prime(z)
	\end{array}
	\right)+\overline{\mathbf W}_{T,i}^{\gamma\alpha}\left(
	\begin{array}{c}
		\widetilde{\alpha}_i(z)-{\alpha}_i\\
		h_1\widetilde{\alpha}_i^\prime(z)
	\end{array}
	\right)+\notag\\
	&&O_P\left(\delta_{NT}\left(\left\Vert\widetilde{\boldsymbol\gamma}_j(z)-{\boldsymbol\gamma}_j(z)\right\Vert+h_1\left\Vert\widetilde{\boldsymbol\gamma}_j^\prime(z)-{\boldsymbol\gamma}_j^\prime(z)\right\Vert\right)\right)+O_P((\delta_{NT}^\alpha)^2),\label{eqC.46}
\end{eqnarray}
and
\begin{eqnarray}
	\overline{\mathbf W}_{T,i}^\alpha(\widetilde{\mathbf B}_i)&=&\overline{\mathbf W}_{T,i}^\alpha({\mathbf B}_{i,0})+\overline{\mathbf W}_{T,i}^{\alpha\gamma}\left(
	\begin{array}{c}
		\widetilde{\boldsymbol\gamma}_j(z)-{\boldsymbol\gamma}_j(z)\\
		h_1\widetilde{\boldsymbol\gamma}_j^\prime(z)-h_1{\boldsymbol\gamma}_j^\prime(z)
	\end{array}
	\right)+\overline{\mathbf W}_{T,i}^{\alpha\alpha}\left(
	\begin{array}{c}
		\widetilde{\alpha}_i(z)-{\alpha}_i\\
		h_1\widetilde{\alpha}_i^\prime(z)
	\end{array}
	\right)+\notag\\
	&&O_P\left(\delta_{NT}\left(\left\Vert\widetilde{\boldsymbol\gamma}_j(z)-{\boldsymbol\gamma}_j(z)\right\Vert+h_1\left\Vert\widetilde{\boldsymbol\gamma}_j^\prime(z)-{\boldsymbol\gamma}_j^\prime(z)\right\Vert\right)\right)+O_P((\delta_{NT}^\alpha)^2)\label{eqC.47}
\end{eqnarray}
for $i\in{\cal G}_j$. By (\ref{eqC.46}) and (\ref{eqC.47}), we have
\begin{eqnarray}
	&&\left[\overline{\mathbf W}_{T,i}^{\gamma\gamma}-\overline{\mathbf W}_{T,i}^{\gamma\alpha}\left(\overline{\mathbf W}_{T,i}^{\alpha\alpha}\right)^{-1}\overline{\mathbf W}_{T,i}^{\alpha\gamma}\right]\left(
	\begin{array}{c}
		\widetilde{\boldsymbol\gamma}_j(z)-{\boldsymbol\gamma}_j(z)\\
		h_1\widetilde{\boldsymbol\gamma}_j^\prime(z)-h_1{\boldsymbol\gamma}_j^\prime(z)
	\end{array}
	\right)\notag\\
	&=&\left[\overline{\mathbf W}_{T,i}^\gamma(\widetilde{\mathbf B}_i)-\overline{\mathbf W}_{T,i}^\gamma({\mathbf B}_{i,0})\right]-\overline{\mathbf W}_{T,i}^{\gamma\alpha}\left(\overline{\mathbf W}_{T,i}^{\alpha\alpha}\right)^{-1}\left[\overline{\mathbf W}_{T,i}^\alpha(\widetilde{\mathbf B}_i)-\overline{\mathbf W}_{T,i}^\alpha({\mathbf B}_{i,0})\right]+\notag\\
	&&O_P\left(\delta_{NT}\left(\left\Vert\widetilde{\boldsymbol\gamma}_j(z)-{\boldsymbol\gamma}_j(z)\right\Vert+h_1\left\Vert\widetilde{\boldsymbol\gamma}_j^\prime(z)-{\boldsymbol\gamma}_j^\prime(z)\right\Vert\right)\right)+O_P((\delta_{NT}^\alpha)^2),\notag
\end{eqnarray}
which, together with (\ref{eqC.44}) and (\ref{eqC.45}) obtained in Step 1, leads to
\begin{eqnarray}
	&&\left[{\boldsymbol\Omega}(K)\otimes {\boldsymbol\Omega}(z; {\cal G}_j)\right]\left(
	\begin{array}{c}
		\widetilde{\boldsymbol\gamma}_j(z)-{\boldsymbol\gamma}_j(z)\\
		h_1\widetilde{\boldsymbol\gamma}_j^\prime(z)-h_1{\boldsymbol\gamma}_j^\prime(z)
	\end{array}
	\right)\notag\\
	&=&\frac{1}{N_j}\sum_{i\in{\cal G}_j}\left[\overline{\mathbf W}_{T,i}^\gamma(\widetilde{\mathbf B}_i)-\overline{\mathbf W}_{T,i}^\gamma({\mathbf B}_{i,0})\right]-\frac{1}{N_j}\sum_{i\in{\cal G}_j}\overline{\mathbf W}_{T,i}^{\gamma\alpha}\left(\overline{\mathbf W}_{T,i}^{\alpha\alpha}\right)^{-1}\left[\overline{\mathbf W}_{T,i}^\alpha(\widetilde{\mathbf B}_i)-\overline{\mathbf W}_{T,i}^\alpha({\mathbf B}_{i,0})\right]+\notag\\
	&&O_P\left((\delta_{NT}+h_1)\left(\left\Vert\widetilde{\boldsymbol\gamma}_j(z)-{\boldsymbol\gamma}_j(z)\right\Vert+h_1\left\Vert\widetilde{\boldsymbol\gamma}_j^\prime(z)-{\boldsymbol\gamma}_j^\prime(z)\right\Vert\right)\right)+O_P((\delta_{NT}^\alpha)^2),\label{eqC.48}
\end{eqnarray}
where ${\boldsymbol\Omega}(z; {\cal G}_j)$ is defined in (4.6) of the main document.

\smallskip

\noindent{\sc Step 3}:\ \ Expand $\frac{1}{N_j}\sum_{i\in{\cal G}_j}\left[\overline{\mathbf W}_{T,i}^\gamma(\widetilde{\mathbf B}_i)-\overline{\mathbf W}_{T,i}^\gamma({\mathbf B}_{i,0})\right]$.

Observe that
\begin{eqnarray}
	\frac{1}{N_j}\sum_{i\in{\cal G}_j}\overline{\mathbf W}_{T,i}^\gamma(\widetilde{\mathbf B}_i)&=&\frac{1}{N_jTh_1}\sum_{i\in{\cal G}_j}\sum_{t=1}^T{\sf E}\left[\eta_{it}(z,{\mathbf B})\left(\overline{Z}_t(z)\otimes {\mathbf X}_{it}\right)K_{h_1}(Z_t-z)\right]\notag\\
	&=&\frac{1}{N_jTh_1}\sum_{i\in{\cal G}_j}\sum_{t=1}^T\eta_{it}(z,{\mathbf B})\left(\overline{Z}_t(z)\otimes {\mathbf X}_{it}\right)K_{h_1}(Z_t-z)-U_{NT}^\gamma(\widetilde{\mathbf B}, {\cal G}_j)\notag\\
	&=&\frac{1}{N_j}\sum_{i\in{\cal G}_j}{\mathbf W}_{T,i}^\gamma(\widetilde{\mathbf B}_i)-U_{NT}^\gamma(\widetilde{\mathbf B}, {\cal G}_j),\label{eqC.49}
\end{eqnarray}
where
{\footnotesize\begin{eqnarray}
	&&U_{NT}^\gamma(\widetilde{\mathbf B}, {\cal G}_j)\notag\\
	&=&\frac{1}{N_jTh_1}\sum_{i\in{\cal G}_j}\sum_{t=1}^T\left\{\eta_{it}(z,\widetilde{\mathbf B}_i)\left(\overline{Z}_t(z)\otimes {\mathbf X}_{it}\right)K_{h_1}(Z_t-z)-{\sf E}\left[\eta_{it}(z,\widetilde{\mathbf B}_i)\left(\overline{Z}_t(z)\otimes {\mathbf X}_{it}\right)K_{h_1}(Z_t-z)\right]\right\}.\notag
\end{eqnarray}}
By (\ref{eqC.49}), we may show that
\begin{eqnarray}
	\frac{1}{N_j}\sum_{i\in{\cal G}_j}\left[\overline{\mathbf W}_{T,i}^\gamma(\widetilde{\mathbf B}_i)-\overline{\mathbf W}_{T,i}^\gamma({\mathbf B}_{i,0})\right]&=&\frac{1}{N_j}\sum_{i\in{\cal G}_j}{\mathbf W}_{T,i}^\gamma(\widetilde{\mathbf B}_i)-\left[U_{NT}^\gamma(\widetilde{\mathbf B}, {\cal G}_j)-U_{NT}^\gamma({\mathbf B}_0, {\cal G}_j)\right]-\notag\\
	&&\left[U_{NT}^\gamma({\mathbf B}_0, {\cal G}_j)-\overline{\mathbf W}_{T,i}^\gamma({\mathbf B}_{i,0})\right],\label{eqC.50}
\end{eqnarray}
where
{\footnotesize\begin{align}
		&U_{NT}^\gamma({\mathbf B}_0, {\cal G}_j)\notag\\
		&=\frac{1}{N_jTh_1}\sum_{i\in{\cal G}_j}\sum_{t=1}^T\left\{\eta_{it}(z,{\mathbf B}_{i,0})\left(\overline{Z}_t(z)\otimes {\mathbf X}_{it}\right)K_{h_1}(Z_t-z)-{\sf E}\left[\eta_{it}(z,{\mathbf B}_{i,0})\left(\overline{Z}_t(z)\otimes {\mathbf X}_{it}\right)K_{h_1}(Z_t-z)\right]\right\}\notag\\
		&=\frac{1}{N_jTh_1}\sum_{i\in{\cal G}_j}\sum_{t=1}^T\left\{\eta_{it}(z)\left(\overline{Z}_t(z)\otimes {\mathbf X}_{it}\right)K_{h_1}(Z_t-z)-{\sf E}\left[\eta_{it}(z)\left(\overline{Z}_t(z)\otimes {\mathbf X}_{it}\right)K_{h_1}(Z_t-z)\right]\right\}.\notag
\end{align}}
By the computational property of the quantile regression estimate, we have
\begin{equation}\label{eqC.51}
	\frac{1}{N_j}\sum_{i\in{\cal G}_j}{\mathbf W}_{T,i}^\gamma(\widetilde{\mathbf B}_i)=O_P\left(\frac{1}{Th_1}\right).
\end{equation}
It is straightforward to show
\begin{equation}\label{eqC.52}
	U_{NT}^\gamma({\mathbf B}_0, {\cal G}_j)-\overline{\mathbf W}_{T,i}^\gamma({\mathbf B}_{i,0})=\frac{1}{N_jTh_1}\sum_{i\in{\cal G}_j}\sum_{t=1}^T\eta_{it}(z)\left(\overline{Z}_t(z)\otimes {\mathbf X}_{it}\right)K_{h_1}(Z_t-z).
\end{equation}
Using Lemma \ref{le:B.5} and following the proof of Lemma \ref{le:B.4}, we may prove that
\begin{equation}\label{eqC.53}
	U_{NT}^\gamma(\widetilde{\mathbf B}, {\cal G}_j)-U_{NT}^\gamma({\mathbf B}_0, {\cal G}_j)=O_P\left(\xi_{NT}^\ast\delta_{NT}^{1/2}\right),
\end{equation}
where $\xi_{NT}^\ast=\left(\frac{\log (T\vee N)}{Th_1}\right)^{1/2}$. Combining (\ref{eqC.50})--(\ref{eqC.53}), we have
\begin{eqnarray}
	\frac{1}{N_j}\sum_{i\in{\cal G}_j}\left[\overline{\mathbf W}_{T,i}^\gamma(\widetilde{\mathbf B}_i)-\overline{\mathbf W}_{T,i}^\gamma({\mathbf B}_{i,0})\right]&=&\frac{1}{N_jTh_1}\sum_{i\in{\cal G}_j}\sum_{t=1}^T\eta_{it}(z)\left(\overline{Z}_t(z)\otimes {\mathbf X}_{it}\right)K_{h_1}(Z_t-z)+\notag\\
	&&O_P\left(\frac{1}{Th_1}\right)+O_P\left(\xi_{NT}^\ast\delta_{NT}^{1/2}\right).\label{eqC.54}
\end{eqnarray}

\smallskip

\noindent{\sc Step 4}:\ \ Expand $\overline{\mathbf W}_{T,i}^\alpha(\widetilde{\mathbf B}_i)-\overline{\mathbf W}_{T,i}^\alpha({\mathbf B}_{i,0})$ uniformly over $i\in{\cal G}_j$.

Using the arguments in Step 3, we may show that
\begin{equation}\label{eqC.55}
	\overline{\mathbf W}_{T,i}^\alpha(\widetilde{\mathbf B}_i)-\overline{\mathbf W}_{T,i}^\alpha({\mathbf B}_{i,0})=\frac{1}{Th_1}\sum_{t=1}^T\eta_{it}(z)\overline{Z}_t(z)K_{h_1}(Z_t-z)+O_P\left(\frac{1}{Th_1}\right)+O_P\left(\xi_{NT}^\ast\delta_{NT}^{1/2}\right)
\end{equation}
uniformly over $i\in{\cal G}_j$. Details are omitted to save the space.

\smallskip

\noindent{\sc Step 5}:\ \  Derive the specific rates for $\delta_{NT}^\gamma$, $\delta_{NT}^\alpha$ and $\delta_{NT}$.

Inserting the expansions (\ref{eqC.54}) and (\ref{eqC.55}) into (\ref{eqC.48}), we readily have that
\begin{eqnarray}
	&&\left[{\boldsymbol\Omega}(K)\otimes {\boldsymbol\Omega}(z; {\cal G}_j)\right]\left(
	\begin{array}{c}
		\widetilde{\boldsymbol\gamma}_j(z)-{\boldsymbol\gamma}_j(z)\\
		h_1\widetilde{\boldsymbol\gamma}_j^\prime(z)-h_1{\boldsymbol\gamma}_j^\prime(z)
	\end{array}
	\right)\notag\\
	&=&\frac{1}{N_jTh_1}\sum_{i\in{\cal G}_j}\sum_{t=1}^T\eta_{it}(z)\left(\overline{Z}_t(z)\otimes {\mathbf X}_{it}\right)K_{h_1}(Z_t-z)-\notag\\
	&&\frac{1}{N_jTh_1}\sum_{i\in{\cal G}_j}\overline{\mathbf W}_{T,i}^{\gamma\alpha}\left(\overline{\mathbf W}_{T,i}^{\alpha\alpha}\right)^{-1}\sum_{t=1}^T\eta_{it}(z)\overline{Z}_t(z)K_{h_1}(Z_t-z)+\notag\\
	&&O_P\left((\delta_{NT}+h_1)\left(\left\Vert\widetilde{\boldsymbol\gamma}_j(z)-{\boldsymbol\gamma}_j(z)\right\Vert+h_1\left\Vert\widetilde{\boldsymbol\gamma}_j^\prime(z)-{\boldsymbol\gamma}_j^\prime(z)\right\Vert\right)\right)+\notag\\
	&&O_P\left((\delta_{NT}^\alpha)^2\right)+O_P\left(\frac{1}{Th_1}\right)+O_P\left(\xi_{NT}^\ast\delta_{NT}^{1/2}\right)\label{eqC.56}
\end{eqnarray}
Following the proof of Lemma \ref{le:B.3}, we may show that
\[\frac{1}{Th_1}\sum_{t=1}^T\eta_{it}(z)\left(\overline{Z}_t(z)\otimes \overline{\mathbf X}_{it}\right)K_{h_1}(Z_t-z)=O_P\left(h_1^2+\xi_{NT}^\ast\right)=O_P\left(\xi_{NT}^\ast\right)
\]
uniformly over $i\in{\cal G}_j$, which, together with (\ref{eqC.44}), (\ref{eqC.45}), indicates that
\begin{equation}\label{eqC.57}
	\left\Vert\widetilde{\boldsymbol\gamma}_j(z)-{\boldsymbol\gamma}_j(z)\right\Vert+h_1\left\Vert\widetilde{\boldsymbol\gamma}_j^\prime(z)-{\boldsymbol\gamma}_j^\prime(z)\right\Vert=O_P\left(\xi_{NT}^\ast\right)+O_P\left((\delta_{NT}^\alpha)^2\right).
\end{equation}
Similarly, by (\ref{eqC.44}), (\ref{eqC.45}), (\ref{eqC.47}), (\ref{eqC.55}) and (\ref{eqC.57}), we have
\begin{equation}\label{eqC.58}
	\max_{i\in{\cal G}_j}\big(\left\vert \widetilde{\alpha}_i(z)-\alpha_i\right\vert+h_1\left\vert \widetilde{\alpha}_i\right\vert\big)=O_P\left(\xi_{NT}^\ast\right).
\end{equation}
With (\ref{eqC.57}) and (\ref{eqC.58}), we may claim that $\delta_{NT}^\gamma$, $\delta_{NT}^\alpha$ and $\delta_{NT}$ all have the rate of $\xi_{NT}^\ast$.

\smallskip

\noindent{\sc Step 6}:\ \ Obtain the Bahadur representation for $\widetilde{\boldsymbol\gamma}_j(z)-{\boldsymbol\gamma}_j(z)$.

Finally, using (\ref{eqC.44}), (\ref{eqC.45}) and (\ref{eqC.56}), and noting that
\[(\delta_{NT}^\alpha)^2+\frac{1}{Th_1}+\xi_{NT}^\ast\delta_{NT}^{1/2}=O\left(\frac{1}{Th_1}+\left(\xi_{NT}^\ast\right)^{3/2}\right)=o\left(\frac{1}{\sqrt{N_jTh_1}}\right)\]
by Assumption 6(ii), we can obtain the Bahadur representation (A.11) for $\widetilde{\boldsymbol\gamma}_j(z)-{\boldsymbol\gamma}_j(z)$, completing the proof of Lemma \ref{le:A.3}. \hfill$\blacksquare$


\renewcommand{\theequation}{C.\arabic{equation}}
\setcounter{equation}{0}

\section*{Appendix C:\ \ Extensions and discussions of the methodology and theory}

In this appendix, we modify the estimation methods developed in the main document to identify the latent group structures in linear panel quantile regression uniformly over $\tau$, extend the asymptotic theory to cover time-varying panel quantile regression, and then construct the point-wise confidence intervals.


\subsection*{Appendix C.1:\ \ Linear quantile regression with uniform latent group structure}

At a given quantile level $\tau$, with the collected panel observations $\left(Y_{it}, {\mathbf X}_{it}\right)$, $i=1,\cdots,N$, $t=1,\cdots,T$, we consider the linear quantile regression model defined by
\begin{equation}\label{eq5.1}
	{\sf Q}_{\tau, i}\left(Y_{it} | {\mathbf X}_{it}, \alpha_i\right)= {\mathbf X}_{it}^{^\intercal} {\boldsymbol\beta}_{i}^\star(\tau)+\alpha_i^\star(\tau),
\end{equation}
where ${\boldsymbol\beta}_{i}^\star(\tau)$ are $d$-dimensional vectors of regression parameters and $\alpha_i^\star(\tau)$ are fixed effects. Here we make the dependence of ${\boldsymbol\beta}_{i}^\star(\tau)$ and $\alpha_i^\star(\tau)$ on $\tau$ explicit. The model (\ref{eq5.1}) can be viewed as a special case of (2.1). As in \cite{KX06}, we may re-write (\ref{eq5.1}) in a functional-coefficient specification:
\begin{equation}\label{eq5.2}
	Y_{it}= {\mathbf X}_{it}^{^\intercal} {\boldsymbol\beta}_{i}^\star(U_t)+\alpha_i^\star(U_t),
\end{equation}
where $U_t$ independently follows ${\sf U}[0,1]$ and the right hand side is monotonically increasing in $U_t$. Model (\ref{eq5.2}) is often called a random-coefficient quantile regression. As in Section 2, we may impose a uniform group structure for the heterogenous regression coefficients, i.e., there exists a partition of $\{1,2,\cdots,N\}$ denoted by $\boldsymbol{{\cal G}}^\star=\left\{{\cal G}_1^\star,\cdots,{\cal G}_{K_0}^\star\right\}$ such that
\begin{equation}\label{eq5.3}
	{\boldsymbol\beta}_i^\star(\tau)={\boldsymbol\gamma}_k^\star(\tau),\ \ 0<\tau<1,\ \ {\rm for}\ \ i\in{\cal G}_k^\star,\ \ k=1,\cdots,K_0.
\end{equation}
The main difference between (\ref{eq5.3}) and (2.3) is that the functional equivalence in (\ref{eq5.3}) holds uniformly over the quantile level $\tau$ whereas that in (2.3) holds uniformly over the support of an observable index variable for a fixed $\tau$. However, the clustering algorithm and the ratio criterion proposed in Sections 3.1 and 3.2 are still applicable (with some modifications) to estimate the group number $K_0$ and the unknown structure $\boldsymbol{{\cal G}}^\star$ by replacing the preliminary local linear quantile estimation by the conventional linear quantile regression estimation.

At each quantile level $0<\tau<1$ and $1\leq i\leq N$, we estimate $\left[\alpha_i^\star(\tau),{\boldsymbol\beta}_{i}^{\star^\intercal}(\tau)\right]^{^\intercal}$ by minimising the following objective function:
\[
{\rm Q}_{\tau,i}^\star(a,{\mathbf b})=\sum_{t=1}^T\rho_\tau\left(Y_{it}-a-{\mathbf X}_{it}^{^\intercal}{\mathbf b}\right),
\]
and obtain
\[
\left[\overline\alpha_i^\star(\tau), \overline{\boldsymbol\beta}_{i}^{\star^\intercal}(\tau)\right]^{^\intercal}=\argmin_{a,{\mathbf b}}{\rm Q}_{\tau,i}^\star(a,{\mathbf b}).
\]
Following the standard arguments (e.g., the empirical process limit theory) and using the techniques in the proofs of Lemmas \ref{le:A.1} and \ref{le:A.2} in Appendix B, we may show that
\begin{equation}\label{eq5.4}
	\max_{1\leq i\leq N}\sup_{\tau\in[\varepsilon, 1-\varepsilon]}\left\Vert \overline{\boldsymbol\beta}_i^\star(\tau)-{\boldsymbol\beta}_i^\star(\tau)\right\Vert=O_P\left(\sqrt{\log (N\vee T)/T}\right),
\end{equation}
where $\varepsilon$ is a small positive number. As in Section 3.1, we may construct $\widetilde{\boldsymbol\Delta}$, an $N\times N$ estimated distance matrix whose $(j,k)$-entry is defined by
\[\widetilde{\Delta}(j,k)=\int_{\varepsilon}^{1-\varepsilon}\left\Vert \overline{\boldsymbol\beta}_j^\star(u)-\overline{\boldsymbol\beta}_k^\star(u)\right\Vert du.\]
With the estimated distance matrix $\widetilde{\boldsymbol\Delta}$, we implement the agglomerative clustering algorithm as in Section 3.1. If the group number $K_0$ is known a priori, following the proof of Theorem 4.1 and using the uniform consistency result (\ref{eq5.4}), we may show that the estimated clusters $\overline{\cal G}_k^\star$, $k=1,\cdots,K_0$, are consistent. Then, the ratio criterion proposed in Section 3.2 can be used to consistently estimate the number $K_0$ after some minor modifications. For example, ${\sf D}(R)$ defined in (3.2) is replaced by
\begin{equation}\label{eq5.5}
	\overline{\sf D}(K)= \frac{1}{K}\sum_{k=1}^K\frac{1}{\left\vert\overline{\cal G}_{k|K}^\star\right\vert}\sum_{j\in\overline{\cal G}_{k|K}}\int_{\varepsilon}^{1-\varepsilon} \left\Vert\overline{\boldsymbol\beta}_j^\star(u)-\overline{\boldsymbol\beta}_{k|K}^\star(u)\right\Vert du,
\end{equation}
where $\overline{\boldsymbol\beta}_{k|K}(u)=\frac{1}{\left\vert\overline{\cal G}_{k|K}^\star\right\vert}\sum_{j\in\overline{\cal G}_{k|K}^\star}\overline{\boldsymbol\beta}_j^\star(u)$, and $\overline{\cal G}_{k|K}^\star$, $k=1,\cdots,K$, are the estimated groups when the group number is assumed to be $K$.


\subsection*{Appendix C.2:\ \ Time-varying coefficient panel quantile regression}

In practical applications, it is often common to choose $Z_t=t/T$, a scaled time point, and study the time-varying coefficient panel quantile regression model in (2.2). As mentioned in the introductory section, the methods developed in Section 3 are still applicable to estimate the latent group structure and the time-varying coefficient functions ${\boldsymbol\gamma}_j(\cdot)$. Similar asymptotic properties can be established by modifying the technical assumptions in Section 4.1.

\renewcommand{\theassumption}{C.\arabic{assumption}}\setcounter{assumption}{0}

\begin{assumption}\label{ass:C.1}

{\em The joint process $\{({\mathbb Y}_{t}, {\mathbb X}_{t}) \}$ is stationary and $\alpha$-mixing dependent with the mixing coefficient decaying at a geometric rate, where ${\mathbb Y}_{t}$ and ${\mathbb X}_{t}$ are defined in Assumption 6(i).}

\end{assumption}
	
\begin{assumption}\label{ass:C.2}

{\em The conditional density function of $e_{it}^\diamond=Y_{it}-{\mathbf X}_{it}^{^\intercal} {\boldsymbol\beta}_{i}(t/T)-\alpha_i$ given ${\mathbf X}_{it}={\mathbf x}$, $f_{ie}^\diamond(\cdot |{\mathbf x})$, exists and is positive with continuous and bounded first-order derivative. Furthermore, $\max_{1\leq i\leq N}\sup_{{\mathbf x}}f_{ie}^\diamond(0|{\mathbf x})\leq\overline{c}_e^\diamond<\infty$.}
	
\end{assumption}
	
\begin{assumption}\label{ass:C.3}
	
{\em (i)\ The matrix 
\[{\boldsymbol\Omega}_{i}^\diamond:={\sf E}\left[f_{ie}^\diamond(0|{\mathbf X}_{it})\left(
\begin{array}{c}
	1\\
	{\mathbf X}_{it}
\end{array}\right)
\left(1,{\mathbf X}_{it}^{^\intercal}\right)\right]\]
is positive definite with eigenvalues bounded away from zero and infinity uniformly over $1\leq i\leq N$. In addition, let $\max_{1\leq i\leq N}{\sf E}\left[\Vert {\mathbf X}_{it}\Vert^{\kappa+\epsilon}\right]<\infty$, where $4<\kappa<\infty$ and $\epsilon>0$ as in Assumption 3(i).
	
(ii)\ The coefficient functions ${\boldsymbol\beta}_i(\cdot)$ and ${\boldsymbol\gamma}_j(\cdot)$ satisfy the smoothness condition in Assumption 3(ii).}

\end{assumption}

\begin{assumption}\label{ass:C.4}
	
{\em Assumptions 4 and 5 in Section 4.1 hold with $f(z)\equiv1$ in Assumptions 4(ii) and 5(ii).}

\end{assumption}

\begin{assumption}\label{ass:C.5}

 {\em (i)\ The matrix ${\boldsymbol\Omega}_\diamond({\cal G}_j^\tau)$ defined in (\ref{eq5.6}) below is positive definite with eigenvalues bounded away from zero and infinity.
 	
 (ii)\ Define
		\[
		\Gamma_{t0}^\diamond({\cal G}_j^\tau)=\sum_{i\in{\cal G}_j^\tau}\eta_{it}^\diamond(z),\ \ {\boldsymbol\Gamma}_{t1}^\diamond({\cal G}_j^\tau)=\sum_{i\in{\cal G}_j^\tau}\eta_{it}^\diamond(z)\left[{\mathbf X}_{it}-{\boldsymbol\Omega}_{i,\diamond}^{\beta\alpha}/\omega_{i,\diamond}^{\alpha}\right],
		\]
		where $\eta_{it}^\diamond(z)=\tau-I\left(e_{it}^\diamond\leq -b_{it}(z)\right)$. There exists $\zeta_\diamond>2$ such that 
	\[{\sf E}\left[\left|\Gamma_{t0}^\diamond({\cal G}_j^\tau)-{\sf E}\left[\Gamma_{t0}^\diamond({\cal G}_j^\tau)\right]\right|^{\zeta_\diamond}\right]=O\left(N_j^{\zeta_\diamond/2}\right),\] \[ {\sf E}\left[\left\Vert{\boldsymbol\Gamma}_{t1}^\diamond({\cal G}_j^\tau)-{\sf E}\left[{\boldsymbol\Gamma}_{t1}^\diamond({\cal G}_j^\tau)\right]\right\Vert^{\zeta_\diamond}\right]=O\left(N_j^{\zeta_\diamond/2}\right).\]
	
 (iii)\ There exists a $d\times d$ matrix ${\boldsymbol\Lambda}_\diamond({\cal G}_j^\tau)$ such that
		\[\frac{1}{N_j}\sum_{t=\lfloor Tz\rfloor-L}^{\lfloor Tz\rfloor +L}{\sf E}\left[\big({\boldsymbol\Gamma}_{\lfloor Tz\rfloor1}^\diamond({\cal G}_j^\tau)-{\sf E}\big[{\boldsymbol\Gamma}_{\lfloor Tz\rfloor1}^\diamond({\cal G}_j^\tau)\big]\big)\big({\boldsymbol\Gamma}_{t1}^\diamond({\cal G}_j^\tau)-{\sf E}\left[{\boldsymbol\Gamma}_{t1}^\diamond({\cal G}_j^\tau)\right]\big)^{^\intercal}\right]\rightarrow {\boldsymbol\Lambda}_\diamond({\cal G}_j^\tau)\]
		as $N_j, L\rightarrow\infty$, where $\lfloor\cdot\rfloor$ denotes the floor function.}

\end{assumption}

\renewcommand{\theprop}{C.\arabic{prop}}\setcounter{assumption}{0}

\begin{prop}\label{prop:C.1}

{\em Suppose that Assumptions \ref{ass:C.1}--\ref{ass:C.4} are satisfied, and $N=O(T^{\kappa/2})$. Then, (4.8)--(4.10) hold for the time-varying coefficient panel quantile regression.}

\end{prop}

\noindent{\bf Proof of Proposition \ref{prop:C.1}}.\ \ Similarly to the proof of Theorem 4.1, we may show that (4.8) holds for the time-varying coefficient quantile regression when the number of groups $R_0$ is pre-specified. Then, as in the proof of Theorem 4.2, we can also show that (4.9) holds. Combining both (4.8) and (4.9), we may show that (4.10) in Corollary 4.1 holds.\hfill$\blacksquare$

\medskip

For $i=1,\cdots,N$, let $f_{ie}^\diamond(\cdot |{\mathbf x})$ be the conditional density of $e_{it}^\diamond=Y_{it}-{\mathbf X}_{it}^{^\intercal} {\boldsymbol\beta}_{i}(t/T)-\alpha_i$ given ${\mathbf X}_{it}={\mathbf x}$, and write
\[
{\boldsymbol\Omega}_{i}^\diamond={\sf E}\left[f_{ie}^\diamond(0|{\mathbf X}_{it})\left(
\begin{array}{c}
	1\\
	{\mathbf X}_{it}
\end{array}\right)
\left(1,{\mathbf X}_{it}^{^\intercal}\right)\right]=\left(
\begin{array}{cc}
	\omega_{i,\diamond}^\alpha&{\boldsymbol\Omega}_{i,\diamond}^{\alpha\gamma}\\
	{\boldsymbol\Omega}_{i,\diamond}^{\gamma\alpha}& {\boldsymbol\Omega}_{i,\diamond}^\gamma
\end{array}
\right),
\]
where $\omega_{i,\diamond}^\alpha$ is univariate and ${\boldsymbol\Omega}_{i,\diamond}^\gamma$ is a $d\times d$ matrix. Define
\begin{equation}\label{eq5.6}
	{\boldsymbol\Omega}_\diamond({\cal G}_j)=\frac{1}{\vert{\cal G}_j\vert}\sum_{i\in{\cal G}_j}\left[{\boldsymbol\Omega}_{i,\diamond}^\gamma-{\boldsymbol\Omega}_{i,\diamond}^{\gamma\alpha}{\boldsymbol\Omega}_{i,\diamond}^{\alpha\gamma}/{\omega}_{i,\diamond}^{\alpha}\right],\ \ j=1,\cdots,R_0.
\end{equation}
The following proposition gives the asymptotic normality for the post-grouping local linear estimates of the time-varying coefficients.

\begin{prop}\label{prop:C.2}

 {\em Suppose that Assumptions \ref{ass:C.1}--\ref{ass:C.5} and 6(ii) are satisfied. For $z$, an interior point of $[0,1]$, and $j=1,\cdots,R_\tau$, we have
	\begin{equation}\label{eq5.7}
		\sqrt{N_jTh_1}\left[\widetilde{\boldsymbol\gamma}_j(z)-{\boldsymbol\gamma}_j(z)-{\mathbf B}_j(z)\right]\stackrel{d}\longrightarrow {\sf N}\left({\bf 0}_d,\ \left[{\boldsymbol\Omega}_\diamond({\cal G}_j^\tau)\right]^{-1}\left[\nu_0{\boldsymbol\Lambda}_\diamond({\cal G}_j^\tau)\right]\left[{\boldsymbol\Omega}_\diamond({\cal G}_j^\tau)\right]^{-1}\right),
	\end{equation}
	where $N_j$ and ${\mathbf B}_j(z)$ are defined as in Theorem 4.3, and ${\boldsymbol\Lambda}_\diamond({\cal G}_j^\tau)$ is defined in Assumption \ref{ass:C.5}(iii).}
	
\end{prop}

\noindent{\bf Proof of Proposition \ref{prop:C.2}}.\ \ The proof is similar to the proof of Theorem 4.3. Details are omitted here to save the space.\hfill$\blacksquare$


\subsection*{Appendix C.3:\ \ Construction of the point-wise confidence intervals}

In this appendix, we assume that the panel observations are independent over subjects (but may be not identically distributed), to derive a more explicit form of the asymptotic variance matrix in Theorem 4.3 and facilitate construction of its estimation. With the cross-sectional independence assumption, we may show that
\begin{equation}\label{eq5.8}
{\boldsymbol\Lambda}(z;{\cal G}_j)=\frac{\tau(1-\tau)}{|{\cal G}_j|}\sum_{i\in{\cal G}_j}{\sf var}\left({\mathbf X}_{it}-{\boldsymbol\Omega}_i^{\gamma\alpha}(z)/\omega_i^\alpha(z)\right).
\end{equation}

 The estimated density function for the index variable $Z_t$ can be obtained by the kernel smoothing method, denoted by $\widetilde f(z)$. Letting $\widetilde{e}_{it}=Y_{it}-{\mathbf X}_{it}\widehat{\boldsymbol\beta}_i(Z_t)-\widehat\alpha_i(Z_t)$, we estimate $f_{ie}(0 |{\mathbf x},z)$ by 
\[
\widetilde{f}_{ie}(0|{\mathbf x},z) =  \frac{ \sum_{t=1}^T K_{b_1}(\widehat{e}_i){\mathbf{K}}_{b_1} ({\mathbf X}_{it}-{\mathbf x}) K_{b_1}(Z_t-z) }{b_1\sum_{t=1}^T {\mathbf{K}}_{b_1} ({\mathbf X}_{it}-{\mathbf x}) K_{b_1}(Z_t-z)},
\]
where $b_1$ is a bandwidth (determined by the rule of thumb) and 
\[
{\mathbf{K}}_{b_1}({\mathbf u})=\prod_{j=1}^dK(u_j/b_1),\ \ {\mathbf u}=(u_1,\cdots,u_d)^{^\intercal}.
\]
Then, for ${\boldsymbol\Omega}_{i}(z)$ defined in (4.1), we estimate it again by the kernel-smoothing method:
\begin{eqnarray}
\widetilde{\boldsymbol\Omega}_{i}(z)&=&\widetilde f(z)\sum_{t=1}^T \widetilde f_{ie}(0|{\mathbf X}_{it},Z_t)\left(
	\begin{array}{c}
		1\\
		{\mathbf X}_{it}
	\end{array}\right)
	\left(1,{\mathbf X}_{it}^{^\intercal}\right)K_{b_2}(Z_t-z)/\sum_{t=1}^TK_{b_2}(Z_t-z)\nonumber\\
	&=&\left(
	\begin{array}{cc}
		\widetilde\omega_i^\alpha(z)&\widetilde{\boldsymbol\Omega}_{i}^{\alpha\gamma}(z)\\
		\widetilde{\boldsymbol\Omega}_{i}^{\gamma\alpha}(z)& \widetilde{\boldsymbol\Omega}_{i}^\gamma(z)
	\end{array}
	\right),\nonumber
\end{eqnarray}
where $b_2$ is a bandwidth. By (4.6) and (\ref{eq5.8}), it is natural to estimate the matrices ${\boldsymbol\Omega}(z;{\cal G}_j)$ and ${\boldsymbol\Lambda}(z;{\cal G}_j)$ by 
\begin{eqnarray}
	&&\widetilde{\boldsymbol\Omega}(z;{\cal G}_j)=\frac{1}{\vert\widetilde{\cal G}_j\vert}\sum_{i\in\widetilde{\cal G}_j}\left[\widetilde{\boldsymbol\Omega}_i^\gamma(z)-\widetilde{\boldsymbol\Omega}_i^{\gamma\alpha}(z)\widetilde{\boldsymbol\Omega}_i^{\alpha\gamma}(z)/\widetilde{\omega}_i^{\alpha}(z)\right],\label{eq5.10}\\
	&&\widetilde{\boldsymbol\Lambda}(z;{\cal G}_j)=\frac{\tau(1-\tau)}{\vert\widetilde{\cal G}_j\vert}\sum_{i\in\widetilde{\cal G}_j}\widetilde{\sf var}\left({\mathbf X}_{it}-\widetilde{\boldsymbol\Omega}_i^{\gamma\alpha}(z)/\widetilde\omega_i^\alpha(z)\right),\label{eq5.11}
\end{eqnarray}
respectively, where $\widetilde{\sf var}\left({\mathbf X}_{it}-\widetilde{\boldsymbol\Omega}_i^{\gamma\alpha}(z)/\widetilde\omega_i^\alpha(z)\right)$ is the sample variance of ${\mathbf X}_{it}-\widetilde{\boldsymbol\Omega}_i^{\gamma\alpha}(z)/\widetilde\omega_i^\alpha(z)$ (over $t$). With (\ref{eq5.10}) and (\ref{eq5.11}), we estimate the asymptotic variance matrix ${\boldsymbol\Sigma}(z;{\cal G}_j)$ in Theorem 4.3 by 
\[
\widetilde{\boldsymbol\Sigma}(z;{\cal G}_j)=\left[\widetilde{\boldsymbol\Omega}(z;{\cal G}_j)\right]^{-1}\left[\nu_0\widetilde{\boldsymbol\Lambda}(z;{\cal G}_j)\right]\left[\widetilde{\boldsymbol\Omega}(z;{\cal G}_j)\right]^{-1}.\]

The asymptotic bias term ${\mathbf B}_j(z)$ in Theorem 4.3 may be estimated by $\widetilde{\mathbf B}_j(z)=\frac{1}{2}h_1^2\mu_2\widetilde{\boldsymbol\gamma}_j^{\prime\prime}(z)$, where $\widetilde{\boldsymbol\gamma}_j^{\prime\prime}(z)$ denotes the estimate of ${\boldsymbol\gamma}_j^{\prime\prime}(z)$ via local quadratic quantile regression \citep[e.g.,][]{CP02}. However, the convergence of $\widetilde{\boldsymbol\gamma}_j^{\prime\prime}(z)$ to ${\boldsymbol\gamma}_j^{\prime\prime}(z)$ is often slow in finite samples. To simplify our construction of the confidence intervals, we ignore the bias term by appropriate undersmoothing. In practice, we may adopt the so-called jackknife bias correction \citep[e.g.,][]{X98} to further reduce the asymptotic bias order and asymptotically ignore the estimation bias.

Let $u_l$ be a $d$-dimensional vector with the $l$-th element being one and the others being zeros. For $\alpha\in(0,1)$, the $100(1-\alpha)\%$ confidence interval of $\gamma_{j,l}(z)=u_l^{^\intercal}{\boldsymbol\gamma}_j(z)$ is constructed as
\[
\left[\widetilde\gamma_{j,l}(z)-c_{1-\alpha/2}\left(\frac{u_l^{^\intercal} \widetilde{\boldsymbol\Sigma}(z;{\cal G}_j)u_l}{N_jTh_1}\right)^{1/2},\ \widetilde\gamma_{j,l}(z)+c_{1-\alpha/2}\left(\frac{u_l^{^\intercal} \widetilde{\boldsymbol\Sigma}(z;{\cal G}_j)u_l}{N_jTh_1}\right)^{1/2}\right],
\]
where $\widetilde\gamma_{j,l}(z)=u_l^{^\intercal}\widetilde{\boldsymbol\gamma}_j(z)$ and $c_{1-\alpha/2}$ is the $(1-\alpha/2)$-quantile of the standard normal distribution.

\bigskip

\renewcommand{\theequation}{D.\arabic{equation}}
\renewcommand{\thefigure}{D.\arabic{figure}}
\setcounter{equation}{0}
\setcounter{figure}{0}

\section*{Appendix D:\ \ Functional-coefficient mean regression based grouping results for UK house price data}

For comparison purposes, in this appendix we provide latent group estimation results based on the functional-coefficient mean regression for the UK house price data. 

\begin{figure}
	\begin{center}
	{\includegraphics[height=8.6cm]{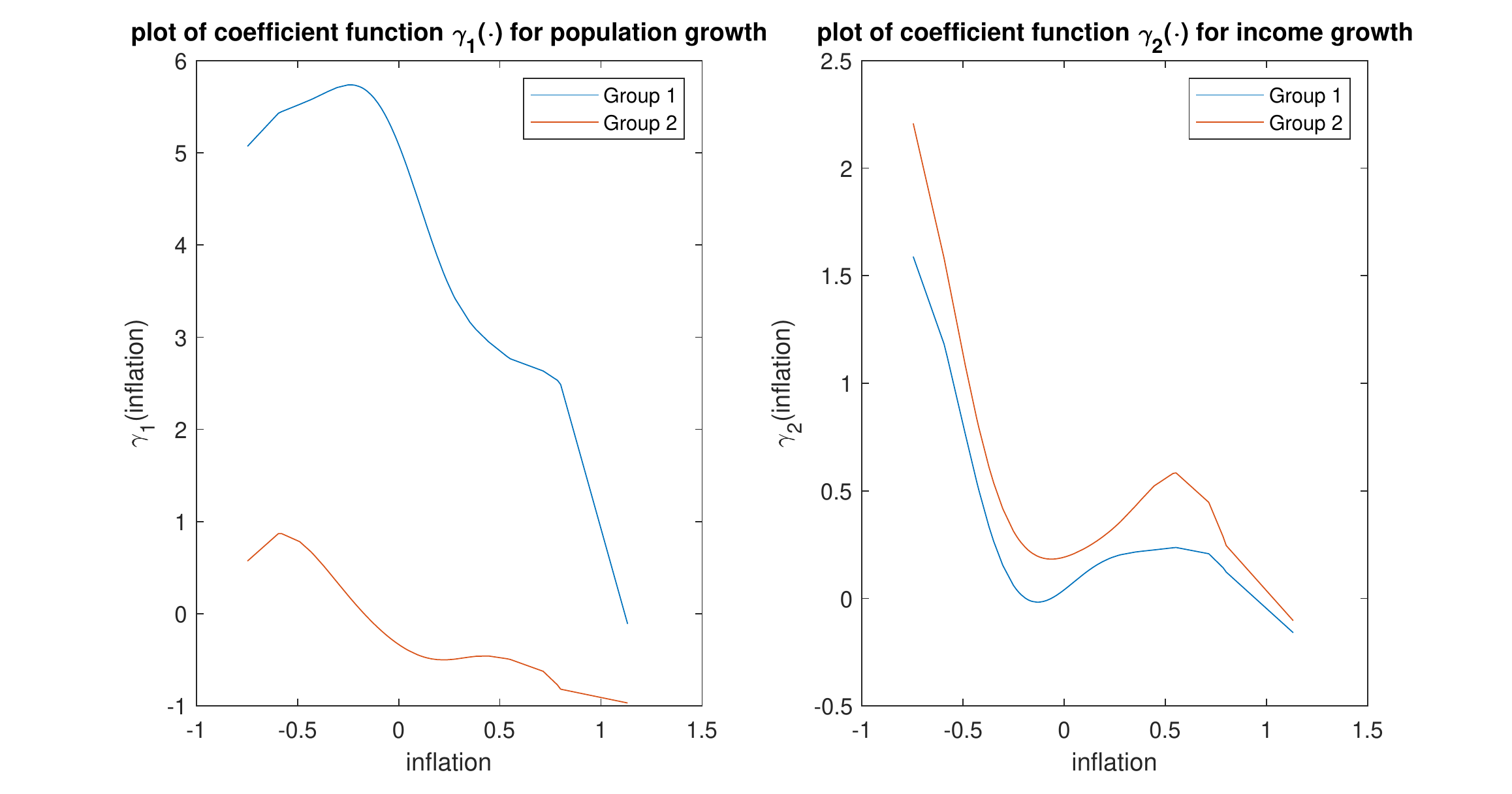}}
	\caption{\small \label{figD.1}Mean-regression based post-grouping local linear estimates of the functional coefficients for population growth ($\gamma_1(\cdot)$, left) and income growth ($\gamma_2(\cdot)$, right): blue line - for Group 1; red line - for Group 2.}
	\end{center}
\end{figure}

By applying the same ratio criterion and the clustering algorithm as detailed in Section 3 of the main text to functional coefficients estimated from the mean regression, we find 2 latent groups with Group 1 containing 97 LADs and Group 2 containing 238 LADs.  The membership of the two groups is similar to that of the two groups discovered from the quantile regression at $\tau=0.50$, although they are not exactly the same (as the results from the mean regression are more susceptible to the influence of outliers). The shapes of the group-specific functional coefficients from the mean regression and the quantile regression at $\tau=0.50$ are also similar. Figure \ref{figD.1} plots the estimated group-specific functional coefficients from the mean regression.

Note that mean regression analysis provides only the average effects of the explanatory variables on the dependent variable. On the other hand, quantile regression allows us to uncover possible different effects of the explanatory variables at different quantiles across the distribution of the dependent variable. Furthermore, quantile regression is more robust than mean regression and hence, its results are less affected by outliers.

\end{document}